\documentclass[]{article}
\usepackage{amsfonts}
\usepackage{bm}
\usepackage[margin=1in]{geometry}
\usepackage[affil-it]{authblk}
\usepackage{graphicx}
\usepackage{subcaption}
\usepackage{bmpsize}
\usepackage{epstopdf}
\usepackage[square, numbers, comma, sort]{natbib}
\usepackage{amsmath}
\usepackage{xcolor}
\usepackage[toc,page]{appendix}
\usepackage{booktabs}
\usepackage{hyperref}

\renewcommand{\vec}[1]{\boldsymbol{#1}}

\renewcommand\appendix{\par
	\setcounter{section}{0}%
	\setcounter{subsection}{0}%
	\setcounter{table}{0}
	\setcounter{figure}{0}
	\gdef\thetable{\Alph{table}}
	\gdef\thefigure{\Alph{figure}}
	\section*{Appendix}
	\gdef\thesection{\Alph{section}}
	\setcounter{section}{1}}
\begin{document}
\title{Tilted resonators in a triangular elastic lattice: chirality, Bloch waves and negative refraction}
\author{Domenico Tallarico%
	\footnote{Corresponding author. Office 405,	Mathematical Sciences Building, The University of Liverpool, L69 7ZL, Liverpool, United Kingdom. E-mail: \texttt{domenico.tallarico@liverpool.ac.uk}}}
\author{Natalia V. Movchan}
\author{Alexander B. Movchan}
\author{Daniel J. Colquitt}
\affil{Department of Mathematical 
	Sciences,
	Mathematical Sciences Building,The University of Liverpool, L69 7ZL, Liverpool, United Kingdom.}
\date{}
\maketitle
\begin{abstract}
We consider a vibrating triangular mass-truss lattice whose unit cell contains a resonator of a triangular shape. The resonators are connected to the triangular lattice by trusses. Each resonator is tilted, \emph{i.e.} it is rotated with respect to the triangular lattice's unit cell through an angle $\vartheta_0$. This geometrical parameter is responsible for the emergence of a resonant mode in the Bloch spectrum for elastic  waves and strongly affects the dispersive properties of the lattice. Additionally, the tilting angle  $\vartheta_0$ triggers the opening  of a band gap at a Dirac-like point.  We provide a physical interpretation of these phenomena and discuss the dynamical implications on elastic Bloch waves. The dispersion properties are used to design a structured interface containing tilted resonators which  exhibit  negative refraction and focussing,  as in a ``flat elastic lens''.  
 \end{abstract}
%!TEX root = draft_comments.tex

\section{Introduction\label{sec:introduction}}
This paper introduces the novel concept of dynamic rotational degeneracy for  Bloch-Floquet waves in periodic multi-scale media.
The model is introduced in a concise, analytically tractable manner, and accompanied by numerical simulations; we place particular emphasis on the vibrational modes that represent rotational standing waves. 
Special attention is given to the phenomena of dynamic anisotropy and negative refraction in the context of multi-scale materials.

A geometric figure is said to be chiral if ``its image in a plane mirror, ideally realised, cannot be brought to coincide with itself''~\cite{Lord_Kelvin}.
Typical examples are the hands which are either ``left-handed'' or ``right-handed''.
Indeed, the etymological origin of the word ``chiral'' derives from the ancient Greek for hand: $\chi\varepsilon\acute{\iota}\rho$.
In an earlier paper~\cite{Brun_2012}, Brun \emph{et al.} introduced an analytical model for chiral elastic media, which takes into account internal rotations induced by gyroscopes.
These structured media act as polarisers of elastic waves and also have very interesting dispersive properties in the context of the Bloch-Floquet waves in multi-scale periodic solids possessing internal rotations.
Further studies of chiral structures have lead to exciting and novel results~\cite{Bigoni_PRB_87_174303_2013,Carta_2014,Wang_PRL_113_014301_2014}. Hexagonal chiral lattices were studied by Spadoni \emph{et al.} \cite{Spadoni_WM_46_435_2009,Spadoni_2012}. Their  numerical results highlighted the influence of chirality  on the dynamic anisotropy of elastic in-plane waves and on the auxetic behaviour of the structure under static loads. Moussavi \emph{et al.} \cite{Mousavi_2015} demonstrated that  breaking the spatial mirror symmetry can be used to study ``topologically protected" elastic waves in metamaterials. 

An elastic lattice, with the inertia concentrated at the nodal points, is a highly attractive object for this studies of Bloch-Floquet waves, as the dispersion equations can be expressed as polynomials with respect to a spectral parameter.
The paper by Martinsson and Movchan \cite{Martinsson_QJMAM_56_45_2003} presented a unified analytical approach for studies of Bloch-Floquet waves in multi-scale truss and frame structures, which included problems of the design of stop bands around pre-defined frequencies. In particular, low-frequency rotational waveforms were identified in some of the standing waves at the boundaries of the stop bands in the spectrum. 

The asymptotic analysis of eigenvalue problems for degenerate and non-degenerate multi-structures was systematically presented in the monograph by Kozlov \emph{et al.} \cite{Kozlov_1999}, where rotational modes were studied in the context of the asymptotic analysis of the eigenvalues and corresponding eigenfunctions of multi-structures consisting of components of different limit dimensions. 

In the last two decades, significant progress has been made in gaining control of elastic waves at the micro-scale.
This advance is primarily due to recent improvements in micro- and nano-fabrication as well as the novel theoretical developments, such as focussing through negative refraction.
The use of negative refraction for focussing was studied theoretically for the first time by Pendry \cite{Pendry_2000}.
Later, Luo \emph{et al.} \cite{Luo_PRB_65_201104_2002} exploited the dynamic anisotropy of a photonic crystal to achieve all-angle negative refraction and focussing without a negative effective index of refraction.
The same idea has been exploited in the context of phononic crystals for both elastic~\cite{Ke_PRB_2005} and water~\cite{Hu_PRE_2004} waves as well as waves in lattice systems \cite{Colquitt2011,Colquitt_2012}.
Finally, we would like to mention the pivotal contribution made by Craster \emph{et al.} \cite{Craster_2010} in the analysis and control of the effective dynamic properties of phononic and photonic crystals through high frequency homogenisation techniques.  
\begin{figure}[h]
	\centering
	\includegraphics[width=1\textwidth]{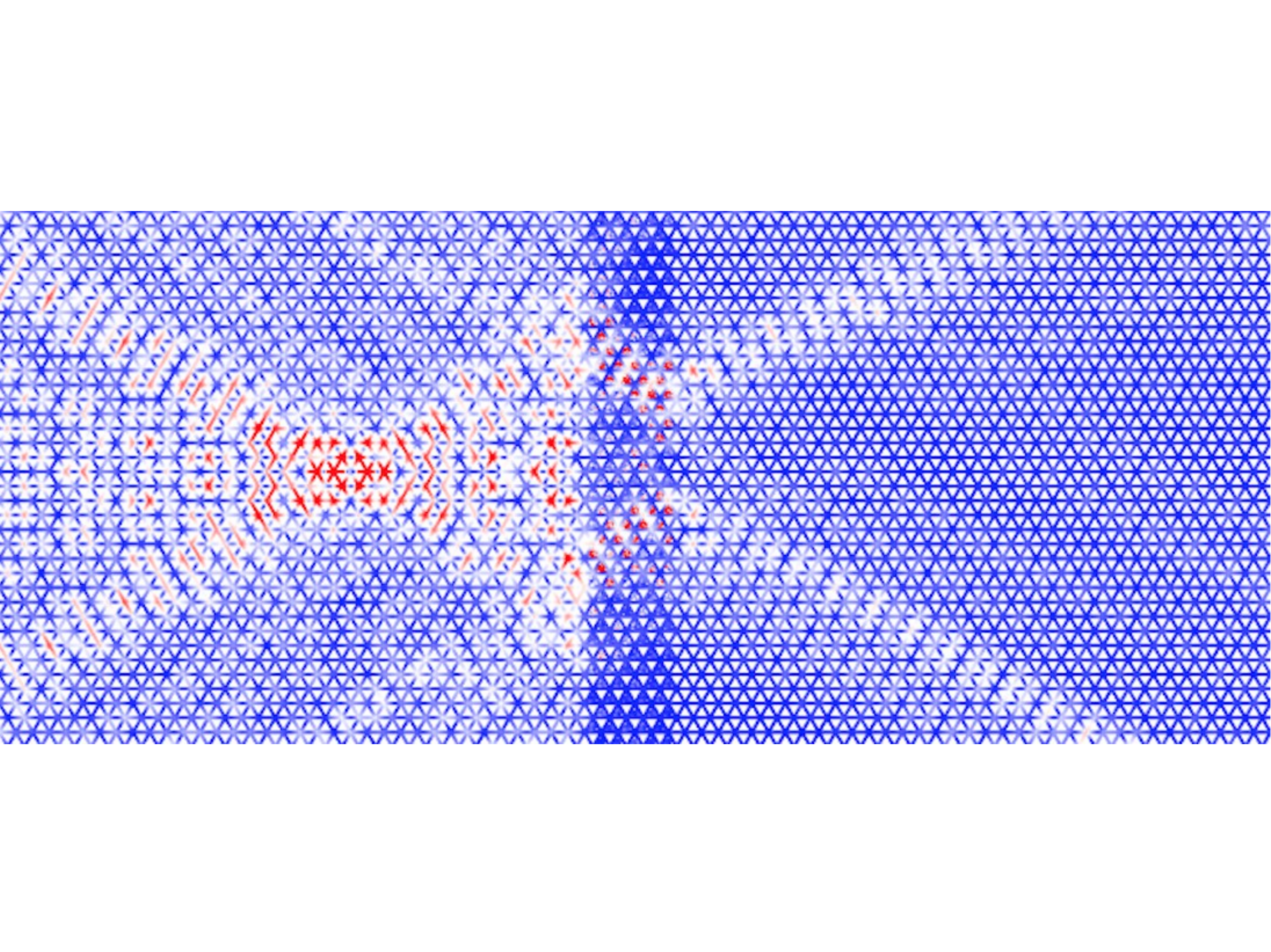}
	\caption{\label{fig:intro} Elastic waves exhibit  negative refraction  at the interfaces between a monoatomic triangular lattice and a triangular lattice containing \emph{tilted inertial resonators}.}
\end{figure}

In the present paper, we study the dynamic properties of both degenerate and non-degenerate multi-scale lattice structures in the context of the modelling and analysis of meta-materials with pre-designed dynamic properties. For example, we will  design and implement structured interfaces that exhibit negative refraction of  elastic waves and can be used to focus mechanical waves, as shown in Fig. \ref{fig:intro}.

The structure of the paper is as follows.
The formulation of the spectral problem together with the  governing equations for Bloch-Floquet waves are described in section \ref{sec:bloch}.
For simplicity and ease of exposition we consider a regular triangular lattice with embedded resonators as shown in Fig. \ref{fig:system_t}(a). 
In section \ref{sec:model_resonator} we specify the chiral geometry of the resonator and analyse its dynamic behaviour. A discussion of the dispersion equations and standing waves, and a detailed analysis of the dispersion diagrams, which show the radian frequency as a function of two components of the Bloch vector is provided in section~\ref{sec:disp_results}.
In particular, we identify  standing waves
and examine the effects of the orientation of the resonator on the dispersive properties of the lattice.
Slowness contours and results concerning  dynamic anisotropy are also presented in section~\ref{sec:disp_results}.
In section~\ref{sec:modulation} we examine the dynamic behaviour of the lattice in which the orientation of the resonators is non-uniform. In particular, the case where adjacent resonators are rotated by angles of opposite signs is considered. 

Transmission problems are of particular interest, as they enable one to see the effects of negative refraction for rotational waveforms in metamaterials.
An example of a chiral interface is illustrated in Fig. \ref{fig:intro}, which incorporates a structured layer with rotational resonators of a special design, and shows negative refraction.
This and other examples are discussed in section \ref{sec:transmission}. Finally, in section \ref{sec:conclusions} we draw together our main conclusions.
%!TEX root = draft_comments.tex

\section{The spectral problem \label{sec:bloch}}

Before studying the dynamical properties of the lattice system, it is necessary to introduce some notation, the geometry and the equations of motion. For the sake of simplicity, we restrict ourselves to the study of triangular lattices; such lattices are particularly interesting as they are statically isotropic but exhibit strong dynamic anisotropy at finite frequencies. Subsequently, we specify the geometry and physical parameters of the resonators embedded in the triangular lattice unit cell - see Fig. \ref{fig:system_sr}. We conclude this section by stating the form of the equation of motion for a general two-dimensional micro-structured lattice made of point masses and massless trusses.    

\subsection{The triangular lattice}
\begin{figure}
	\includegraphics[width=0.5\textwidth]{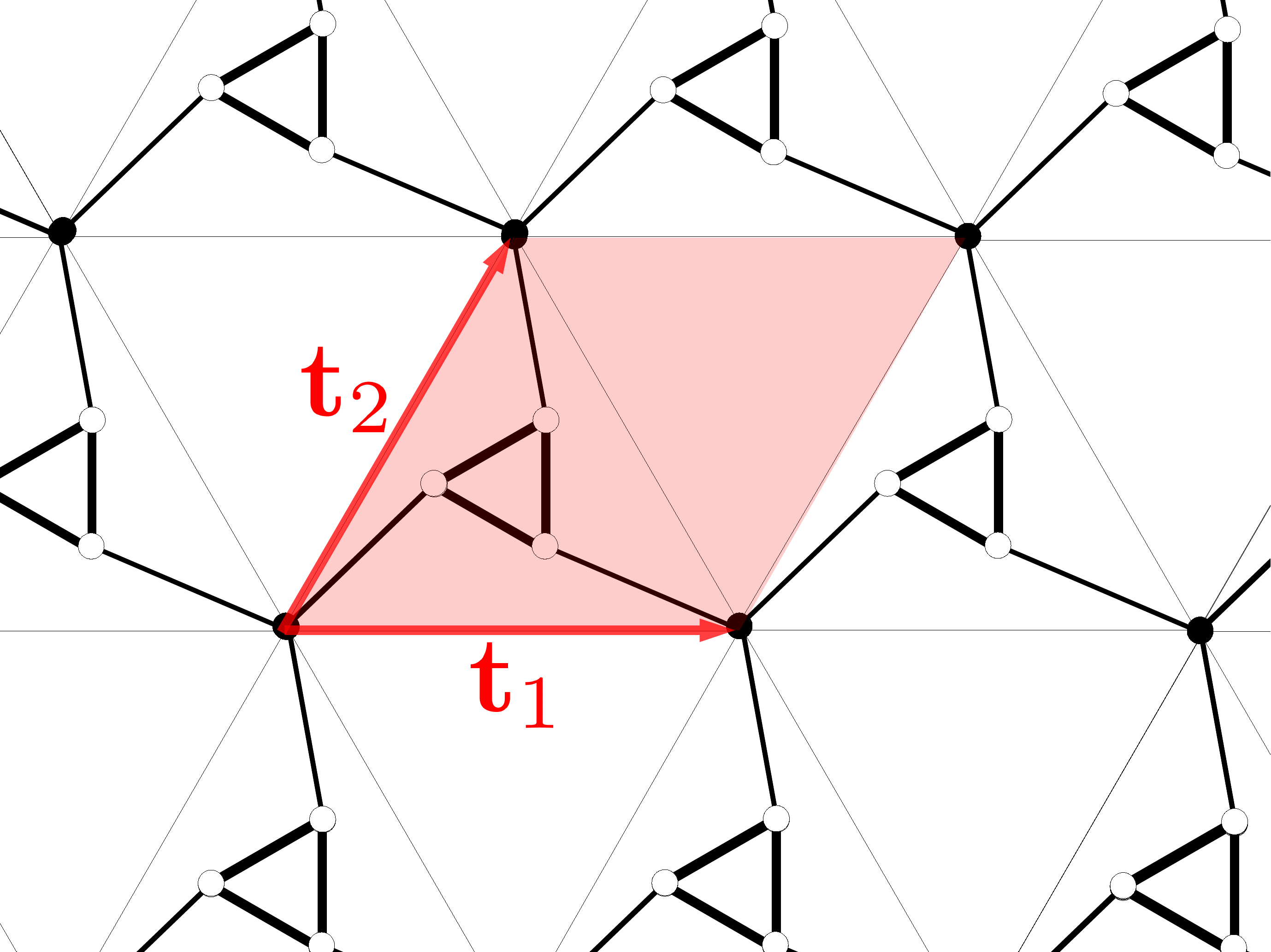}\hfill
	\centering 
	\includegraphics[width=0.5\textwidth]{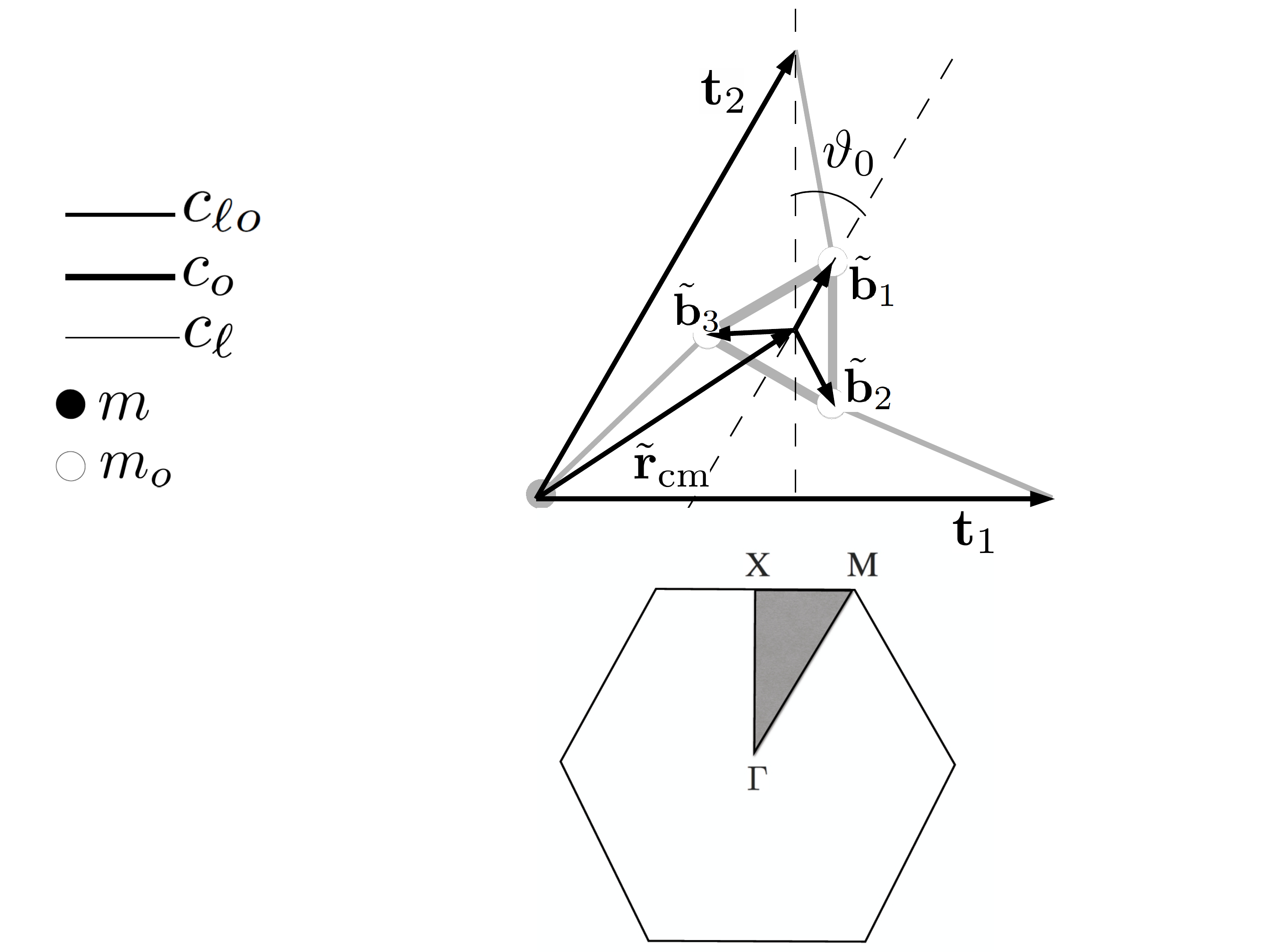}\hfill
	$\rm \,\,\,\,\,\,\,\,\,\,\,\,\,\,\,\,\,\,\,\,\,\,\,\,\,\,\,\,\,\,\,\,\,\,\,\,\,\,\,\,\,\,\,\,\,\,\,\,\,\,\,\,\,\,\,\,\,\,\,\,\,\,\,\,\,\,\,\,\,\,\,\,\,\,\,\,\,\,\,\,\,\,\,\,\,\,\,\,\,\,\,\,\,\,\,\,\,\,\,\,\,\,\,\,\,\,\,\,\,\,\,(a)\,\,\,\,\,\,\,\,\,\,\,\,\,\,\,\,\,\,\,\,\,\,\,\,\,\,\,\,\,\,\,\,\,\,\,\,\,\,\,\,\,\,\,\,\,\,\,\,\,\,\,\,\,\,\,\,\,\,\,\,\,\,\,\,\,\,\,\,\,\,\,\,\,\,\,\,\,\,\,\,\,\,\,\,\,\,\,\,\,\,\,\,\,\,\,\,\,\,\,\,\,\,\,\,\,\,\,\,\,\,\,\,(b)\,\,\,\,\,\,\,\,\,\,\,\,\,$ 
	\centering
	\caption{\label{fig:system_t}Panel (a): A schematic representation of the triangular elastic lattice  containing resonators; the unit cell of the lattice is highlighted in red. Panel (b), top:  The vectors ${\bm t}_1$ and ${\bm t}_2$ are the primitive vectors of the triangular mass-truss lattice; the vector $\tilde{\bf{r}}_{\rm cm}$ describes the rest position of the centre of mass of the resonator; the tilting angle $\vartheta_0$ describes the rotation of the resonator about $\tilde{\bm r}_{\rm cm}$; the vectors $\tilde{\bf  b}_1$,$\tilde{\bf  b}_2$ and $\tilde{\bf  b}_3$ identify the rest positions for the masses $m_o$, relatively to $\tilde{\bf{r}}_{\rm cm}$. Panel (b), bottom: The first Brillouin zone for the triangular lattice and irreducible fraction (grey shaded  region).}
\end{figure}
%\begin{figure}
%\centering
%\subfloat[]{\includegraphics[width=0.5\textwidth]{surf_BT_osc_no_stif.pdf}}
%\subfloat[]{\includegraphics[width=0.5\textwidth]{surf_BT_osc_no_mass.pdf}}
%\caption{\label{fig:disp_t_osc_no_stif}Triangular lattice with $m_o=m=1$, $c=1$ and $c_o=0$.}
%\end{figure}
A schematic representation of the periodic lattice is shown in Fig. \ref{fig:system_t}(a), which corresponds to the unit cell depicted in Fig. \ref{fig:system_t}(b).
The triangular lattice (TL) has the primitive vectors
\begin{equation}\label{eq:t_unit_vects}
	{\bm t}_1=
	\begin{pmatrix}
		1\\
		0
	\end{pmatrix} L\,\,\,\,{\rm and }\,\,\,\,
	{\bm t}_2=
	\begin{pmatrix}
		1\\
		\sqrt{3}
	\end{pmatrix}\frac{L}{2},
\end{equation}
whose norm is $L$.
It is convenient to introduce the auxiliary vector ${\bm t}_3={\bm t}_1-{\bm t}_2$.

At equilibrium, the point masses $m$ (black solid  dots in Fig. \ref{fig:system_t} (a)) occupy the triangular lattice sites 
\begin{equation}\label{eq:u_eq_lat_t}
	\tilde{\bf{x}}^{({\bm n})}_{0}=n_1{\bm t}_1+n_2{\bm t}_2,
\end{equation} 
where ${\bm n}=(n_1,n_2)^{\rm T}$ is a pair of integers.
The basis vectors for the reciprocal lattice are 
\begin{equation}\label{eq:G_unit_vects}
	{\bf G}_1=\frac{2\pi}{L}
	\begin{pmatrix}
		1\\
		-1/\sqrt{3}
	\end{pmatrix} \,\,\,\,{\rm and }\,\,\,\,
	{\bf G}_2=\frac{2\pi}{L}
	\begin{pmatrix}
		0\\
		2/\sqrt{3}
	\end{pmatrix},
\end{equation}
and the associated first Brillouin zone is shown in Fig. \ref{fig:system_t}(b), together with the high symmetry points, which lie at
\begin{equation}\label{eq:GammaMX}
	{\rm \Gamma}= 
	\begin{pmatrix}
		0 \\
		0 \\
	\end{pmatrix}
	\,\,\,\,\,
	{\rm M}= 
	\frac{2\pi}{\sqrt{3}L}\begin{pmatrix}
		1/\sqrt{3}\\
		1\\
	\end{pmatrix}
	\,\,\,{\rm and}\,\,\,
	{\rm X}= 
	\frac{2\pi}{\sqrt{3}L}\begin{pmatrix}
		0\\
		1\\
	\end{pmatrix}.
\end{equation}
In this paper, unless otherwise stated, Bloch frequency surfaces are presented as a function the Bloch wave vector such that
\begin{equation}\label{eq:k_BZ}
	{\bm k}=
	\begin{pmatrix}
		k_x\\
		k_y
	\end{pmatrix}
	\in\left[-\frac{4\pi}{3L},\frac{4\pi}{3L}\right]^2.
	%=k_1{\bf G}_1+k_2{\bf G}_2
\end{equation} 
The square region defined by Eq. (\ref{eq:k_BZ})  encloses the first Brillouin zone represented in Fig. \ref{fig:system_t}(b).  
\subsection{Geometry of a resonator} 
The centre of mass of the triangular resonator in Fig. \ref{fig:system_t}(b) is located  at the point with the position vector
\begin{equation}\label{eq:rcm_eq_t}
	\tilde{\bf{r}}_{\rm cm}=
	\begin{pmatrix}
		1\\
		1/\sqrt{3}
	\end{pmatrix}\frac{L}{2}.
\end{equation}
%schematically reported in Fig. \ref{fig:system_t}. relative to the nodal point $\tilde{\bf{x}}^{({\bm n})}_{0}$. 
The vector $\tilde{\bf{b}}_i$ with $i=\{1,2,3\}$ shown in  Fig. \ref{fig:system_t}(b), is the position vector  of the $i^{\rm th }$ mass relative to the centre of mass in Eq. (\ref{eq:rcm_eq_t}). The explicit expression is
\begin{equation}\label{eq:b_i_eq}
	\tilde{\bf{b}}_{i}=b \hat{\cal R}_i \tilde{\pmb{\beta}}_{1}=
	b \hat{\cal R}_i\begin{pmatrix}
		\sin \vartheta_0\\
		\cos\vartheta_0
	\end{pmatrix},\,\,\,\,\,{\rm with}\,\,\,\,\,\hat{\cal R}_i=\left.\hat{\cal R}_{\vartheta}\right|_{\vartheta=2\pi(i-1)/3},
\end{equation} 
where  $\vartheta_{0}$ is the tilting angle, $b=\ell/\sqrt{3}$, and
\begin{equation}\label{eq:r_matrix}
	\hat{\cal R}_{\vartheta}=
	\begin{pmatrix}
		\cos{\vartheta} & \sin{\vartheta} \\
		-\sin{\vartheta} & \cos{\vartheta} \\
	\end{pmatrix}
\end{equation}
is the clockwise rotation matrix. It follows that the $i ^{\rm th}$ mass belonging to the \emph{tilted inertial resonator}  (TIR), embedded in an arbitrarily chosen  unit cell ${\bm n}$, is located at  
\begin{equation}\label{eq:u_eq_res_t}
	{\bf  \tilde{x}}_{i}^{({\bm n})}=\tilde{\bf{x}}_{0}^{({\bm n})}+\tilde{{\bf r}}_{\rm cm}+\tilde{{\bf b}}_{i},
\end{equation}
where the vectors on the right-hand side of the equation, are given in Eqs (\ref{eq:u_eq_lat_t}), (\ref{eq:rcm_eq_t}) and  (\ref{eq:b_i_eq}). 

The vector linking the triangular lattice to the $i^{\rm th}$ mass of the TIR in the reference cell ${\bm n}={\bm 0}$ is
\begin{equation}\label{eq:alpha_i}
	\tilde{\bm \alpha}_i=\hat{\cal R}_i
	\tilde{\pmb{\alpha}}_{1},\,\,\,\,{\rm with}\,\,\,\,\tilde{\pmb{\alpha}}_{1}=\bm{t}_2 - \tilde{\bm{x}}_1^{({\bm 0})}= 
	\begin{pmatrix}
		b \sin \vartheta_0 \\
		-(B-b\cos\vartheta_0 )
	\end{pmatrix},
\end{equation}
where $B=L/\sqrt{3}$ and the matrix $\hat{\cal R}_i$ is introduced in Eq. (\ref{eq:b_i_eq}). For every $i=\{1,2,3\}$, the norm of the vectors in Eq. (\ref{eq:alpha_i}) is
\begin{equation}\label{eq:link_len}
	\ell_r = ||\tilde{\bm \alpha}_i||= \frac{1}{\sqrt{3}} \sqrt{L^2+\ell^2-2\ell L\cos(\vartheta_0)}.
\end{equation}

Given the set of vectors  Eqs (\ref{eq:t_unit_vects}), (\ref{eq:b_i_eq}) and (\ref{eq:alpha_i}), we introduce the corresponding projector matrices 
\begin{align}\label{eq:projectors}
	\hat{\tau}_1&=\frac{1}{L^2} {\bm t}_1 {\bm t}_1^{\rm T},\,\,\,\,\,\,
	\hat{\tau}_2=\frac{1}{L^2} {\bm t}_2 {\bm t}_2^{\rm T},\,\,\,\,\,\,
	\hat{\tau}_3=\frac{1}{L^2}\left( {\bm t}_1- {\bm t}_2\right) \left( {\bm t}_1- {\bm t}_2\right)^{\rm T}, \,\,\,\,\,\,\nonumber\\
	\hat{\pi}_1 &= \frac{1}{\ell^2}
	\left( \tilde{\bm{b}}_3-\tilde{\bm{b}}_2 \right) 
	\left( \tilde{\bm{b}}_3-\tilde{\bm{b}}_2 \right)^{\rm T},\,\,\,\,\,\,
	\hat{\pi}_2 = \frac{1}{\ell^2}
	\left( \tilde{\bm{b}}_1-\tilde{\bm{b}}_3 \right) 
	\left( \tilde{\bm{b}}_1-\tilde{\bm{b}}_3 \right)^{\rm T},\,\,\,\,\,\,
	\hat{\pi}_3 = \frac{1}{\ell^2}
	\left( \tilde{\bm{b}}_1-\tilde{\bm{b}}_3 \right) 
	\left( \tilde{\bm{b}}_1-\tilde{\bm{b}}_3 \right)^{\rm T},\nonumber \\
	\hat{\Pi}_i &=\frac{1}{\ell_r^2} \tilde{\bm{\alpha}}_i\tilde{\bm{\alpha}}_i^{\rm T},
\end{align}
where $i=\{1,2,3\}$.  The notation ${\bm v}{\bm u}^{\rm T}$ in Eqs (\ref{eq:projectors})  denotes the dyadic product ${\bm v}\otimes{\bm u}$ of two vectors ${\bm u}$ and ${\bm v}$.
\subsection{The truss structure}

The TL is formed from an array of point masses located at $\tilde{\vec{x}}_0^{(\vec{n})}$ and connected to neighboring masses by thin elastic rods.
The rods are assumed to be massless and extensible, with longitudinal stiffness $c_{\ell}$, but not flexible, i.e. they are trusses.
The links are represented in Fig. \ref{fig:system_t}(a) by thin solid lines.
Additionally, each mass is connected to three resonators by rods of the same type as the ambient lattice and longitudinal stiffness $c_{\ell o}$;
these links are denoted by the solid lines of intermediate thickness in Fig. \ref{fig:system_t}(a).

The resonators themselves are composed of three point masses of magnitude $m_o$, located at vertices of a smaller equilateral triangle, and connected by massless trusses of longitudinal stiffness $c_o$;
these trusses are indicated by the thick solid lines in Fig. \ref{fig:system_t}(a). The TIRs are constrained such that they do not cross the links of the exterior lattice or those connecting the resonator to the ambient lattice. It is easy to demonstrate that the constraints  
\begin{equation}\label{eq:geom_cond}
	0<\frac{\ell}{L}<\frac{1}{2}\,\,\,\,\,\,\,\,\,{\rm and }\,\,\,\,\,\,\,\,\,|\vartheta_0|<\vartheta_{\rm max}={\rm arcos}\left(\frac{\ell}{L}\right),
\end{equation}
are sufficient to ensure the aforementioned conditions.\color{black}

The solution to the  Bloch-Floquet problem associated with a general lattice unit cell was studied by Martinsson and  Movchan \cite{Martinsson_QJMAM_56_45_2003} in 2003.
For the case of massless trusses studied here, the Bloch-Floquet problem for time-harmonic waves of angular frequency $\omega$ can be reduced \cite{Martinsson_QJMAM_56_45_2003}  to the algebraic system 
%\begin{equation}
%\omega^2M^\kappa {\bm u}^{(\bm{n},\kappa)}= - \sum_{{\bm m}\in{ {\cal B}_\kappa }}\left[ \cal{A}_{11}^{\kappa,{\bm m},\lambda}{\bm u}^{({\bm n},\kappa)} + {\cal A}_{11}^{\kappa,{\bm m},\lambda}{\bm u}^{({\bm n},\kappa)} \right]
%\end{equation} 
\begin{equation}\label{eq:secular_prime}
	\left[ \hat{\Sigma}'_{\bm k}-\omega^2 \hat{{\cal M}}'\right]{\bm U}'_{\bm k}=0,
\end{equation}
where $\hat{\Sigma}'_{\bm k}$ is a Hermitian positive semi-definite matrix containing information about the geometry of the lattice and stiffness of the links and depends on the Bloch wave-vector $\vec{k}$.
The eigenvector ${\bm U}'_{\bm k} \in\mathbb{C}^{d\cdot q}$ defines the displacement amplitudes of the masses; for a lattice with truss-like links $d$ is the spatial dimension and $q$ is the number of masses per unit cell.
The matrix $\hat{{\cal M}}'$ is diagonal with entries corresponding to the magnitude of the masses in the unit cell.
Finally, the multiplicative factor $e^{-i\omega t}$ is omitted but understood.

For the geometry considered here, and shown in Fig. \ref{fig:system_t}(a), $d=2$ and $q=4$ and we therefore have an eight-dimensional algebraic problem~\eqref{eq:secular_prime}.
The column vector ${\bm U}'_{\bm k}$, introduced in~\eqref{eq:secular_prime} may be written in the form
\begin{equation}\label{eq:U_prime_k}
	{\bm U}'_{\bm k} = 
	\begin{pmatrix}
		{\bm u}^{\rm T}_{0}(\bm k),  & {\bm u}^{\rm T}_{1}(\bm k),  & {\bm u}^{\rm T}_{2}(\bm k),  & {\bm u}^{\rm T}_{3}(\bm k)  \\
	\end{pmatrix}^{\rm T},
\end{equation} 
where  ${\bm u}_i({\bm k})$ with $i=\{0,1,2,3\}$ are the displacement amplitude vectors  of the masses.

\section{A model resonator\label{sec:model_resonator}}
\begin{figure}
	\centering
	\begin{subfigure}[t]{0.4\linewidth}
		\includegraphics[width=\linewidth]{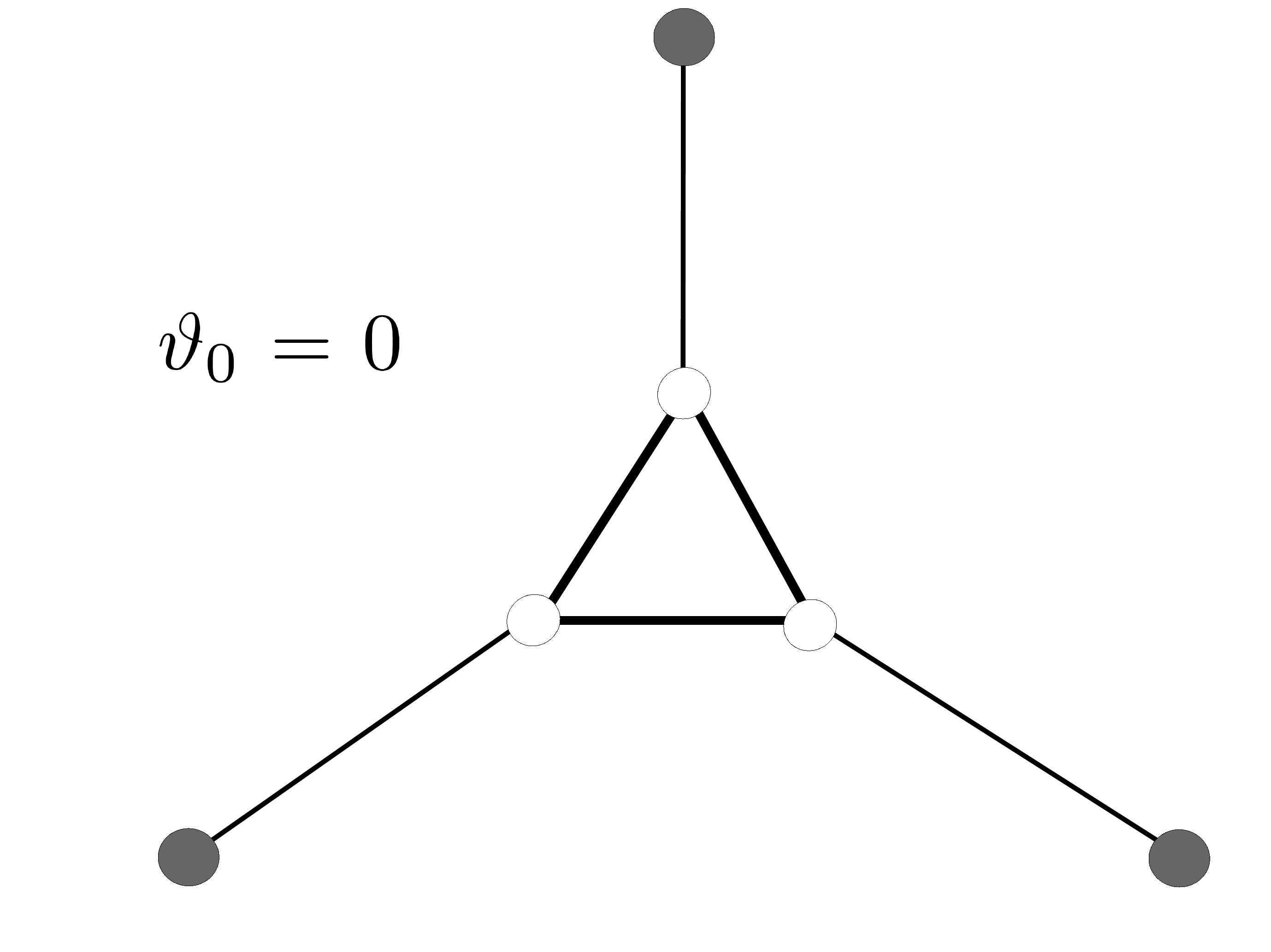}
		\caption{}
	\end{subfigure}
	\qquad
	\begin{subfigure}[t]{0.4\linewidth}
		\includegraphics[width=\linewidth]{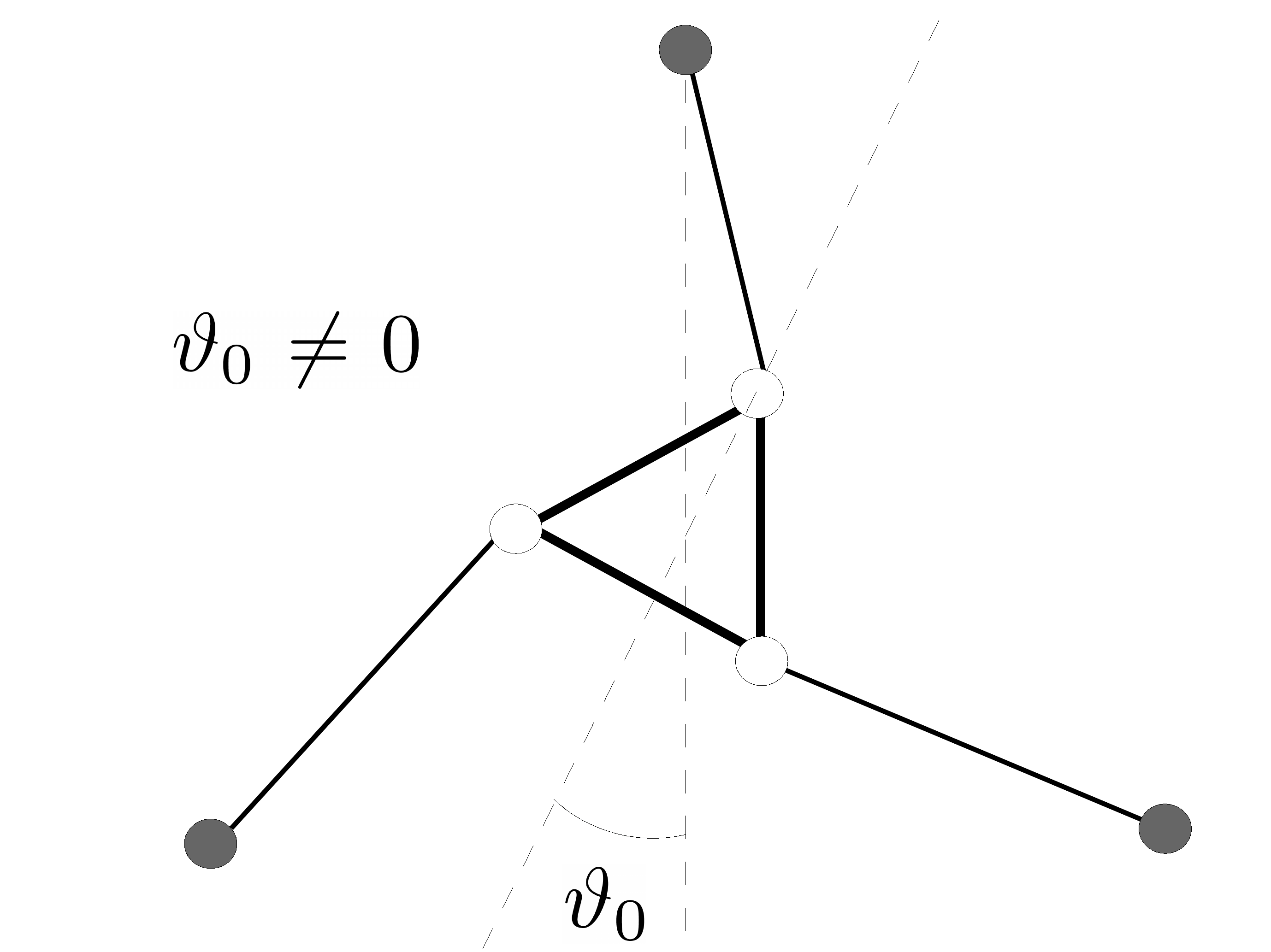}
		\caption{}
	\end{subfigure}
	\caption{\label{fig:system_sr}Schematic representations of a single hinged resonator. Panel (a): The degenerate case corresponding to $\vartheta_0=0$.  Panel (b): The non-degenerate case corresponding to $\vartheta_0\neq0$.}
\end{figure}
In this section, we analyse a single hinged  resonator  (see Fig. \ref{fig:system_sr}). In this context, we identify a degeneracy associated with $\vartheta_0=0$, shown in Fig \ref{fig:system_sr}(a).  We show that the degeneracy disappears when a non-zero tilting angle is considered - see the chiral geometry in Fig. \ref{fig:system_sr}(b). We assume that the resonator is a rigid-body  and  present the corresponding natural frequencies. We conclude this section with a physical interpretation of the degeneracy corresponding to non-tilted resonators.

%The stiffness matrix for Bloch waves  as introduced in Eq. (\ref{eq:secular_prime}) is then presented in its full analytical form. It is demonstrated that the degeneracy, identified for the eigenfrequencies of a single resonator, persists in the dispersion diagram of Bloch waves. 

We consider a single TIR and assume hinged conditions at the vertices of an equilateral triangle of side length $L$ (see Fig. \ref{fig:system_sr}(a)). The equations of motion for the time-harmonic displacements of the masses at the vertices of the TIR are
\begin{equation}\label{eq:secular_hinged}
	\left[\hat{\sigma}'-\omega^2m_0 \hat{I}_6 \right]\bm{U}=\bm{0},
\end{equation}
where $\bm{U}=(\bm{u}_1,\bm{u}_2,\bm{u}_3)^{\rm T}$ contains the aforementioned displacement amplitudes, $m_o$ is the mass at every vertex, $\omega$ is the angular frequency and $\hat{\sigma}'$ is the stiffness matrix and $\hat{I}_6$ is  the $6\times6$ identity matrix.  In Eq. (\ref{eq:secular_hinged}), the stiffness matrix is 
\begin{align}\label{eq:sigma_prime_hinged}
	\hat{\sigma}'=
	\begin{pmatrix}
		c_{\ell o}\hat{\Pi}_1 + c_{o} \hat{\pi}_2+ c_o \hat{\pi}_3 &- c_o \hat{\pi}_3 &  -c_{ o} \hat{\pi}_2 \\
		-c_{ o} \hat{\pi}_3 & c_{\ell o }\hat{\Pi}_2+c_o\hat{\pi}_1+c_o\hat{\pi}_3 &  -c_{ o} \hat{\pi}_1 \\
		-c_{ o} \hat{\pi}_2 & -c_{o} \hat{\pi}_1 &  c_{\ell o}\hat{\Pi}_3+ c_o\hat{\pi}_2+c_o\hat{\pi}_1
	\end{pmatrix},
\end{align} 
where the projectors $\hat{\Pi}_i$ and $\hat{\pi}_i$, are given in Eq. (\ref{eq:projectors}). A derivation of Eqs (\ref{eq:secular_hinged}) and (\ref{eq:sigma_prime_hinged}) is provided in Appendix \ref{sec:app_soft}.\\

\subsection{The non-degenerate case}
\begin{figure}
	\centering
	\begin{subfigure}[t]{0.32\linewidth}
		\includegraphics[width=\linewidth]{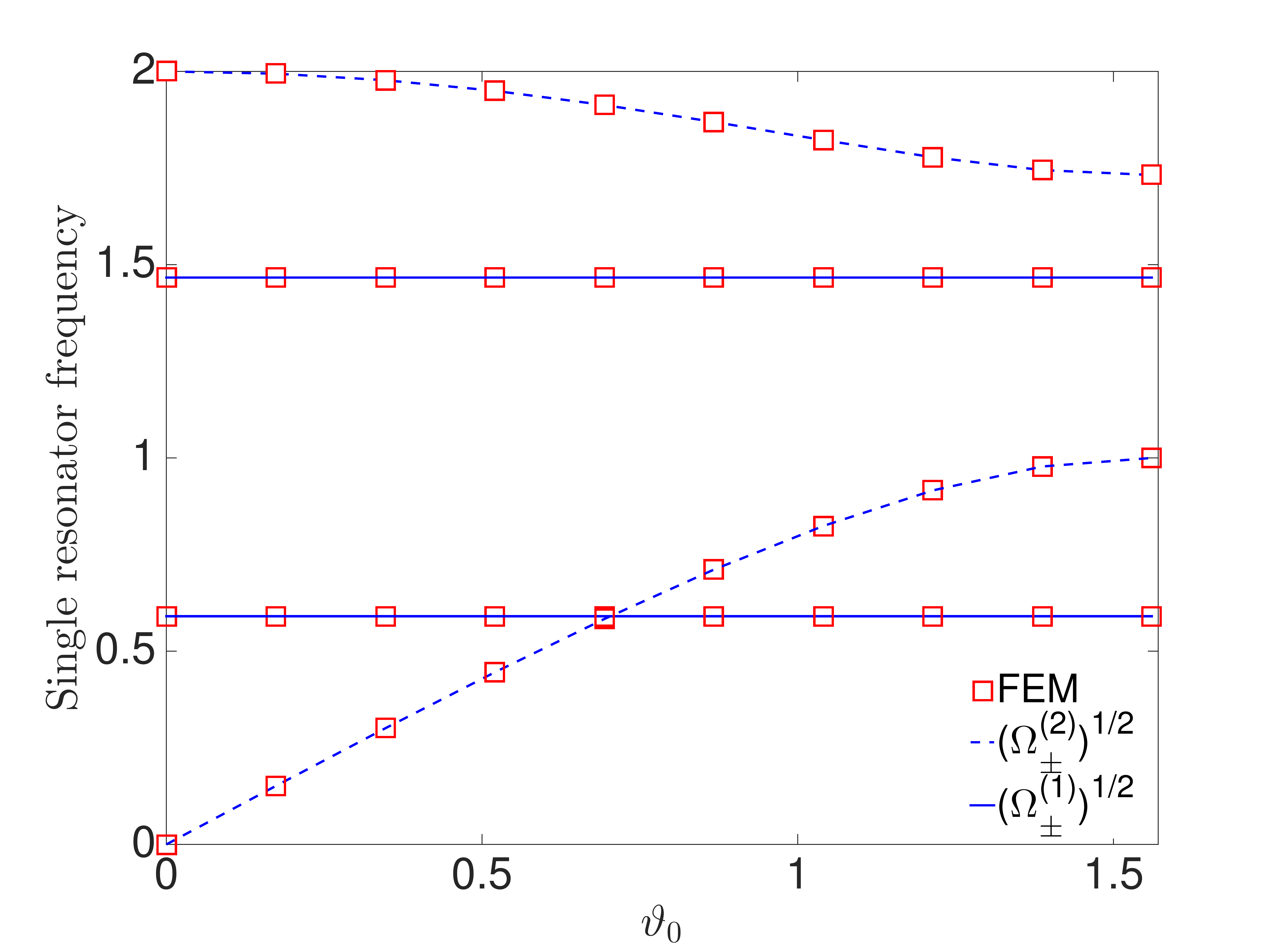}
		\caption{}
	\end{subfigure}
	\begin{subfigure}[t]{0.32\linewidth}
		\includegraphics[width=\linewidth]{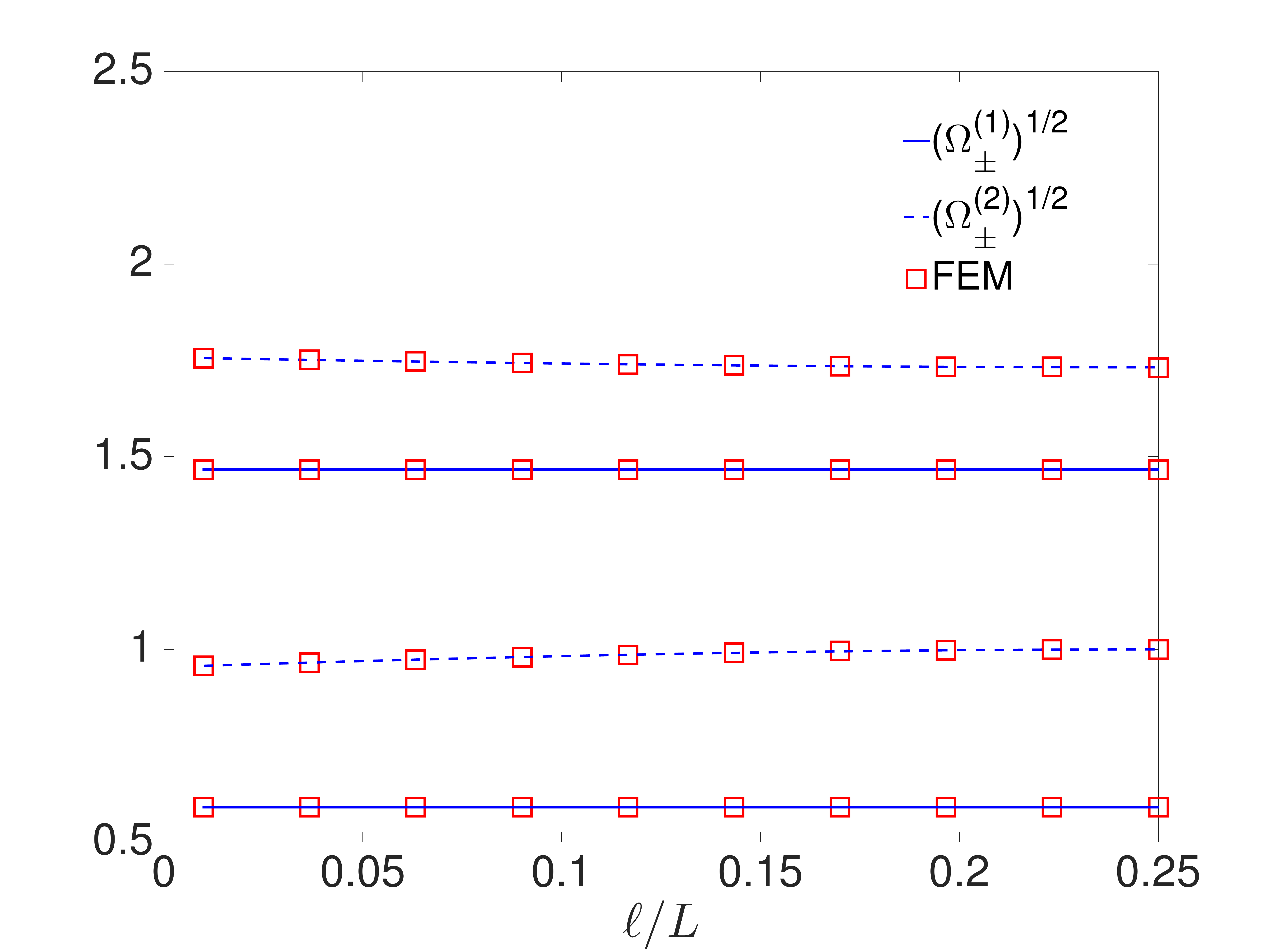}
		\caption{}
	\end{subfigure}
	\begin{subfigure}[t]{0.32\linewidth}
		\includegraphics[width=\linewidth]{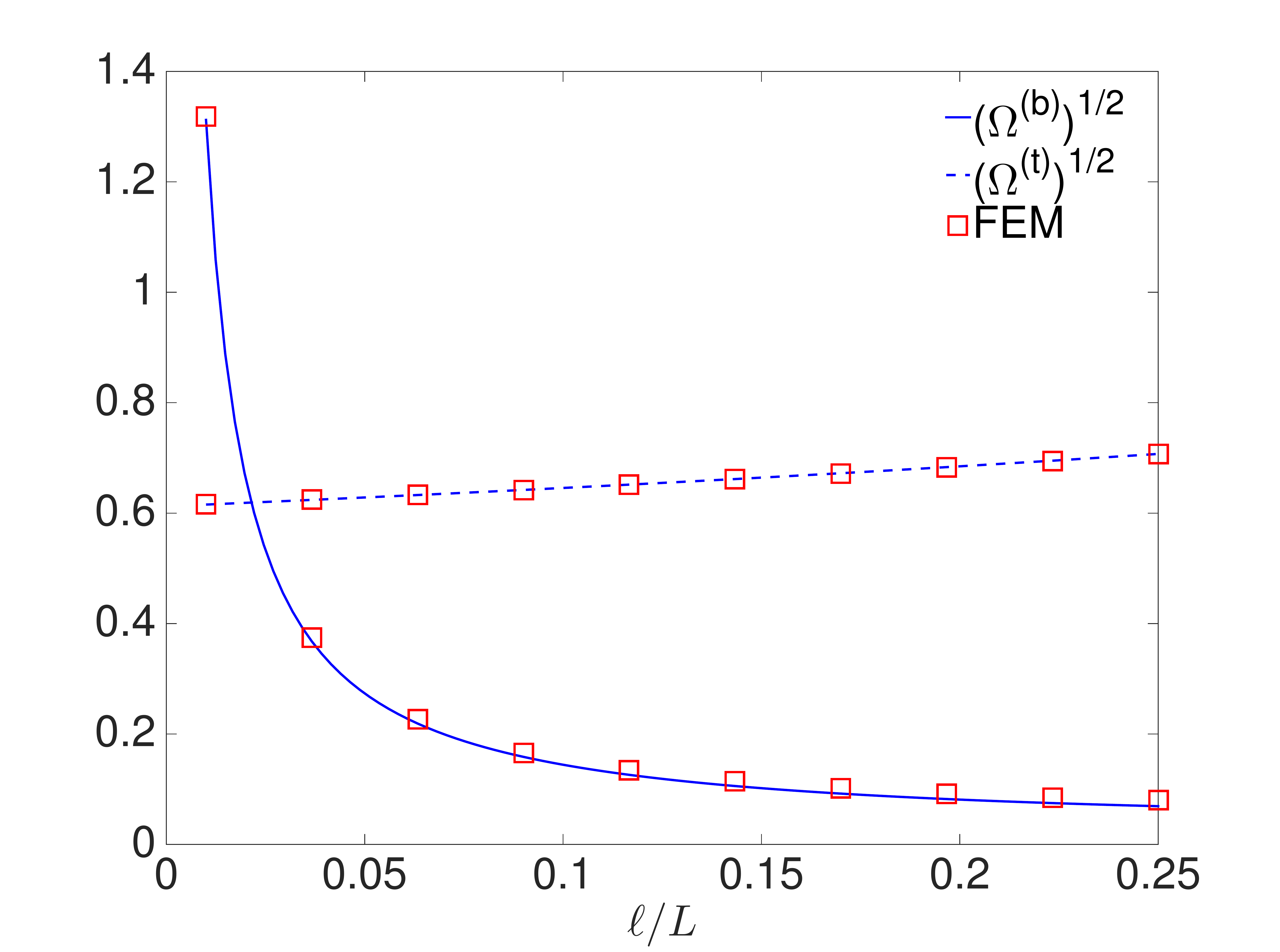}
		\caption{}
	\end{subfigure}
	\caption{\label{fig:single_resonator} The natural frequencies of single resonator. In panels (a) and (b) we compare Eqs. (\ref{eq:eigf_squared}) (blue lines) with  FEM results (red squares). In panel (a), we fix $\ell\approx0$ and vary $\vartheta_0$. In panel (b), we fix $\vartheta_0\approx1.32$ and  vary $\ell$. The remaining parameters are $m_o=1$, $L=1$, $c_o=c_{\ell o}=1$. In panel (c) we compare the analytical predictions (\ref{eq:eigf_squared_bending}) and (\ref{eq:eigf_bending_trasl}) (solid blue and dashed blue line, respectively) with FEM results (red squares) as a function of $\ell$. We fix the Young's modulus to be $E_0\approx1.38~10^3$ and the cross-sectional radius  $r\approx0.01$. The other parameters are $m_o=1$ and $L=1$.}
\end{figure}
We use the notation $\Omega=\omega^2$. For $\vartheta_0\neq0$, solving the eigenvalue problem (\ref{eq:secular_hinged}) with stiffness matrix (\ref{eq:sigma_prime_hinged}) gives 
\begin{align}\label{eq:eigf_squared}
	\Omega^{(1)}_{\pm} & =\frac{1}{2}\frac{c_{\ell o}}{m_o}\left[1+\frac{3}{2}\frac{c_o}{c_{\ell o}} \pm \sqrt{1+\frac{9}{4}\left(\frac{c_o}{c_{\ell o}}\right)^2 } \right],\nonumber\\
	\Omega^{(2)}_{\pm} & = \frac{1}{2}\frac{c_{\ell o} + 3 c_o}{m_o}\left[ 1\pm\sqrt{1-4 \frac{L^2}{\ell_r^2} \frac{c_o c_{\ell o}}{(c_{\ell o}+3 c_o)^2}\sin^2 \vartheta_0} \right],
\end{align}
where the first two eigenvalues have multiplicity two, and the last two eigenvalues are of multiplicity one; $\ell_r$ is given by \eqref{eq:link_len}. This demonstrates that for $\vartheta_0\neq0$ both the structure and the corresponding stiffness matrix are non-degenerate.  For very small resonators  $\ell/L\rightarrow0$, $\Omega_\pm^{(2)}$ has the limit  %the eigenvalue  does not affect $\Omega^{(1)}_\pm$ in Eq. (\ref{eq:eigf_squared}). 
\begin{equation}\label{eq:eigf_squared_l_small}
	\lim_{\ell/L\rightarrow0}\Omega^{(2)}_{\pm} = \frac{1}{2}\frac{c_{\ell o} + 3 c_o}{m_o}\left[ 1\pm\sqrt{1- \frac{12c_o c_{\ell o}}{(c_{\ell o}+3 c_o)^2}\sin^2 \vartheta_0} \right],
\end{equation}
whereas $\Omega^{(1)}_\pm$ is independent of $\ell/L$. 
The absence of degeneracy at non-zero tilting angles and as $\ell$ is varied, is illustrated by Figs \ref{fig:single_resonator}(a) and \ref{fig:single_resonator}(b), respectively. In Fig. \ref{fig:single_resonator}(a), we compare the frequencies (\ref{eq:eigf_squared}) (blue lines) with FEM numerical results obtained using ${\rm COMSOL\, Multiphysics}^{\circledR}$ (red squares), as a function of $\vartheta_0$  and for $\ell\simeq0$.  In Fig. \ref{fig:single_resonator}(b), a similar comparison is done by fixing $\vartheta_0\simeq1.32$ and varying $\ell$. 
\subsection{The degenerate case}

For $\vartheta_0=0$, we observe that $\Omega^{(2)}_-=0$, as illustrated in Fig. \ref{fig:single_resonator}(a) . Therefore, the structure represented in Fig. \ref{fig:system_sr}(a) is degenerate, that is 
\begin{equation}\label{eq:degeneracy}
	{\rm det}\left(\hat{\sigma}'\right)=0,
\end{equation}
where the stiffness matrix $\hat{\sigma}'$ is introduced in Eq. (\ref{eq:sigma_prime_hinged}). Degeneracy is obtained also in the trivial limits for a very soft resonator, \emph{i.e.} $c_o/c_{\ell o}\rightarrow0$, and non-zero tilting angle. In this case, we get  $\Omega_{-}^{(1)}=\Omega_{-}^{(2)}=0$.
%\begin{equation}
%\lim_{c_o/c_{\ell o}\rightarrow0}\Omega_{-}^{(1)}=\lim_{c_o/c_{\ell o}\rightarrow0}\Omega_{-}^{(2)}=0
%\end{equation}

\subsection{The rigid-body approximation\label{subsec:rba}}
%\begin{figure}
%	\centering
%\includegraphics[width=0.5\textwidth]{omega_theta_res.pdf}\hfill
%\includegraphics[width=0.5\textwidth]{theta_star.pdf}
%		$\rm (a)\,\,\,\,\,\,\,\,\,\,\,\,\,\,\,\,\,\,\,\,\,\,\,\,\,\,\,\,\,\,\,\,\,\,\,\,\,\,\,\,\,\,\,\,\,\,\,\,\,\,\,\,\,\,\,\,\,\,\,\,\,\,\,\,\,\,\,\,\,\,\,\,\,\,\,\,\,\,\,\,\,\,\,\,\,\,\,\,\,\,\,\,\,\,\,\,\,\,\,\,\,\,\,\,\,\,\,\,\,\,\,\,\,\,\,\,\,\,\,\,\,\,\,\,\,\,\,\,\,(b)$ 
%	\caption{\label{fig:omega_theta_res} Panel (a): ratio of the  eigenfrequencies in Eq. (\ref{eq:omega_res})  as a function of the equilibrium angle $\vartheta_{0}$.  Panel (b): representation of Eq. (\ref{eq:theta_star}) as a function of $\ell/L$.}
%\end{figure}
Here we assume that  
\begin{equation}\label{eq:cond_stiff}
	c_o/c_{\ell o} \gg 1\,\,\,\,\,{\rm and}\,\,\,\,\, c_o/c_{\ell} \gg 1,
\end{equation}
\emph{i.e.} the TIR's trusses are inextensible. In turn, this implies that the lengths of the links connecting  the vertices of the TIR are fixed to $\ell$. Specifically, three spatial degrees of freedom, corresponding to the relative motion of the three vertices of the TIRs, do not contribute to the propagation of Bloch waves. Therefore, in the limit (\ref{eq:cond_stiff}), Eq. (\ref{eq:secular_hinged}) can be further simplified. In fact, the motion of a single hinged and rigid resonator can be described by a time-harmonic in-plane displacement of the centre of mass and an angular displacement about the centre of mass. The natural frequencies of the system are 
\begin{equation}\label{eq:eigs_rigid}
	\Omega_{\rm cm}=\frac{3}{2} \frac{c_{\ell o}}{M}{\rm \,\,\,and \,\,\,\,}\Omega_{\vartheta}=\frac{c_{\ell o}\ell^2}{I}\frac{\sin^2\vartheta_{0}}{1+\ell^2/L^2-2\ell/L\cos{\vartheta_0}} . 
\end{equation} 
where $M=3m_o$ and $I=m_o \ell^2$. In Eq. (\ref{eq:eigs_rigid}), $ \Omega_{\rm cm}$ corresponds to the displacement of the centre of mass and is consistent with earlier results where, instead of the structured resonator used in this article, a concentrated mass $M=3m_o$ is assumed \cite{Martinsson_QJMAM_56_45_2003}. The role of the resonator's microstructure emerges in the eigenfrequency $\Omega_{\vartheta}$, which describes the angular displacement about the centre of mass. We observe that 
\begin{equation}\label{eq:limits_eigs_squared}
	\Omega_{\rm cm}=\lim_{c_{ o}\rightarrow+\infty}\Omega_\pm^{(1)},  {\rm \,\,\,\,\,\,\,} \Omega_{\rm \vartheta}=\lim_{c_{ o}\rightarrow+\infty}\Omega_-^{(2)}{\rm \,\,\,and\,\,\,\,}\lim_{c_{ o}\rightarrow+\infty}\Omega_+^{(2)}=+\infty,
\end{equation}
where $\Omega_{\pm}^{(1)}$ and $\Omega_\pm^{(2)}$ are eigenvalues given in Eq. (\ref{eq:eigf_squared}) for the hinged single resonator  with soft ligaments. Eqs (\ref{eq:limits_eigs_squared}) show that the eigenvalues for a single rigid hinged resonator can be retrieved as a limiting case from the eigenvalues in Eqs \eqref{eq:eigf_squared}. The additional diverging eigenvalue corresponds to the limiting case of high stiffness between the TIR's vertices and thus can be disregarded in the dynamics of the problem. In addition, we note that the degeneracy at zero tilting angle persists, as $\Omega_{\vartheta}=0$ for $\vartheta_0=0$. This observation further emphasise the fact that a  chiral geometry of the ligaments is a necessary condition to avoid degeneracy, if thin and non-bendable links are assumed.  We finally observe that the eigenvalues (\ref{eq:eigs_rigid}) coincide for the tilting angle 
\begin{equation}\label{eq:theta_star}
	\vartheta^{*}_{0}= \arccos{\frac{1}{2}\left(\frac{\ell}{L}+1\right)}.
\end{equation}  
We would like to remark that the degeneracy corresponding to Eq. (\ref{eq:degeneracy}), arises from  neglecting bending deformations of the hinged links. Conversely, a finite bending stiffness $B=E {\cal I}$, where $E$ represents the Young's modulus and ${\cal I}$ is the second moment of inertia of the beam, would lead to (see Appendix \ref{sec:app_single_bending})
\begin{equation}\label{eq:eigf_squared_bending}
	\Omega^{\rm (b)} = 9 \sqrt{3} \frac{L}{\ell^2(L-\ell)^2}\frac{B}{m_o}.
\end{equation}  
Eq. (\ref{eq:eigf_squared_bending}) is a non-zero frequency at zero tilting angle, corresponding to the rotational motion of the resonator in Fig. \ref{fig:system_sr}(a) with flexible hinged links. Fig. \ref{fig:single_resonator}(c)  shows agreement between Eq. (\ref{eq:eigf_squared_bending}) (solid blue line) and FEM numerical results (red squares),  as $\ell$ is varied. The translational frequency for a rigid resonator with hinged flexible links of Young's modulus $E$  and cross-sectional surface $S=\pi r^2$ can be directly obtained from the first of Eqs (\ref{eq:eigs_rigid}). Its expression is 
\begin{equation}\label{eq:eigf_bending_trasl}
	\Omega^{\rm (t)}=\frac{\sqrt{3}}{2} \frac{S}{L-\ell}\frac{E}{m_o},
\end{equation}
and it is compared  in Fig. \ref{fig:single_resonator}(c) (blue dashed line) against FEM computations (red squares) as a function of $\ell$. 
%!TEX root = draft_comments.tex

\section{Bloch waves in a lattice with tilted resonators\label{sec:disp_results}}
Having examined the behaviour of a single resonator in the previous section, we now proceed to consider the dispersive properties of a two-dimensional triangular lattice of resonators; the geometry is shown in Fig. \ref{fig:system_t}(a).
The primary object of study is the algebraic system~\eqref{eq:secular_prime}, which can be derived using the framework of Martinsson and Movchan~\cite{Martinsson_QJMAM_56_45_2003}; details are provided in Appendix~\ref{sec:app_rigid}.
With reference to~\eqref{eq:secular_prime}, the $8\times8$ stiffness matrix has the form
\begin{small}
	\begin{equation}\label{eq:Sigma_k_prime}
	\hat{\Sigma}'_{\bm k}=
	\begin{pmatrix}
	-\sum_{i=1}^3\left[2c_\ell ( \cos{\bm k}\cdot{\bm t}_i-1)\hat{\tau}_i-c_{\ell o} \hat{\Pi}_i\right]
	&-c_{\ell o} \hat{\Pi}_1e^{-i{\bm k}\cdot{\bm t}_2} &-c_{\ell o} \hat{\Pi}_2e^{-i{\bm k}\cdot{\bm t}_1} &-c_{\ell o} \hat{\Pi}_3 \\
	-c_{\ell o} \hat{\Pi}_1e^{i{\bm k}\cdot{\bm t}_2}&c_{\ell o}\hat{\Pi}_1 + c_{o} \hat{\pi}_2+ c_o \hat{\pi}_3 &- c_o \hat{\pi}_3 &  -c_{ o} \hat{\pi}_2 \\
	-c_{\ell o} \hat{\Pi}_2e^{i{\bm k}\cdot{\bm t}_1}&-c_{ o} \hat{\pi}_3 & c_{\ell o }\hat{\Pi}_2+c_o\hat{\pi}_1+c_o\hat{\pi}_3 &  -c_{ o} \hat{\pi}_1 \\
	-c_{\ell o} \hat{\Pi}_3&-c_{ o} \hat{\pi}_2 & -c_{o} \hat{\pi}_1 &  c_{\ell o}\hat{\Pi}_3+ c_o\hat{\pi}_2+c_o\hat{\pi}_1
	\end{pmatrix},
	\end{equation} 
\end{small}
where we emphasise that each element is a $2\times2$ block matrix, as defined in \eqref{eq:projectors}.
We note that the submatrix formed by removing the the first column and row of block matrices of~\eqref{eq:Sigma_k_prime} is precisely the stiffness matrix of a single hinged resonator~\eqref{eq:sigma_prime_hinged}.
Expanding the determinant of~\eqref{eq:Sigma_k_prime} over the first column results in a sum of four terms: three of them being proportional to ${\rm det}(\hat{\Pi}_i)$ and the fourth one to ${\rm det}(\hat{\sigma}')$.
Recalling that the matrices $\hat{\sigma}'$ and $\hat{\Pi}_i$ (c.f.~\eqref{eq:projectors} and \eqref{eq:sigma_prime_hinged}) are degenerate when $\vartheta_0=0$, it is clear that~\eqref{eq:sigma_prime_hinged} is also degenerate when $\vartheta_0=0$.

\subsection{Rigid resonators}

The limit case of rigid resonators, that is $c_o/c_{\ell o} \gg 1$ and $c_o/c_\ell \gg1$, has been analysed in detail in section~\ref{subsec:rba}.
It was shown that, in this limit, it is sufficient to treat the resonator as a rigid body.
Under this assumption, the equation of motion for Bloch-Floquet waves propagating through a triangular lattice with \emph{rigid} TIRs is
\begin{equation}\label{eq:secular_BF}
\left[ \hat{\Sigma}_{\bm k}-\omega^2 \hat{{\cal M}}\right]{\bm U}_{\bm k}={\bm 0}, \quad\text{with}\quad{\bm U}_{\bm k} =
\begin{pmatrix}
{\bm u}_0^{\rm T}({\bm k}),&{\bm u}^{\rm T}_{\rm cm}({\bm k}),&\vartheta({\bm k})
\end{pmatrix}^\mathrm{T},
\end{equation}
and where $\vartheta({\bm k})$  and  ${\bm u}_{\rm cm}({\bm k})$ denote the amplitude of rotation and displacement of the TIR's centre of mass, respectively.
The vector of displacements ${\bm U}_{\bm k}$ can be written in terms of the displacements of the masses ${\bm u}_{i}({\bm k})$ with $i=\{1,2,3 \}$ and, hence, ${\bm U}^\prime_{\bm k}$ as introduced in Eq.~\eqref{eq:U_prime_k}.
In particular, 
\begin{align}\label{eq:var_cm_t}
{\bm u}_{\rm cm}({\bm k})&=\frac{1}{3}\left(   {\bm u}_{1}({\bm k}) +{\bm u}_{2}({\bm k})+{\bm u}_{3}({\bm k})   \right),\nonumber\\
{\bm u}_{i}(\bm k)&={\bm u}_{\rm cm}({\bm k})+\left( \hat{ {\cal R}}_{\frac{2\pi}{3}(i-1)+\vartheta({\bm k})} -\hat{\cal R}_{\frac{2\pi}{3}(i-1)}\right)\tilde{{\bm b}}_1 \nonumber\\
&\approx{\bm u}_{\rm cm}({\bm k}) + \vartheta({\bm k})\left(\left. \frac{d\hat{{\cal R}}_\vartheta}{d \vartheta}  \right|_{ \vartheta=2\pi (i-1) / 3}\right)\tilde{\bm b}_1,
\end{align}
where $\hat{{\cal R}}$ is given in Eq.  (\ref{eq:r_matrix}), $\tilde{\bm b}_1$ is introduced in Eq. (\ref{eq:b_i_eq});
and we have linearised the final expression for small angular displacements $\vartheta(\vec{k})$.
The stiffness matrix~\eqref{eq:Sigma_k_prime} can then be expressed as
\begin{equation}\label{eq:Sigma_k}
\hat{\Sigma}_{\bm k}=\sum_{i=1}^3
\begin{pmatrix}
-2 c_\ell(\cos({\bm k}\cdot{\bm t}_i)-1)\hat{\tau}_i+c_{\ell o}\hat{\Pi}_i& -c_{\ell o}\varphi_{i}({\bm k})\hat{\Pi}_i&-c_{\ell o} \varphi_{i}({\bm k})\hat{\Pi}_i\hat{\cal R}'_{i}\tilde{\bm b}_1\\
\\
-c_{\ell o} \varphi^*_{i}({\bm k} )\hat{\Pi}_i & c_{\ell o}\hat{\Pi}_i &  c_{\ell o} \hat{\Pi}_i\hat{\cal R}'_{i}\tilde{\bm b}_1\\
\\
-c_{\ell o} (\varphi_{i}({\bm k})\hat{\Pi}_i\hat{\cal R}'_{i}\tilde{\bm b}_1)^\dagger&c_{\ell o}(\hat{\Pi}_i\hat{\cal R}'_{i}\tilde{\bm b}_1)^{\dagger}& c_{\ell o}\tilde{\bm b}_1^{\rm T}\cdot (\hat{\cal R}_i'\hat{\Pi}_i\hat{\cal R}_i'\tilde{\bm b}_1)
\end{pmatrix},
\end{equation}
where we introduce the functions  $\varphi_1({\bm k})={\rm exp}(-{\bm k}\cdot{\bm t}_2)$, $\varphi_2({\bm k})={\rm exp}(-{\bm k}\cdot{\bm t}_1)$ and $\varphi_3({\bm k})=1$; and ${\cal R}'_i$, $i=\{1,2,3\}$, denote the derivatives of the rotation matrix in Eq. \eqref{eq:var_cm_t}. 
The inertia matrix which appears in \eqref{eq:secular_BF} is 
\begin{equation}\label{eq:mass_BF_rigid}
\hat{\cal M}={\rm diag}(m,m,M,M,I)
\end{equation}
where $M=3m_o$ is the total mass of the TIR, $I=m_o\ell^2$ is its moment of inertia, and $m$ is the mass of the nodal points in the ambient lattice.
The solvability condition for the algebraic system~\eqref{eq:secular_BF}, then yields the dispersion equation for Bloch-Floquet waves in the triangular lattice with TIR
\begin{equation}\label{eq:dispersion_eq}
{\cal D}({\bm k},\omega)={\rm det}\left(\hat{\Sigma}_{\bm k} - \omega^2 \hat{\cal M}\right)=0.
\end{equation}
The remainder of this section will be devoted to the analysis and interpretation of the roots of this equation.
\subsection{The Bloch frequencies at $\Gamma$}
At ${\bm k}={\bm 0}$, the roots of fifth-degree polynomial equation in $\Omega=\omega^2$ \eqref{eq:dispersion_eq} can be found in their closed forms.  Introducing the notation $\Omega^{(i)}_{\Gamma}=\left.\Omega^{(i)}_{\bm k}\right|_{{\bm k}={\bm 0}}$, with $i$ indexing the root, we find 
\begin{equation}\label{eq:eigf_gamma}
\Omega^{(1)}_{\Gamma}=0,\,\,\,\Omega^{(2)}_{\Gamma}=\Omega_{\rm cm}\left(1+\frac{3m_o}{m}\right)\,\,\,{\rm and}\,\,\,\Omega^{(3)}_{\Gamma}=\Omega_{\vartheta}=\Omega_{\rm cm}\frac{2 \sin^2\vartheta_0}{1+\ell^2/L^2-2\ell/L\cos\vartheta_0},
\end{equation}
where the frequencies $\Omega_{\rm cm}$ and $\Omega_{\vartheta}$ are given in  Eq. \eqref{eq:eigs_rigid}. The first and second of Eqs. \eqref{eq:eigf_gamma} have multiplicity two, and the last one has multiplicity one. 

Given  $\bar{\ell}=\ell/L$ and $\vartheta_0$ in the intervals \eqref{eq:geom_cond}, we observe that it is possible to obtain a triple-root eigenvalue corresponding to $\Omega^{(2)}_{\Gamma}=\Omega^{(3)}_{\Gamma}$, if there exists 
\begin{equation}
\bar{m}= \frac{\cos 2\vartheta_0 + \bar{\ell}^2- 2 \bar{\ell}\cos \vartheta_0}{2\bar{\ell}\cos\vartheta_0- \bar{\ell}^2-1}>0,
\end{equation}
with $\bar{m}=3m_o/m$.
\subsection{Effective group velocity}
\begin{figure}
	\centering
	\begin{subfigure}[t]{0.48\linewidth}
		\includegraphics[width=\linewidth]{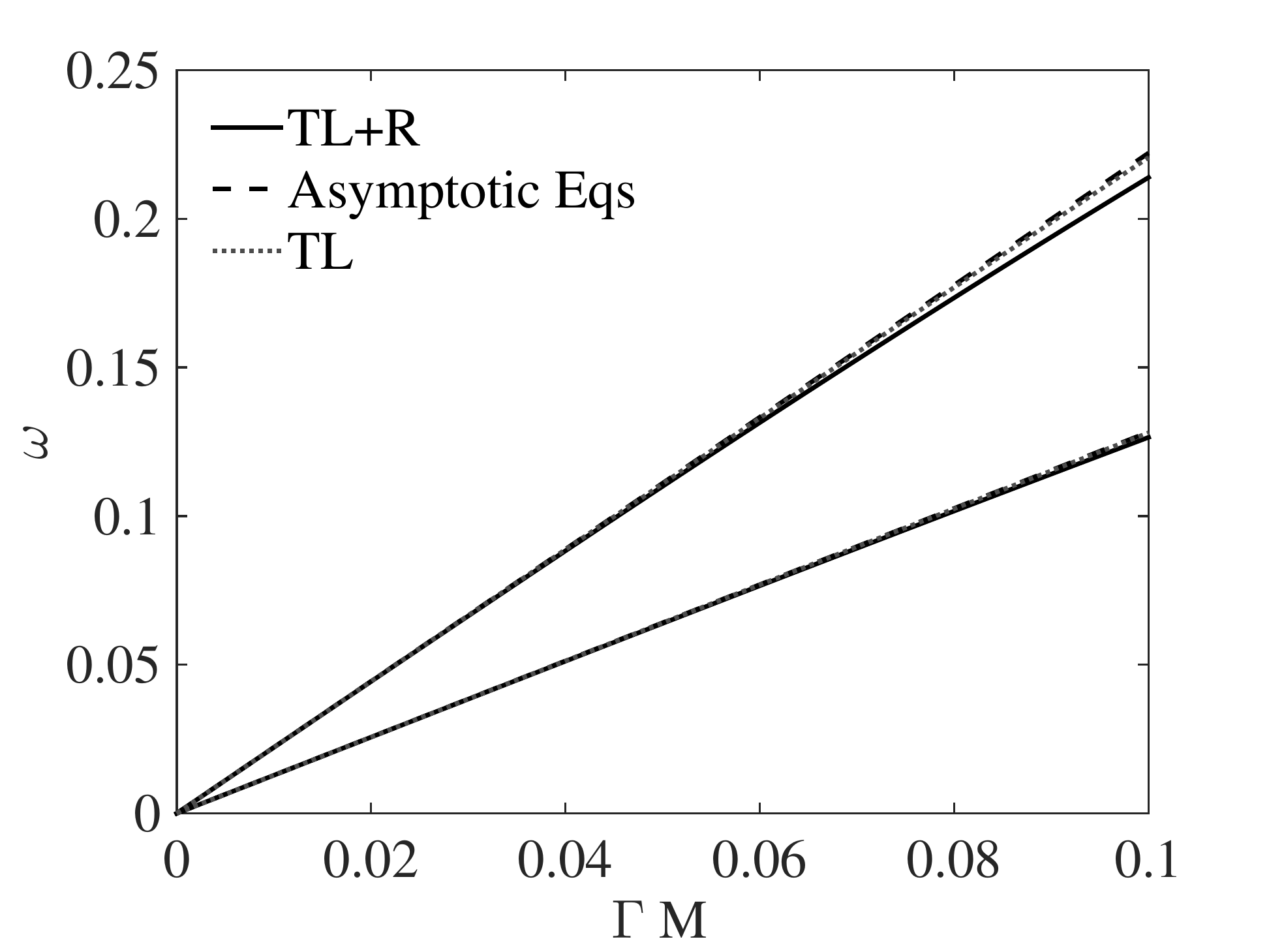}
		\caption{}
	\end{subfigure}
	\begin{subfigure}[t]{0.48\linewidth}
		\includegraphics[width=\linewidth]{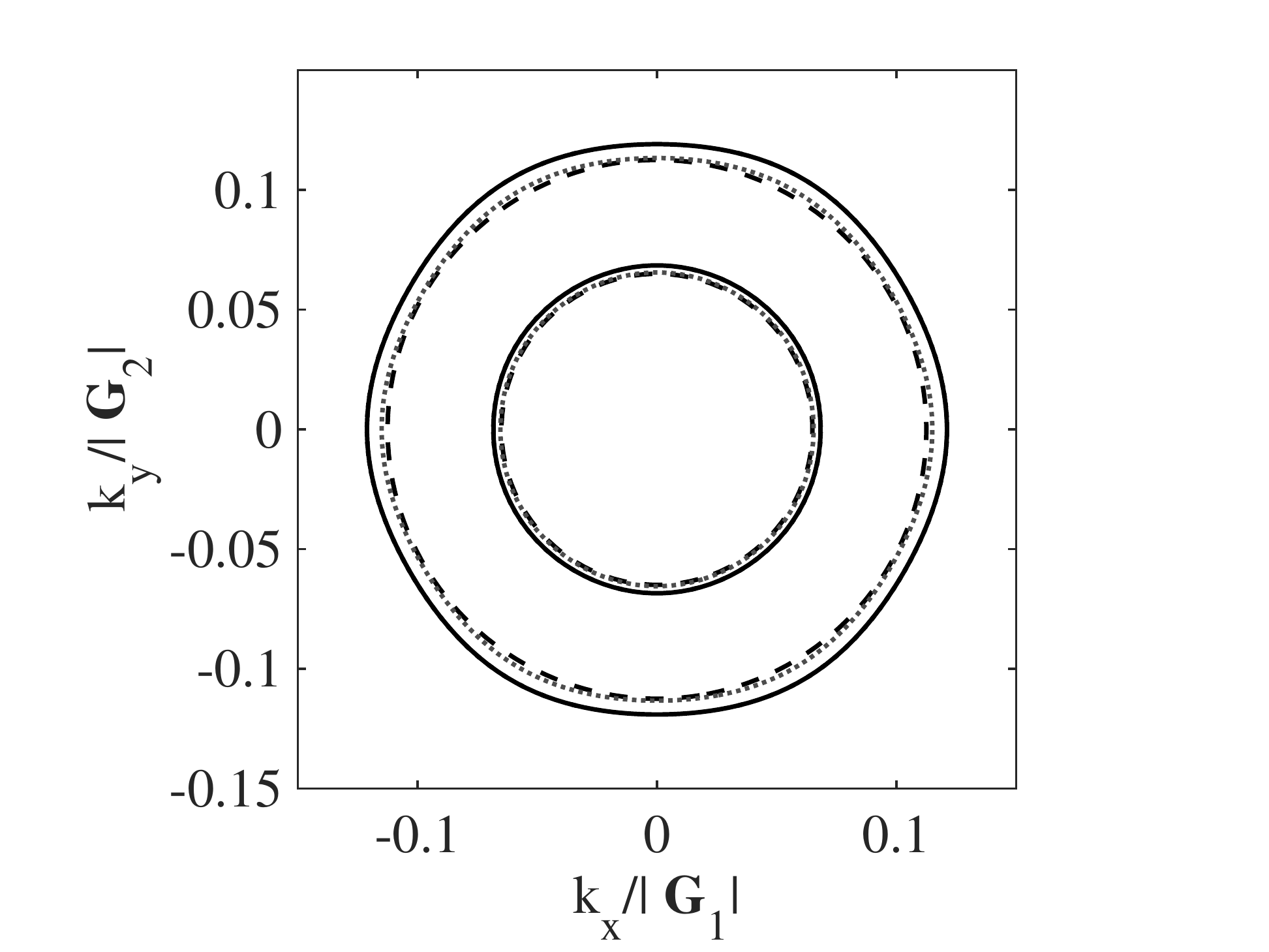}
		\caption{}
	\end{subfigure}
	\caption{\label{fig:curves_asymptotics}Panel (a) shows dispersion diagrams and panel  (b) the corresponding slowness contours. Black solid lines refer to a triangular lattice with tilted TIRs  (see Eq. (\ref{eq:secular_BF}))  for $\vartheta_0=\vartheta_{\rm max}$ with the remaining parameters as listed in Table \ref{tab:parameters}; black dashed lines correspond to the effective group velocities (\ref{eq:eff_v}); grey dotted lines  refer to a monatomic triangular lattice (see Eq. (\ref{eq:disp_TL})) mass per site as in Eq. \eqref{eq:renormalised_mass} and other parameters as listed in Table \ref{tab:parameters}.}
\end{figure}

It is interesting to examine the dynamic behaviour of the lattice containing TIR in the long-wave regime.
In particular, the effect of the resonator on the quasi-static response will be examined.
We begin by considering the non-degenerate case of $\vartheta_0 \neq0$ and expand $\mathcal{D}(\vec{k},\omega)$ in a MacLaurin series assuming that both $0 < \omega \ll1$ and $0 < L|\vec{k}| \ll1$.  In the low frequency regime, Eq. \eqref{eq:dispersion_eq} can be written as
\begin{equation}\label{eq:o_omega}
\mathcal{D}(\bm{k},\omega)=\mathcal{D}_\Gamma(\bm{k},\omega)+o({\omega}^6).
\end{equation}
We then search for solutions of the form $\omega_{\bm{k}}= v k$, where $v$ is the effective group velocity. Consequently, in Eq. \eqref{eq:o_omega} we can write
\begin{equation}\label{eq:disp_eq_taylor}
\mathcal{D}_\Gamma(\bm{k},\omega)=\mathcal{D}^{(4)}_\Gamma(\bm{k})\omega^4+\mathcal{D}^{(2)}_\Gamma(\bm{k})\omega^2+\mathcal{D}^{(0)}_\Gamma(\bm{k})+o(|{\bm k}|^6).
\end{equation}
Full details of the procedure are provided in Appendix \ref{app:taylor}. 
The long-wave and low-frequency pressure $v_{\mathrm{p}}$ and shear $v_{\mathrm{s}}$ \emph{group velocities} for Bloch-Floquet waves in a triangular lattice containing TIRs are
\begin{equation}\label{eq:eff_v}
v_{\rm p}=\sqrt{3} v_{\rm s}\,\,\,\,{\rm and}\,\,\,\,v_{\rm s}=\frac{1}{2} \sqrt{\frac{3}{2}}  L \sqrt{\frac{c_\ell}{m+3m_o}}.
\end{equation}

According to equations \eqref{eq:eff_v}, the low-frequency and long-wave dispersion surfaces are conical and isotropic.
In addition, we remark that the group velocities depend on the total mass of the cell, on the inter-cell stiffness $c_{\ell}$ and on the triangular lattice nearest-neighbour distance $L$.
The intra-cell physical parameters $\vartheta_{0}$, $c_{\ell o}$ and $\ell$, do not appear in this regime because they characterise sub-wavelength structures.
Moreover, equations (\ref{eq:eff_v}) suggest that the low-frequency and long-wave dispersion of elastic waves in a lattice with TIRs is equivalent to the behaviour associated with a simple monatomic triangular lattice with a mass per unit cell of
\begin{equation}\label{eq:renormalised_mass}
m_{\rm TL}=m+3m_o,
\end{equation} 
and ligaments with stiffness $c_\ell$.
These effects are illustrated in Fig. \ref{fig:curves_asymptotics}, where we examine the  dispersion curves in the low-frequency and long-wave regime.
The solid lines correspond to the dispersion curves for a triangular lattice with TIRs. The numerical parameters have been chosen as detailed in Table (\ref{tab:parameters}) and the tilting angle is $\vartheta_{0}=\vartheta_{\rm max}$.
The dashed lines correspond to the  asymptotic dispersion relation $\omega_{\bm{k}} \sim v k$, evaluated using the effective group velocities in Eq. (\ref{eq:eff_v}).
The dotted lines refer to a triangular lattice, whose dispersion curves have been obtained solving 
\begin{equation}\label{eq:disp_TL}
{\rm det}\left(\hat{\Sigma}_{\bm k}^{\rm (TL)}-m_{\rm TL}\omega^2 \hat{I}_2\right)=0,\,\,\,{\rm with}\,\,\,\,\,\,\,\,
\hat{\Sigma}_{\bm k}^{\rm (TL)}=-\sum_{i=1}^32 c_\ell(\cos({\bm k}\cdot{\bm t}_i)-1)\hat{\tau}_i,
\end{equation}
and the mass $m_{\rm TL}$ at the nodal points as in  Eq. (\ref{eq:renormalised_mass}). The matrix $\hat{I}_2$ is the $2\times2$ identity matrix  and the remaining parameters are listed in Table \ref{tab:parameters}. We observe that Eq. \eqref{eq:disp_TL} can be obtained from the stiffness matrix \eqref{eq:Sigma_k} in the limit $c_{\ell o}\rightarrow0$.

Fig. \ref{fig:curves_asymptotics}(a) shows that the dispersion curves for a triangular lattice containing non-degenerate ($\vartheta_0=\vartheta_{\rm max}$) TIRs are linear and equivalent to a monatomic triangular lattice with renormalised mass in the long-wave low-frequency regime.
Fig. \ref{fig:curves_asymptotics}(b) shows slowness contours for the same three examples at a fixed frequency of $\omega=0.25$.
We observe that the slowness contours for the triangular lattice with TIRs (solid lines) are approximately circular, indicating that the lattice with TIRs is isotropic in the long-wave, low-frequency regime. 

In the non-degenerate case,  the unit cell in Fig.~\ref{fig:system_t}(a) is chiral, in a similar sense as considered by Spadoni \emph{et al.} in~\cite{Spadoni_WM_46_435_2009}.
However, we note that the geometry and the effective dispersion properties derived here
are different from those studied in~\cite{Spadoni_WM_46_435_2009}. 
The difference arises due to the fact that the chirality, associated with the topology of the TIRs, emerges at length scales shorter than the unit cell typical width $L$.
Therefore, the chiral effects are likely to emerge at higher frequencies, as it is illustrated below.

\begin{table}[h]
	\centering
	\begin{tabular}{@{}llllllr@{}} \toprule%\cmidrule(r){1-6}
		$c_\ell$ & $c_{\ell o}$ & $L$ & $\ell$ & $m$ &$ m_o$& $\vartheta_{\rm max}$\\ \midrule
		1            &                  1& 1     &$1/4$ &  1     &         1   & 1.318\\ \bottomrule
	\end{tabular}\caption{\label{tab:parameters}Parameters used in Eq. (\ref{eq:secular_BF}) in order to model the triangular lattice with TIRs.}
\end{table}

\subsection{The effect of the tilting angle $\vartheta_0$ on the dispersion properties}

\begin{figure}
	\centering
	\begin{subfigure}[t]{0.48\linewidth}
		\includegraphics[width=\linewidth]{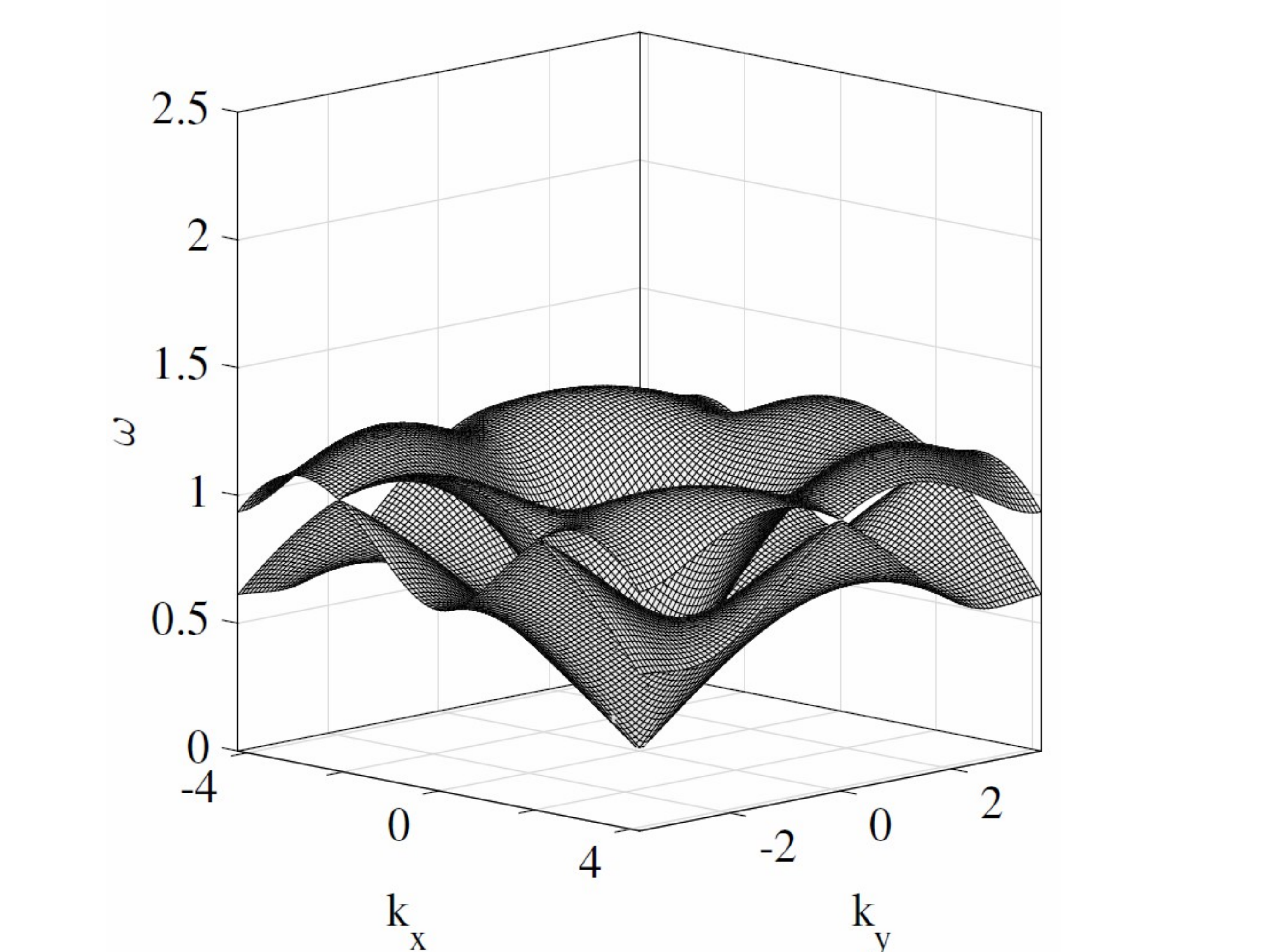}
		\caption{}
	\end{subfigure}
	\begin{subfigure}[t]{0.48\linewidth}
		\includegraphics[width=\linewidth]{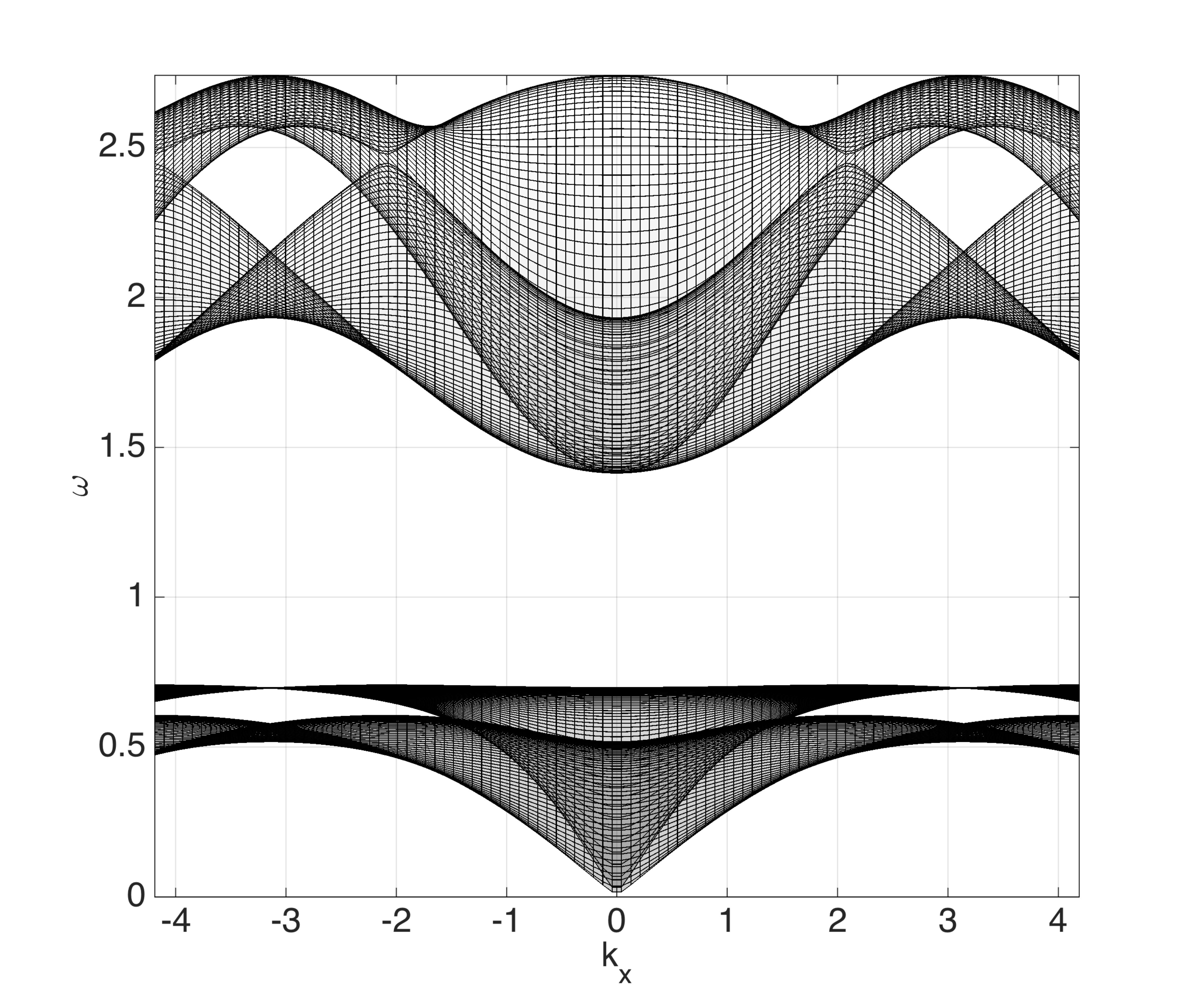}
		\caption{}
	\end{subfigure}
	\begin{subfigure}[t]{0.48\linewidth}
		\includegraphics[width=\linewidth]{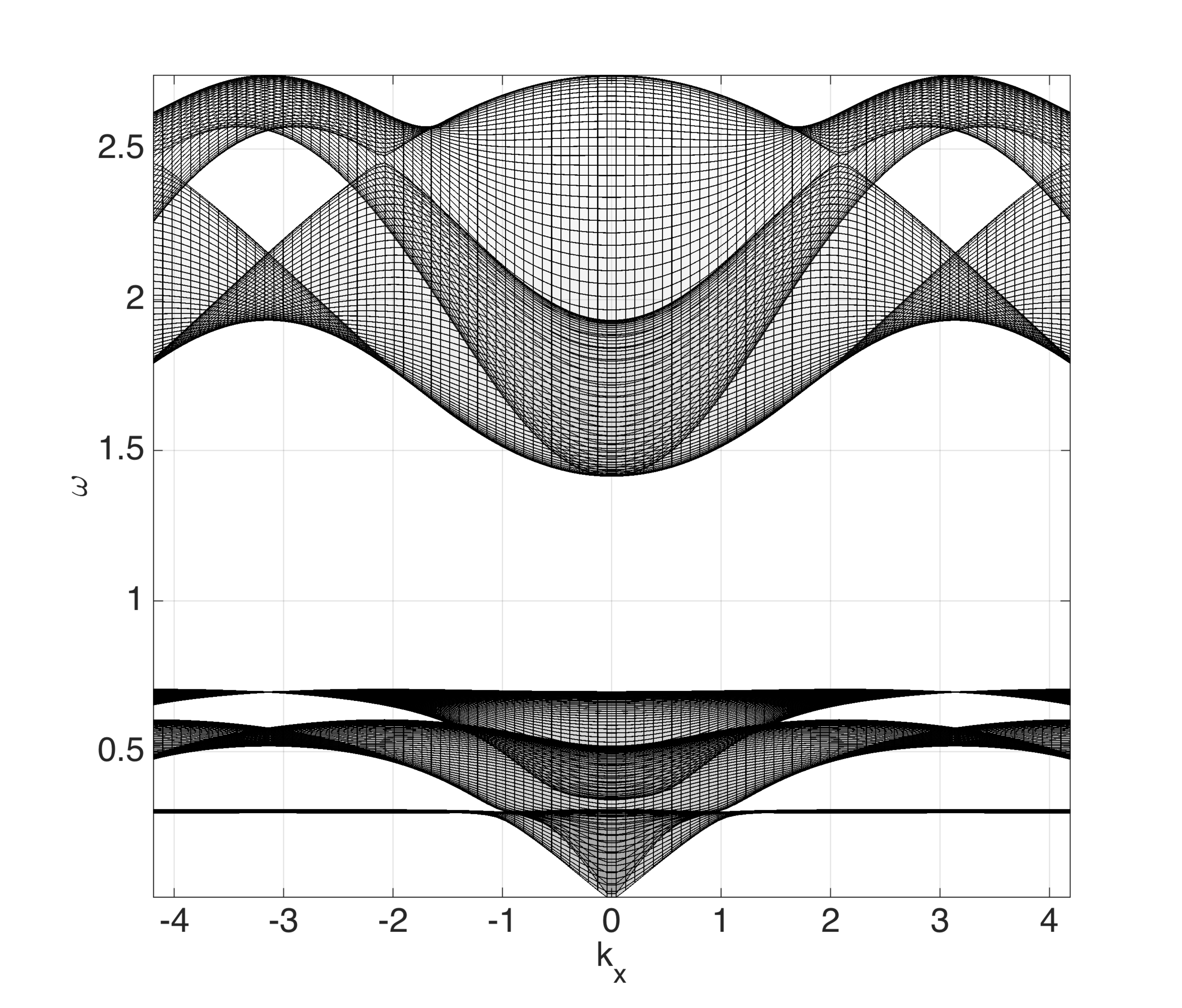}
		\caption{}
	\end{subfigure}
	\begin{subfigure}[t]{0.48\linewidth}
		\includegraphics[width=\linewidth]{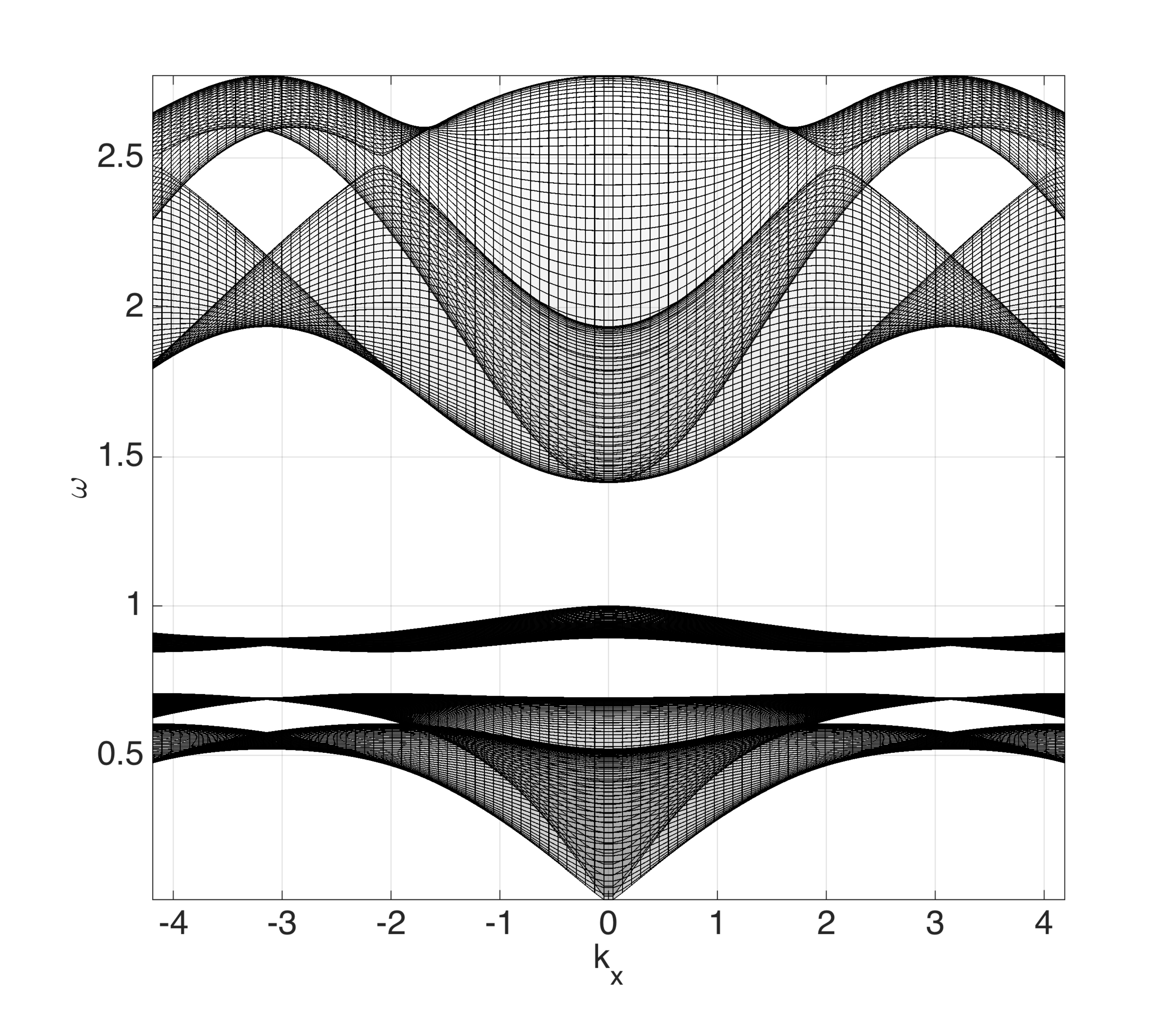}
		\caption{}
	\end{subfigure}
	\caption{\label{fig:surf_theta_0}Bloch dispersion surfaces over a wave vector region which includes the first Brillouin zone for the triangular lattice - see Fig. (\ref{fig:system_t})(c). Panel (a): The dispersion surfaces for a monatomic triangular lattice whose parameters are the same as for  Fig.  \ref{fig:curves_asymptotics},  grey dotted lines. Panel (b): The dispersion surfaces for the triangular lattice with resonator and  $\vartheta_0=0$. Panel (c): The dispersion surfaces for the triangular lattice with TIR at  $\vartheta_0=0.2~\vartheta_{\rm max}$, as defined in Eq. (\ref{eq:geom_cond}). Panel (d): The triangular lattice with TIR at  $\vartheta_0=\vartheta_{\rm max}$. The remaining parameters are listed in Table \ref{tab:parameters}.}
\end{figure}

\begin{figure}
	\centering
	\begin{subfigure}[t]{0.3\linewidth}
		\includegraphics[width=\linewidth]{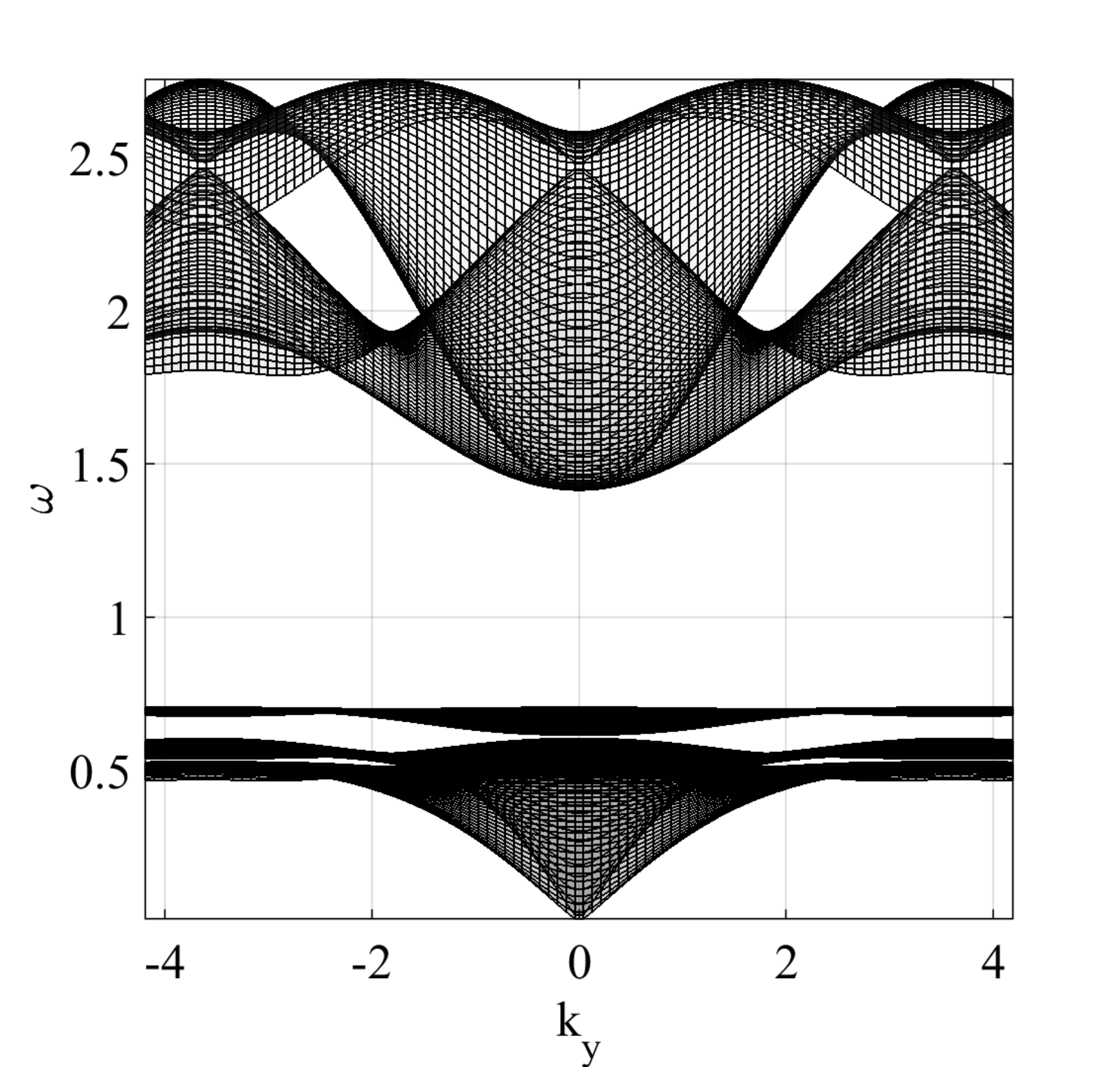}
		\caption{}
	\end{subfigure}
	\begin{subfigure}[t]{0.3\linewidth}
		\includegraphics[width=\linewidth]{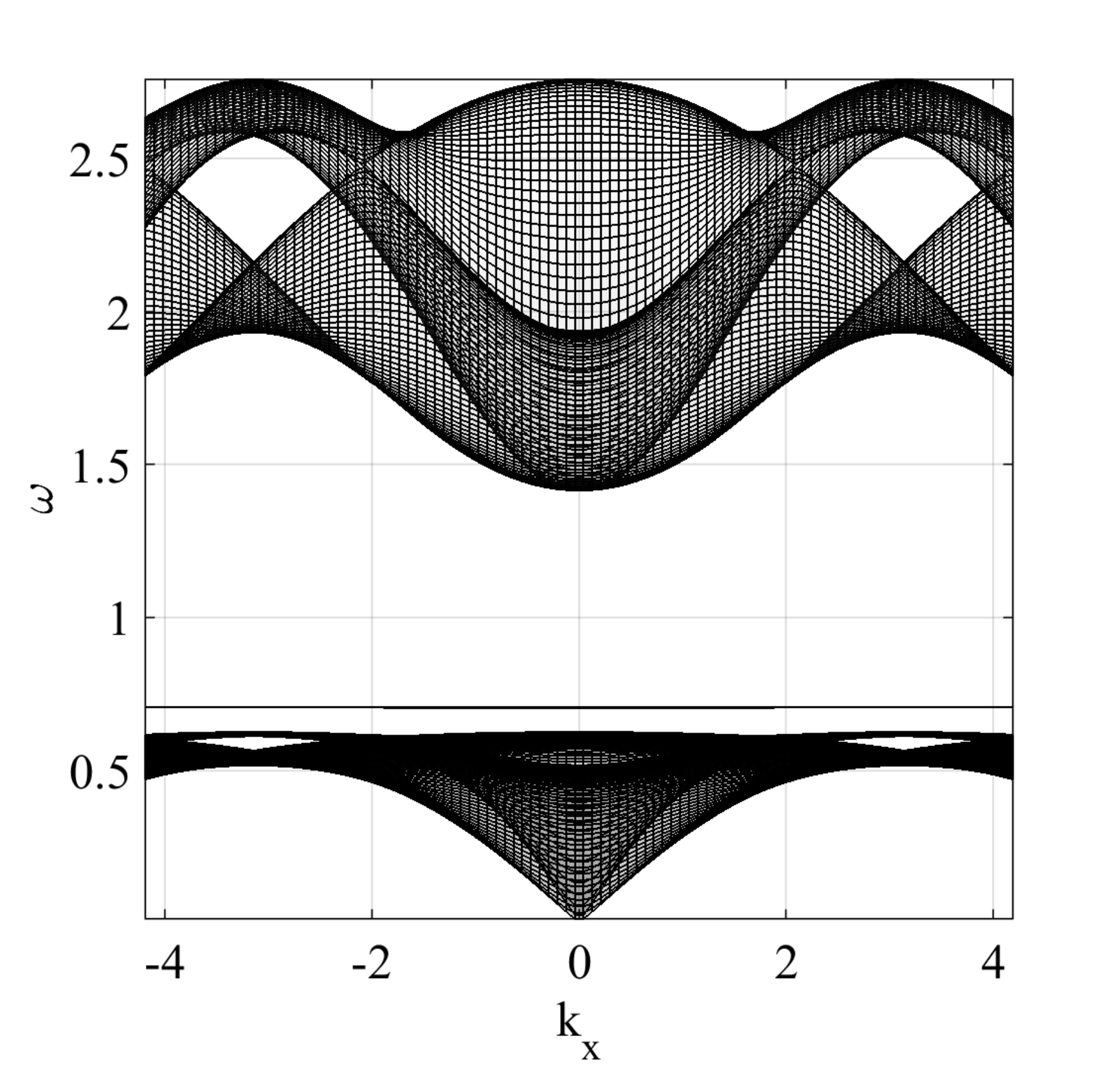}
		\caption{}
	\end{subfigure}
	\begin{subfigure}[t]{0.3\linewidth}
		\includegraphics[width=\linewidth]{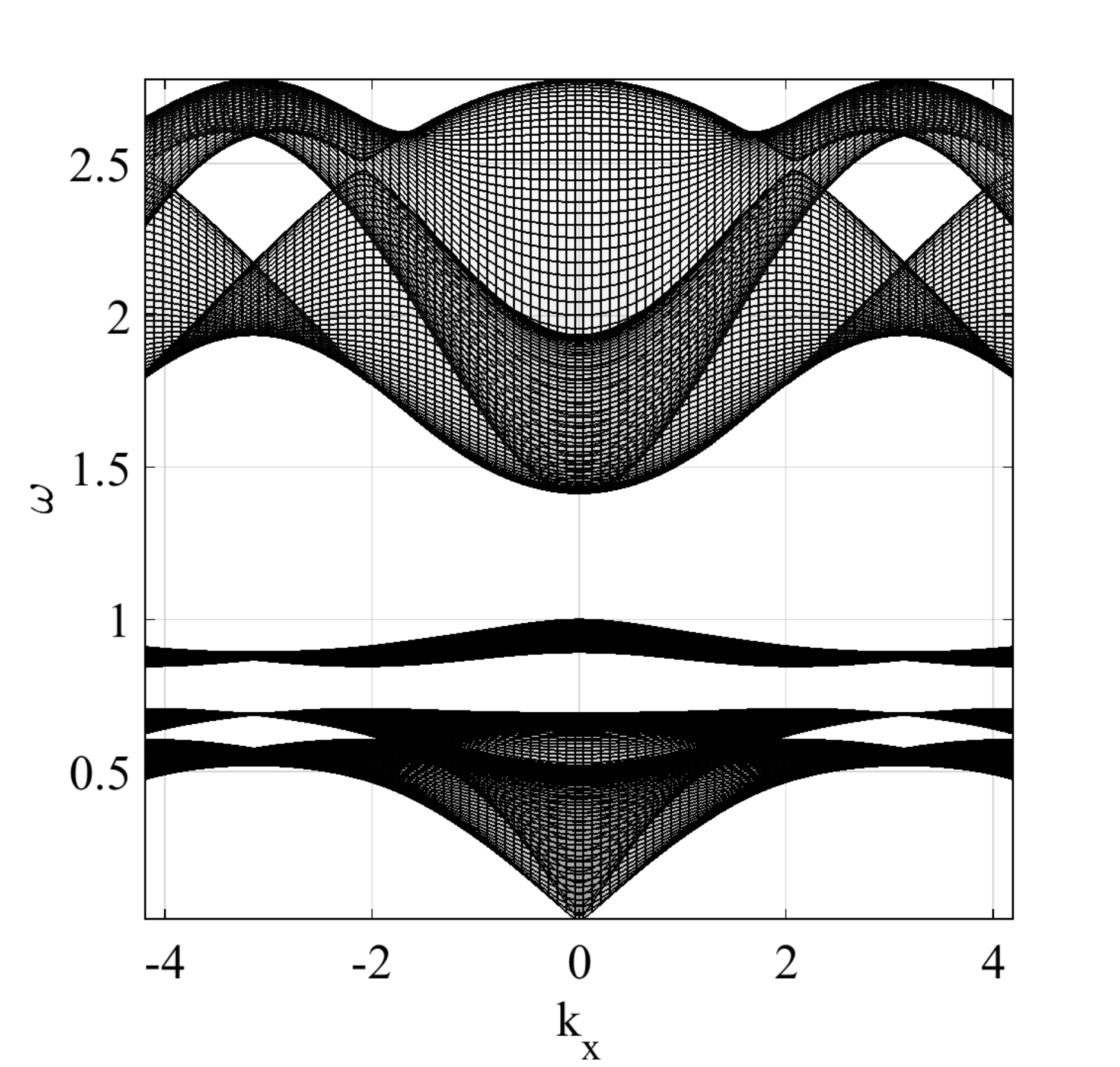}
		\caption{}
	\end{subfigure}
	\begin{subfigure}[t]{0.3\linewidth}
		\includegraphics[width=\linewidth]{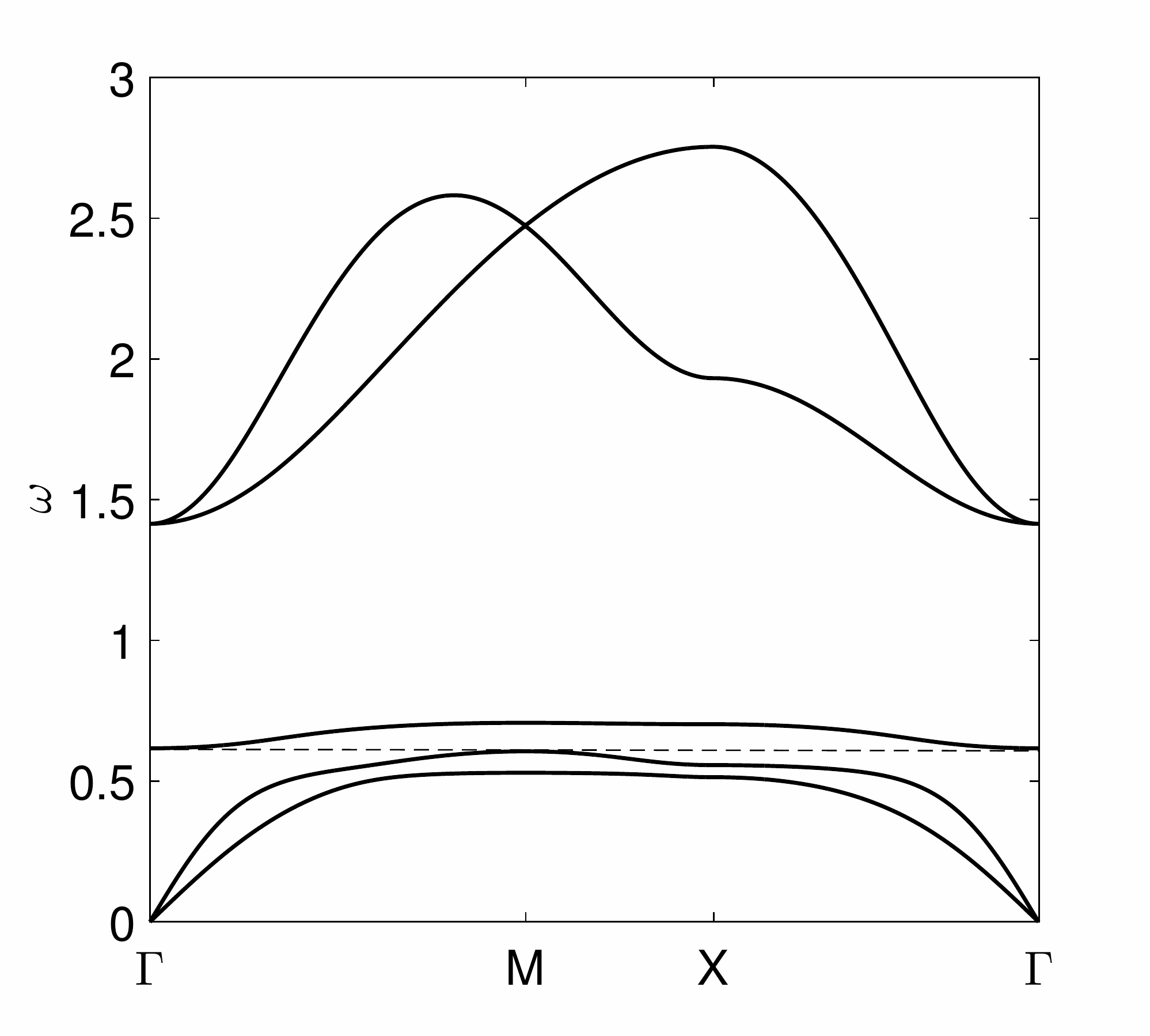}
		\caption{}
	\end{subfigure}
	\begin{subfigure}[t]{0.3\linewidth}
		\includegraphics[width=\linewidth]{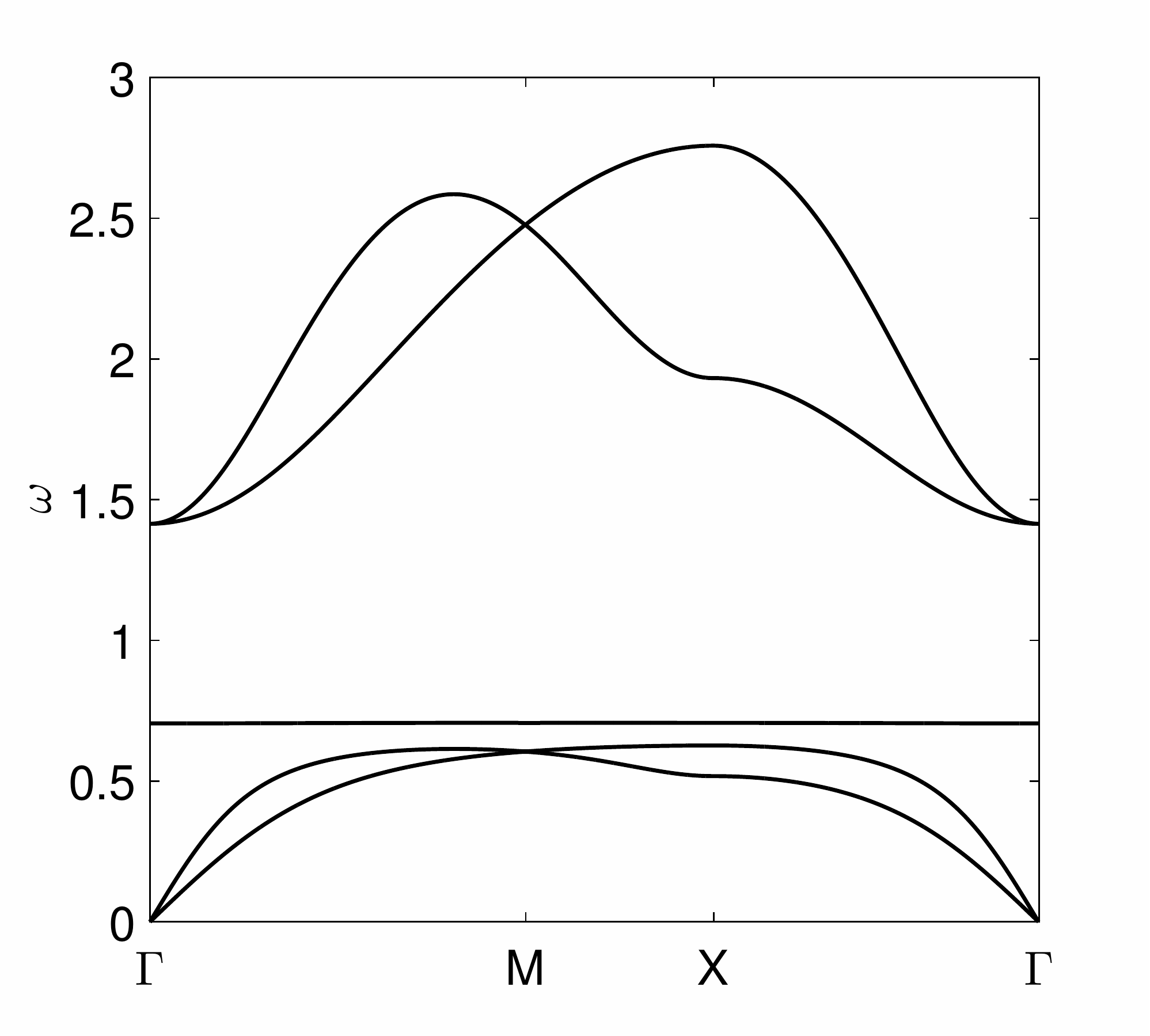}
		\caption{}
	\end{subfigure}
	\begin{subfigure}[t]{0.3\linewidth}
		\includegraphics[width=\linewidth]{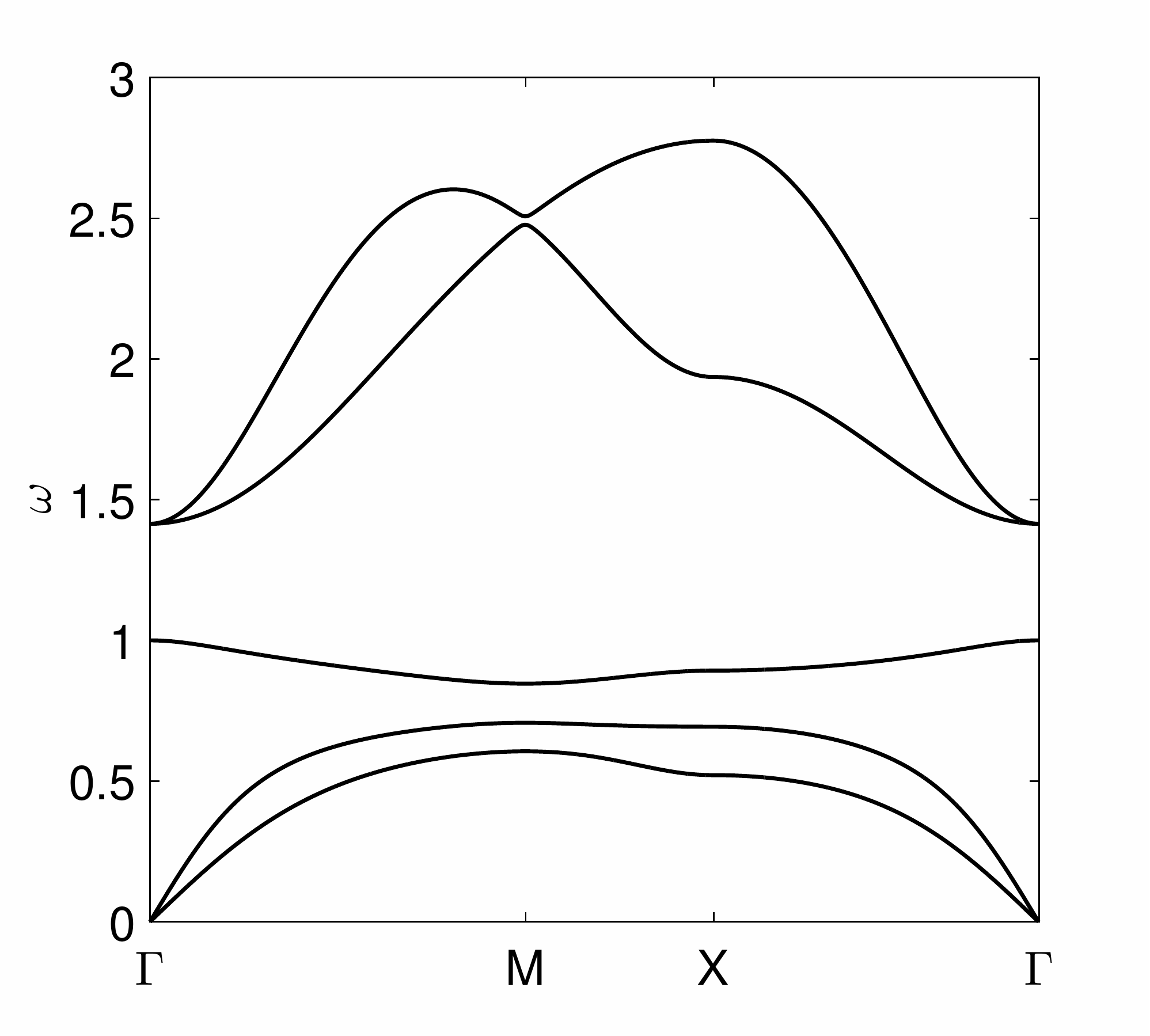}
		\caption{}
	\end{subfigure}
	\begin{subfigure}[t]{0.3\linewidth}
		\includegraphics[width=\linewidth]%,clip,trim={30 140 190 20}]
		{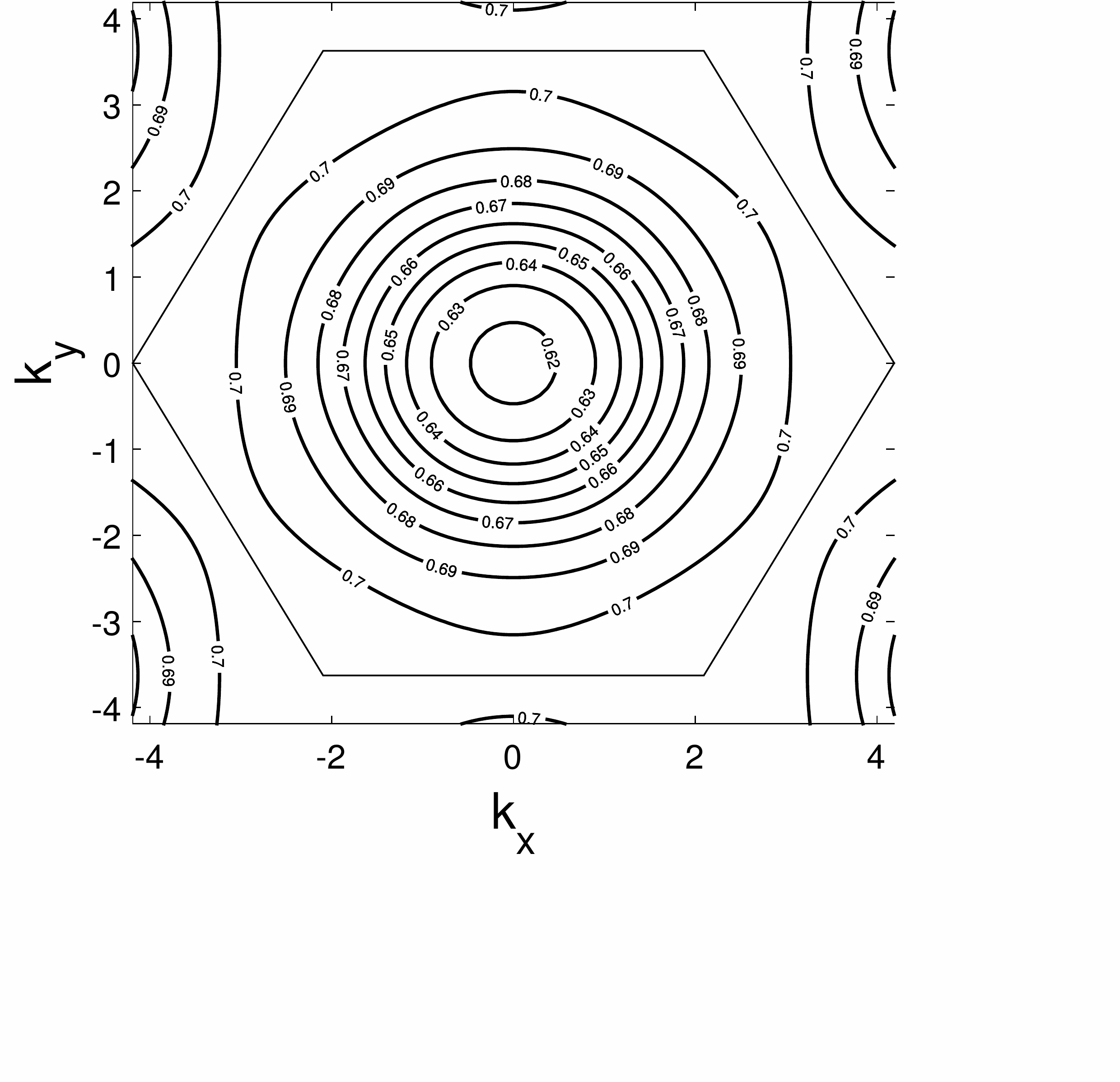}
		\caption{}
	\end{subfigure}
	\qquad
	\begin{subfigure}[t]{0.3\linewidth}
		\includegraphics[width=\linewidth]%,clip,trim={30 140 190 20}]
		{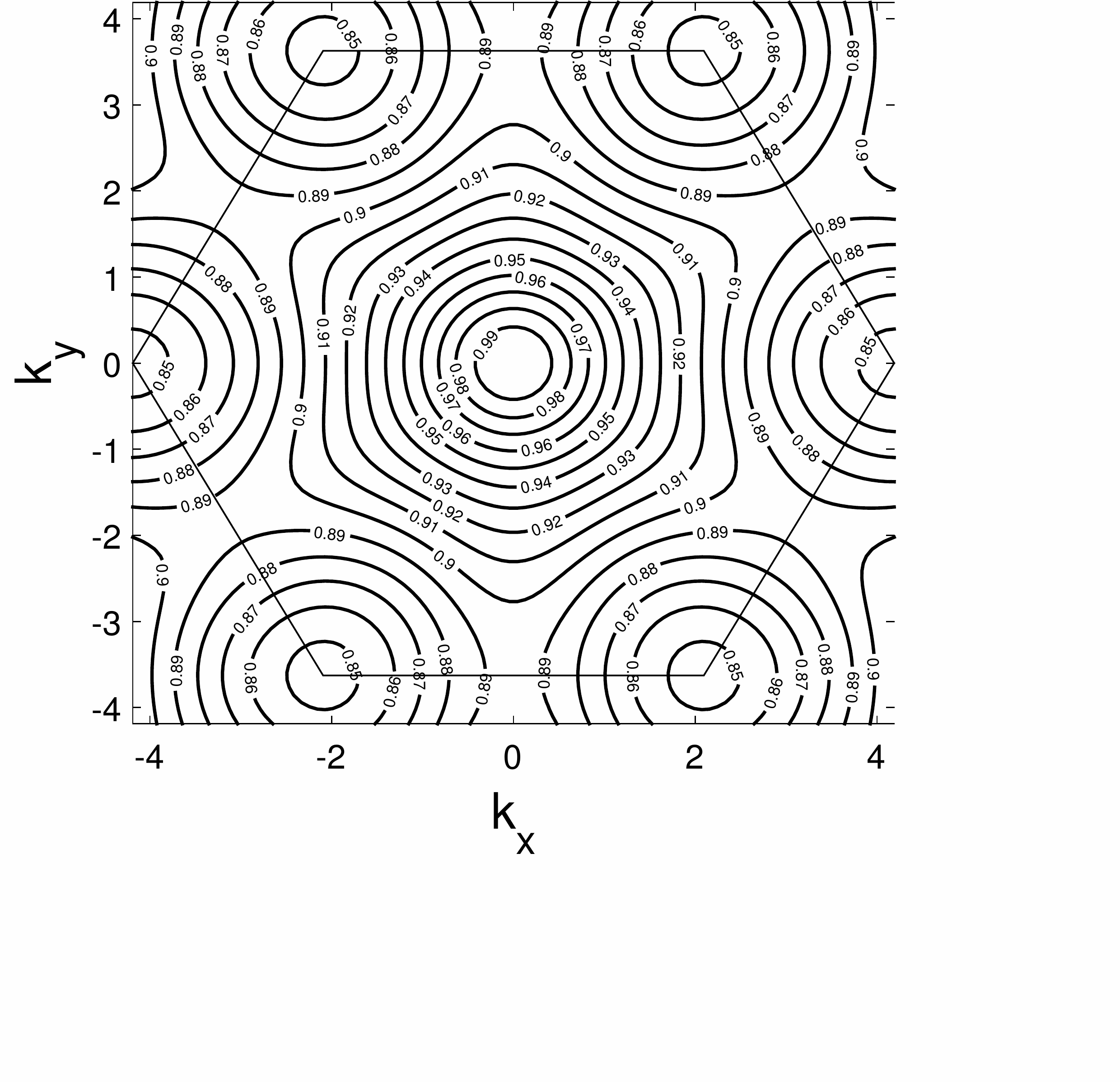}
		\caption{}
	\end{subfigure}
	
	\caption{\label{fig:curve_theta_0}Panels (a), (b) and (c) show dispersion surfaces corresponding to the tilting angles $\vartheta_0=0.51,0.61,1.32~\rm rad$, respectively. In the same order, figures (d), (e) and (f) show the  corresponding  dispersion curves plotted over the the boundary of the irreducible Brillouin zone. 
		The isofrequency diagrams in Figs (g) and (h) show the ``slowness contour" corresponding to the dispersion surfaces (a) and (c), for the tilting angles  $\vartheta_0=0.51, 1.32~{\rm rad}$ respectively. 
	}
\end{figure}
In this subsection, we study the role of the rotational parameter $\vartheta_0$ on the dispersion of Bloch-Floquet elastic waves in a triangular lattice containing TIRs.

\subsubsection{Band gaps in the dispersion diagrams for Bloch waves}

In Fig. \ref{fig:surf_theta_0} we compare the dispersion surfaces, over the set of wave vectors defined in Eq. (\ref{eq:k_BZ}), for four different configurations.
Fig.  \ref{fig:surf_theta_0}(a) corresponds to a monatomic triangular lattice whose physical parameters coincide with the ones used  in Fig. \ref{fig:curves_asymptotics}.  Fig.  \ref{fig:surf_theta_0}(b), shows the dispersion surfaces for the degenerate ($\vartheta_0=0$) triangular lattice with resonators.
Fig.  \ref{fig:surf_theta_0}(c) and Fig.  \ref{fig:surf_theta_0}(d)  correspond to the the choices of the tilting angles $\vartheta_{0}=\vartheta_{\rm max}/5$ and $\vartheta_{0}=\vartheta_{\rm max}$, respectively.
The remaining parameters for the aforementioned examples are listed in Table~\ref{tab:parameters}.
The monatomic triangular lattice  (panel (a)) exhibits two \emph{acoustic branches}  corresponding to the degrees of freedom for the in-plane displacement of the mass in the unit cell.
The insertion of a \emph{non-tilted} resonator, which preserves the symmetry of the elementary cell, results in the two \emph{optical branches} shown in panel (b).
These additional surfaces are associated with the motion of the centre-of-mass of the resonator and are separated from the acoustic branches by a complete band gap.
Tilting the resonator, and thus breaking the symmetry of the elementary cell, induces a new dispersion surface, as shown in panels (c) and (d).
This new surface is associated with the rotational motion of the resonator.
For a sufficiently small $\vartheta_0$ (see panel (c)), the novel dispersion surface intersects the acoustic branches of the dispersion diagram.
Interestingly, for the maximum angle achievable $\vartheta_{0}=\vartheta_{\max}$, panel (d) shows that the novel resonant branch completely decouples from the acoustic branches. Both the position and shape of this surface, associated with the rotational motion of the resonator, can be controlled by tuning the \emph{tilting angle} $\vartheta_0$ as it is further illustrated in the next section.
\subsubsection{Localisation and standing waves}
Fig. \ref{fig:curve_theta_0} shows the dispersion surfaces, diagrams and slowness contours for a range of tilting angles $\vartheta_0$.
The parameters used in this set of computations are listed in Table (\ref{tab:parameters}).
The pairs of panels (a) and (d), (b) and (e), and (c) and (f) correspond to $\vartheta_0\approx0.51~{\rm rad}$, $\vartheta_0\approx0.61~{\rm rad}$ and $\vartheta_0\approx1.32~{\rm rad}$, respectively.
This choice illustrates the influence exerted by the angle $\vartheta_0$ on the dispersive properties of the Bloch resonant mode.
In particular, for $\vartheta_0\approx0.51$, the frequency band gap between the acoustic and the rotational mode closes (c.f. panels (a) and (d)).
On the other hand, for $\vartheta_0=\vartheta_{\rm max}\approx1.32$, the band gap between the rotational mode and  the highest acoustic branch is maximal - see panels (c) and (f).
Moreover, for  $\vartheta_0\approx0.61$, there is a resonant rotational mode, and the corresponding dispersion surface becomes flat.

It is clear that the dispersive properties of the rotational mode depend on the titling angle $\vartheta_0$.
This is illustrated in panels (g) and (h), where the slowness contours for the rotational band in panels (a) and (c) are shown.
In panel (g), the group velocities points outwards in the vicinity of $\Gamma$; this can be identified with with a positive effective mass in the sense of~\cite{Huang_2009}.
Conversely, in panel (h), the group velocity points inwards in the neighbourhood of $\Gamma$, corresponding to a negative group velocity.
Finally, we observe that the frequency of the standing wave reported in panel (b) exactly coincides with the single resonator rotational frequency, already introduced in Eq. (\ref{eq:eigs_rigid}) (see also panel (e)).
We also observe that the resonances of the rotational mode at $\Gamma$, coincide with the single resonator frequency;
for panel (a), this corresponds to the lower boundary of the rotational band, whilst in (f) it corresponds the the upper boundary of the rotational band.
\begin{figure}
	\centering
	\begin{subfigure}[t]{0.48\linewidth}
		\includegraphics[width=\linewidth]{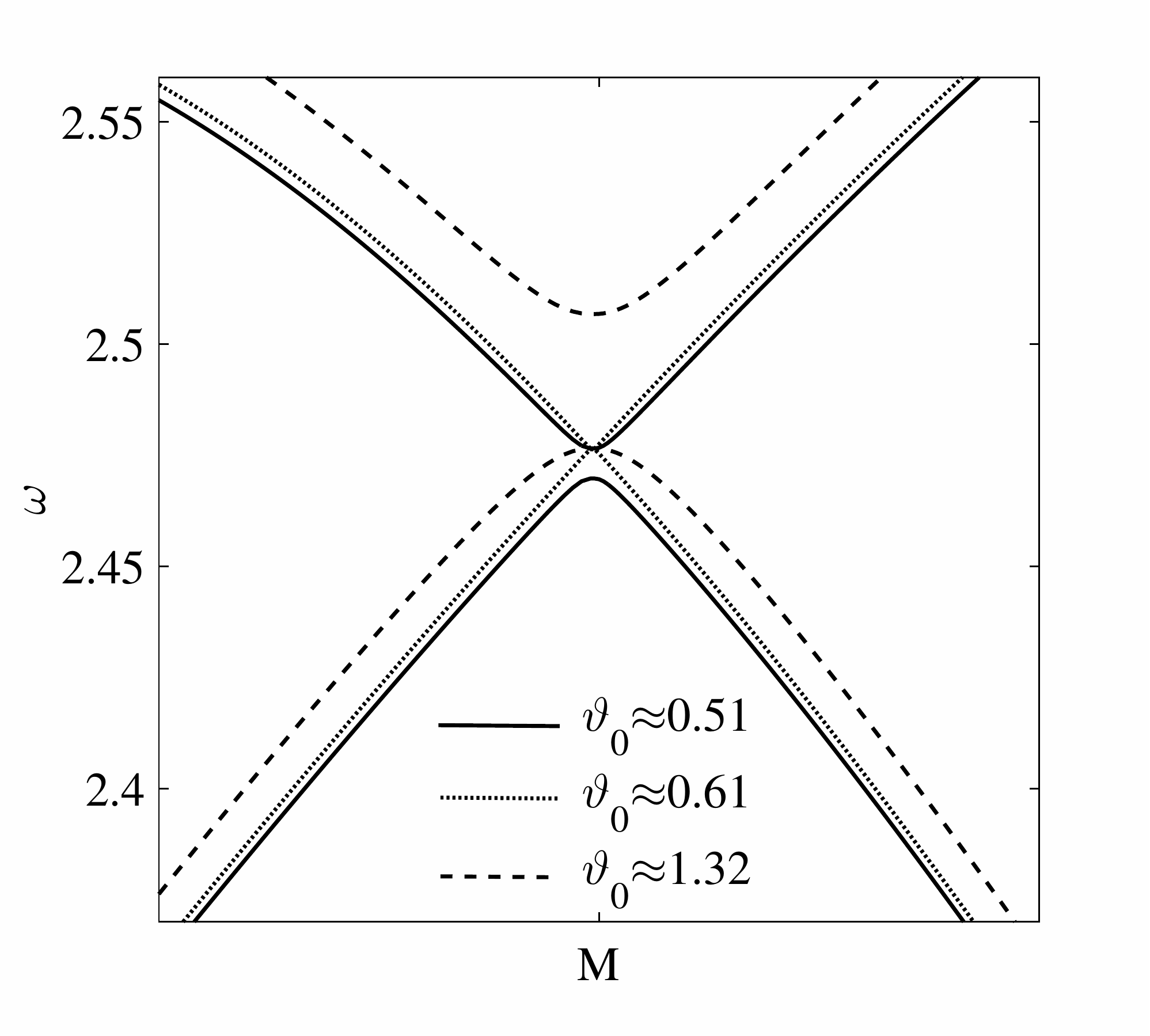}
		\caption{}
	\end{subfigure}
	\begin{subfigure}[t]{0.48\linewidth}
		\includegraphics[width=\linewidth]{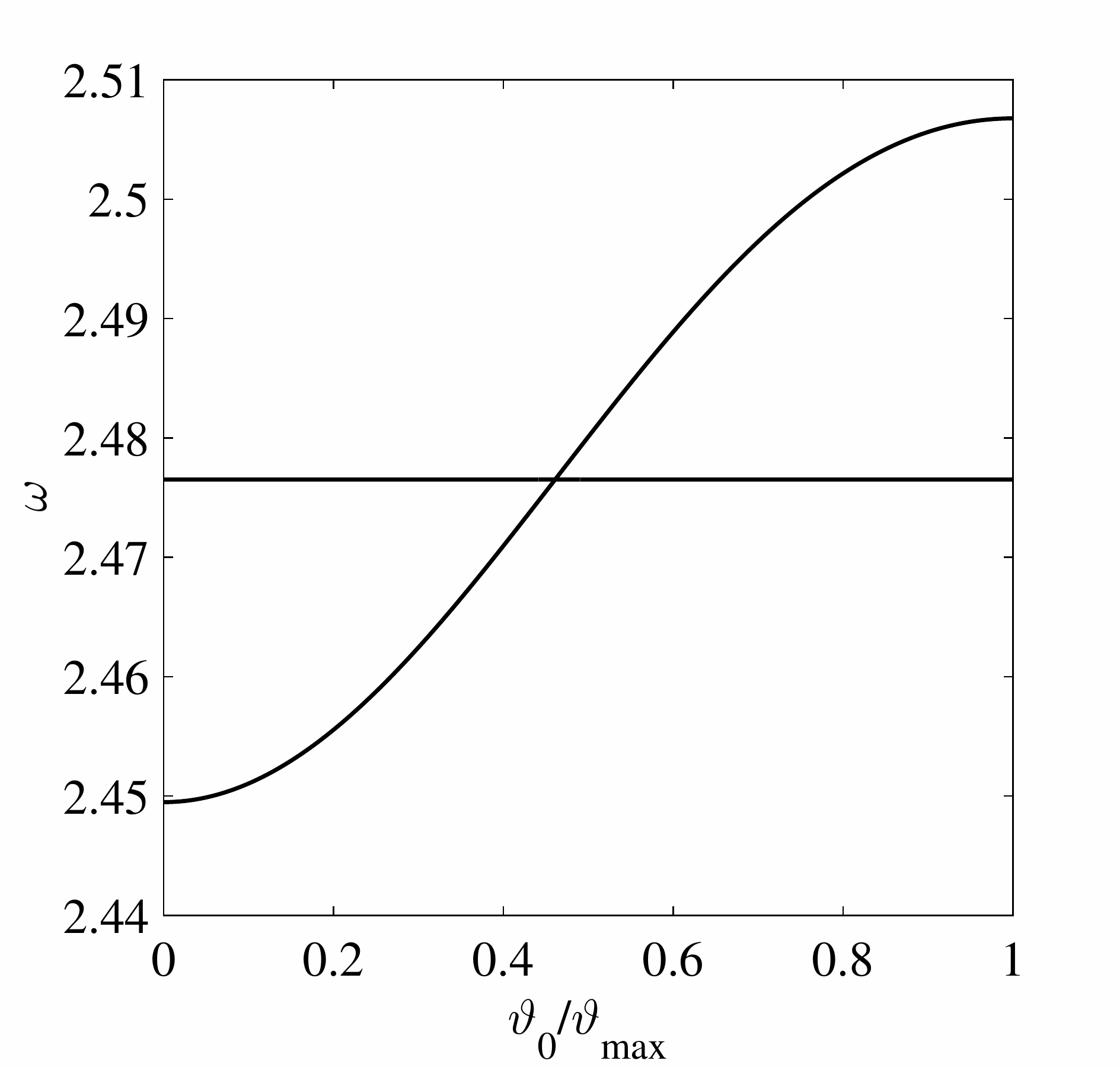}
		\caption{}
	\end{subfigure}
	\caption{\label{fig:topological_band_gap}Panel (a): The dispersion diagram in the vicinity of M for different choices of the tilting angle. Panel (b): The optical mode eigenfrequencies evaluated at point M in the first Brillouin zone as a function of $\vartheta_0$.}
\end{figure}

\begin{figure}
	\centering
	\begin{subfigure}[t]{0.48\linewidth}
		\includegraphics[width=\linewidth]{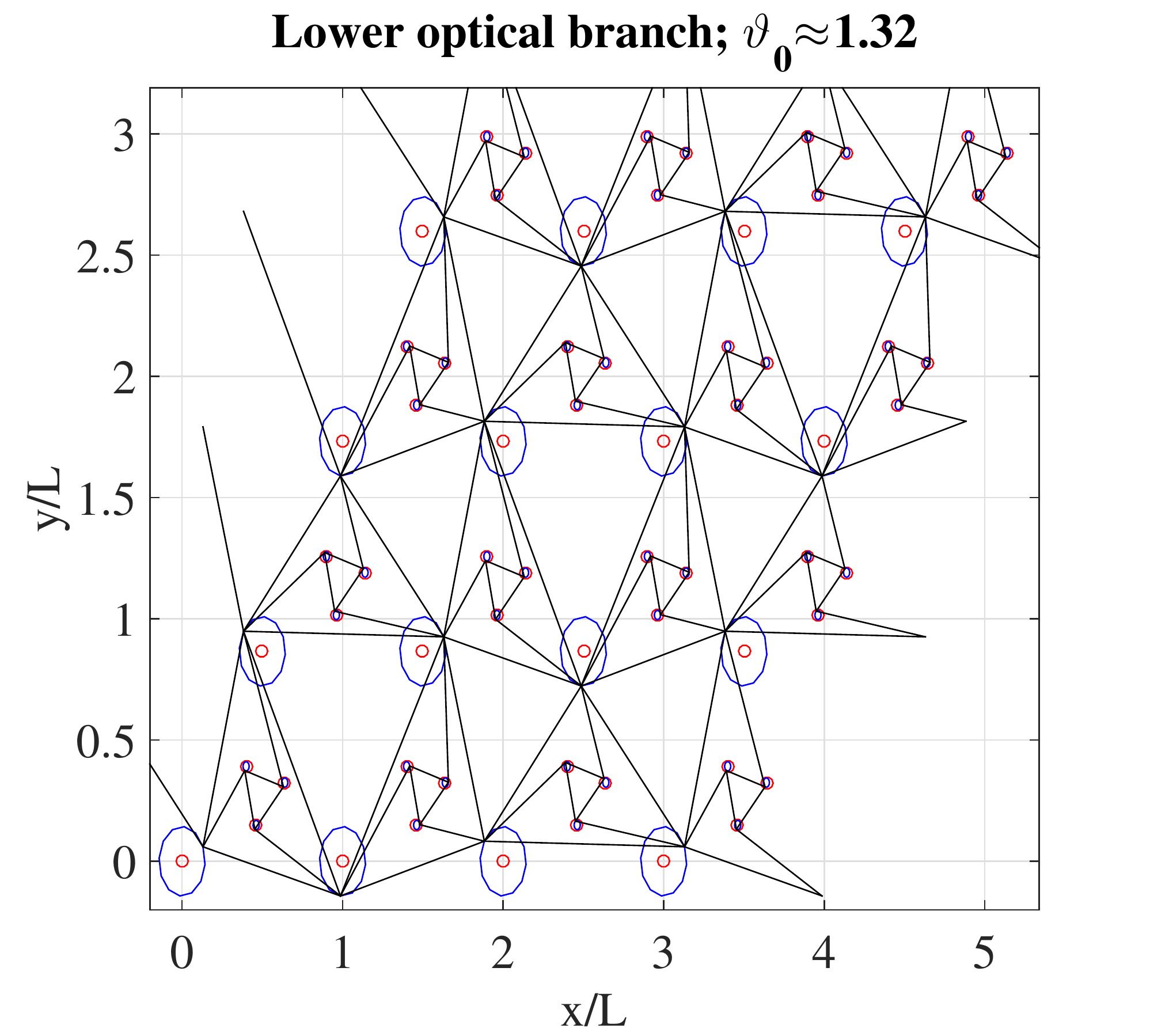}
		\caption{}
	\end{subfigure}
	\begin{subfigure}[t]{0.48\linewidth}
		\includegraphics[width=\linewidth]{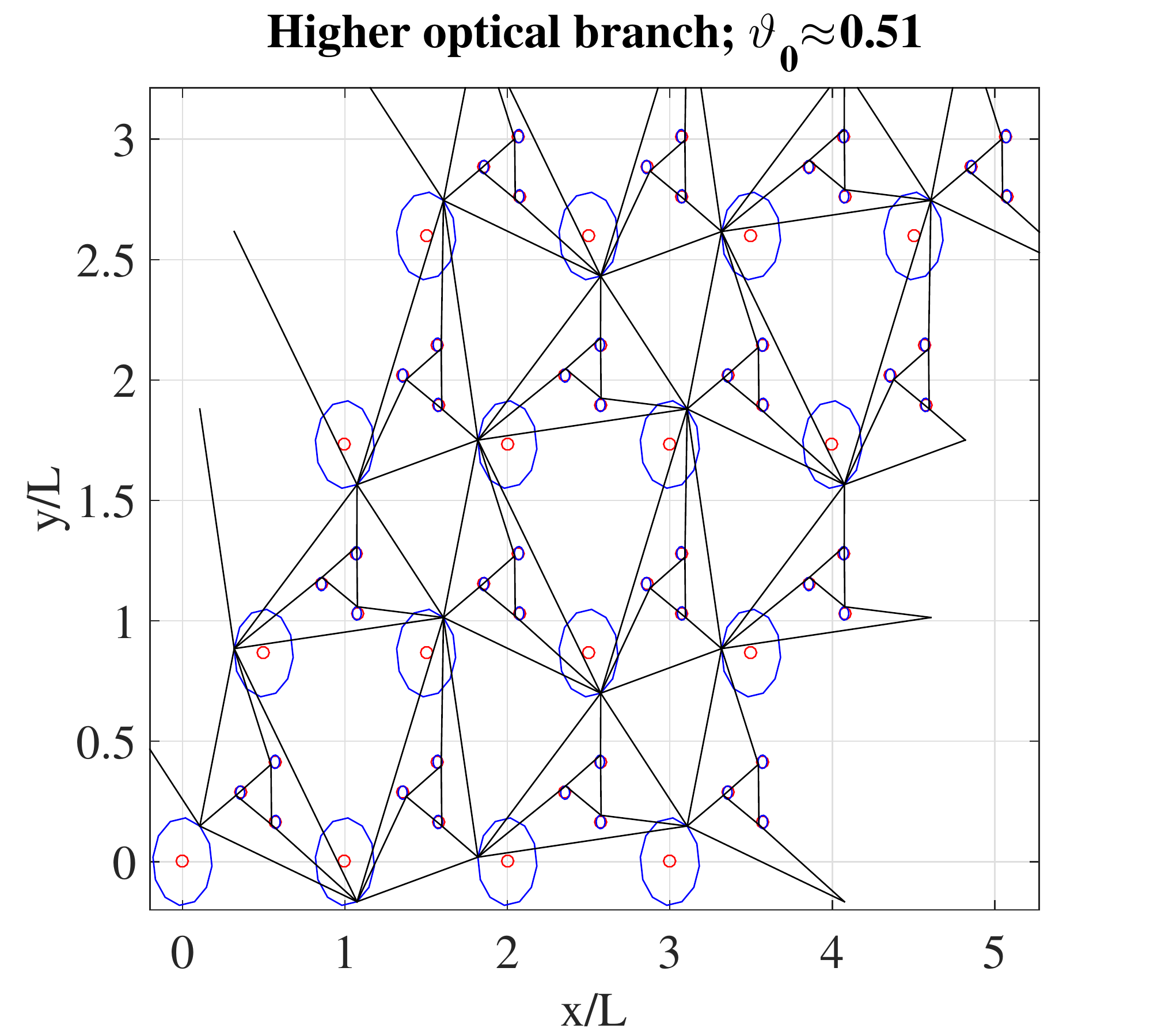}
		\caption{}
	\end{subfigure}
	\caption{\label{fig:mode_shapes}Illustrations of the Bloch eigenmodes at M. Red empty dots are the equilibrium positions of the masses in the lattice; blue lines represent the orbits of the masses in the lattice and are obtained solving the Bloch-Floquet problem. Black solid lines are trusses. Panels (a) and (b) refer to $\vartheta_0\approx0.51$ and $\vartheta_0\approx\vartheta_{\rm max}\approx1.32$, respectively. }
\end{figure}

\subsection{Dispersion and eigenmodes in the neighbourhood of a Dirac cone \label{subsec:M}}

We now examine the Dirac cones and the opening of a partial band gap at the high-symmetry point $\rm M$.
Figure~\ref{fig:topological_band_gap} (a) shows the dispersion curves, in the vicinity of $\rm M$, for the triangular lattice with TIRs;
the solid, dotted and dashed lines correspond to $\vartheta_0\approx0.51~{\rm rad}$, $\vartheta_0\approx0.61~{\rm rad}$ and $\vartheta_0\approx1.32~{\rm rad}$, respectively, and the material parameters are detailed in Table~\ref{tab:parameters}.
The two dispersion surfaces intersect and form a pair of Dirac cones  for the special value $\vartheta\approx0.61~{\rm rad}$, which was shown to give rise to a flat resonance in the dispersion diagrams  \ref{fig:curve_theta_0}(b) and  \ref{fig:curve_theta_0}(e);
these surfaces separate and form a partial band gap (see, also, Fig. \ref{fig:curve_theta_0}) for values of the tilting angle greater or less than the special value $\vartheta_0\approx0.61~{\rm rad}$. Fig. \ref{fig:topological_band_gap}(b) shows the dependence of the optical frequencies at M as a function of the tilting angle $\vartheta_{0}$; the special value of $\vartheta_0\approx0.61$ corresponds to the angle at which the curves intersect. The figure highlights the fact that one of the two optical frequencies, at $M$, does not depend on the tilting angle.  This observation suggests that, at $M$, there exists an eigenmode where the resonator does not contribution to the motion of the lattice and waves propagate purely through the ambient triangular lattice.
Fig.~\ref{fig:mode_shapes} shows the displacement amplitude fields for the triangular lattice with TIRs for two different tilting angles.
We note that the two modes are identical, up to an arbitrary phase shift, and the displacements of the resonators are small compared with those of the ambient lattice.
The red empty circles indicate the equilibrium positions of the masses, whilst the blue solid lines denote their orbit; the black solid lines indicate the trusses.
The displacement fields are obtained by finding the eigenvalues of~\eqref{eq:secular_BF} and the corresponding eigenvectors.

\section{A non-uniform vortex-type lattice
	\label{sec:modulation}}

\begin{figure}
	\centering
	\includegraphics[width=0.5\textwidth]{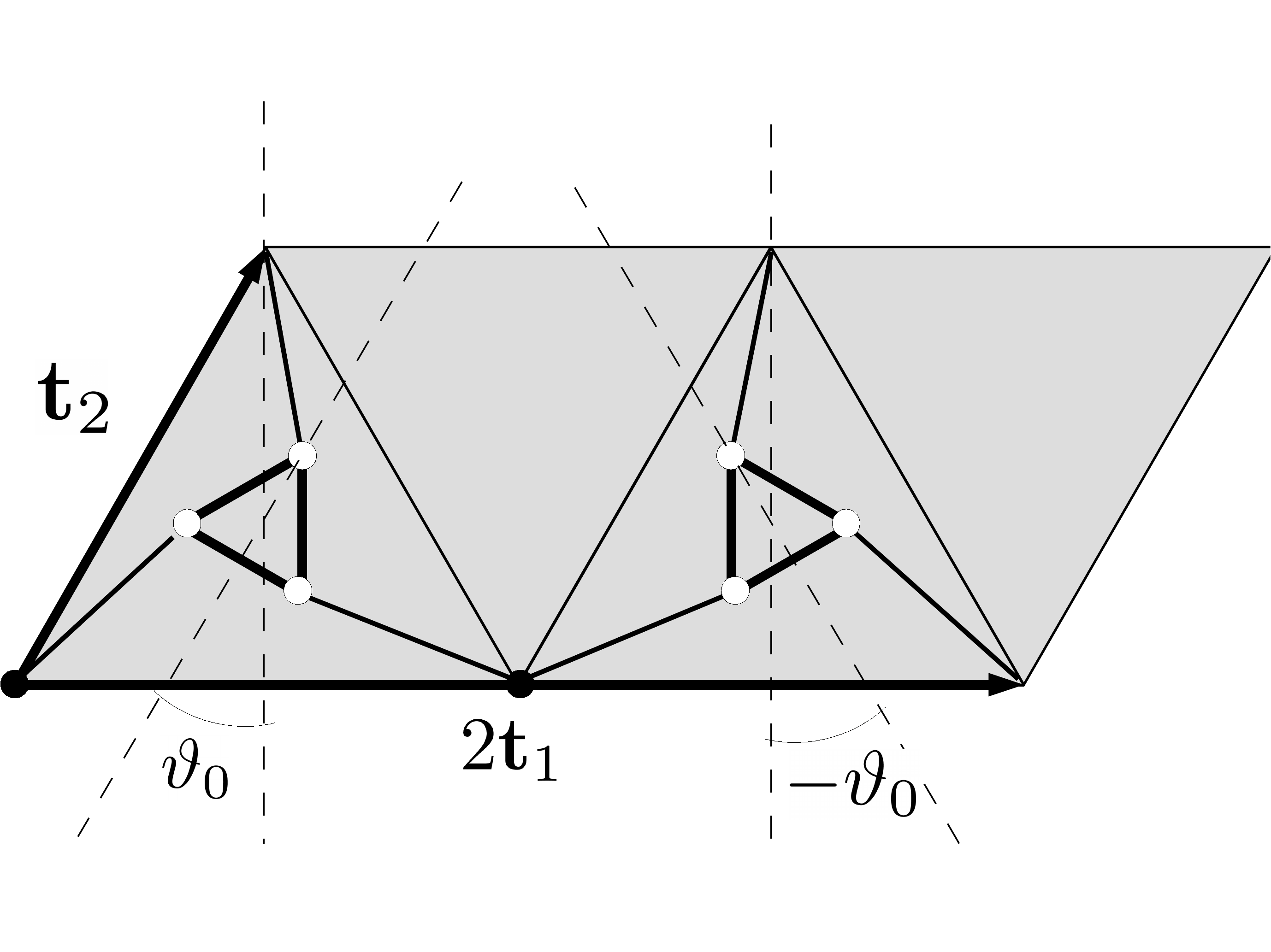}\hfill
	%	\includegraphics[width=0.5\textwidth]{fig_mc_BZ.pdf}\hfill \\
	%	$\rm (a)\,\,\,\,\,\,\,\,\,\,\,\,\,\,\,\,\,\,\,\,\,\,\,\,\,\,\,\,\,\,\,\,\,\,\,\,\,\,\,\,\,\,\,\,\,\,\,\,\,\,\,\,\,\,\,\,\,\,\,\,\,\,\,\,\,\,\,\,\,\,\,\,\,\,\,\,\,\,\,\,\,\,\,\,\,\,\,\,\,\,\,\,\,\,\,\,\,\,\,\,\,\,\,\,\,\,\,\,\,\,\,\,\,\,\,\,\,\,\,\,\,\,\,\,\,(b)\,\,\,\,\,\,\,\,\,\,\,\,\,\,\,\,\,\,\,\,$ 
	\caption{\label{fig:system_MC}A schematic representation of a macrocell where a specific variation of the tilting angle is introduced.}
	%	\caption{\label{fig:system_MC}Panel (a) shows the geometry of a macro-cell which comprises resonators rotated of opposite angles; the primitive vectors of the lattice are shown - see Eq. (\ref{eq:t_unit_vects}) for further reference. Panel (b) illustrates the reciprocal lattices: red dots are the lattice sites in reciprocal space stemming from the unit cell in Panel (a)  while blue circles denote lattice sites in reciprocal space for unit cell in Fig.\ref{fig:system_t} (a) ; the primitive vectors in reciprocal space for the macro cell are drawn - see Eq. (\ref{eq:G_unit_vects}) for further reference; given two nearest-neighbours lattice sites belonging to the same lattice, the thin dotted lines are the loci of points equidistant from the considered sites.  }
\end{figure}

\begin{figure}
	\centering
	\begin{subfigure}[t]{0.48\linewidth}
		\includegraphics[width=\linewidth]{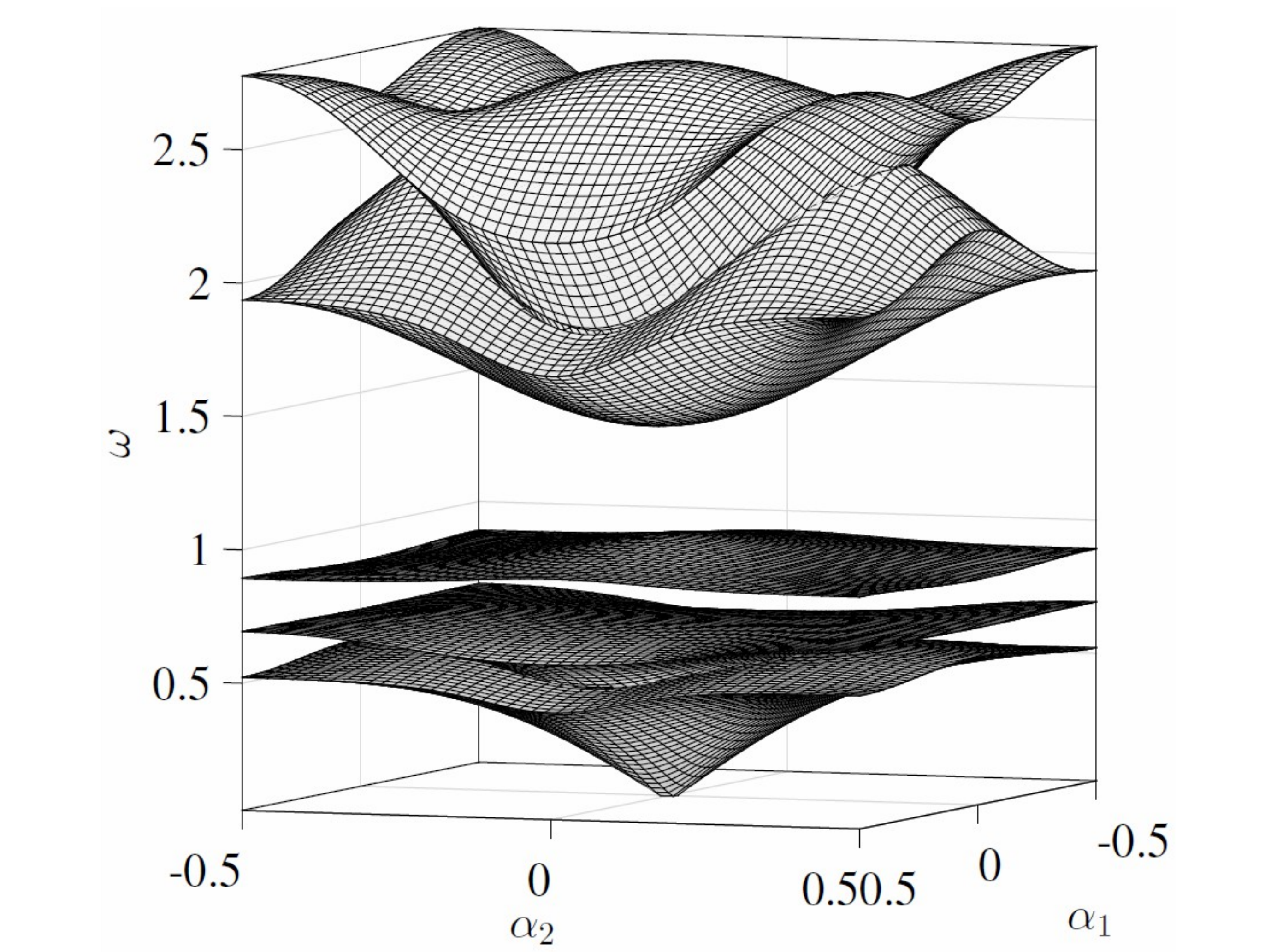}
		\caption{}
	\end{subfigure}
	\begin{subfigure}[t]{0.48\linewidth}
		\includegraphics[width=\linewidth]{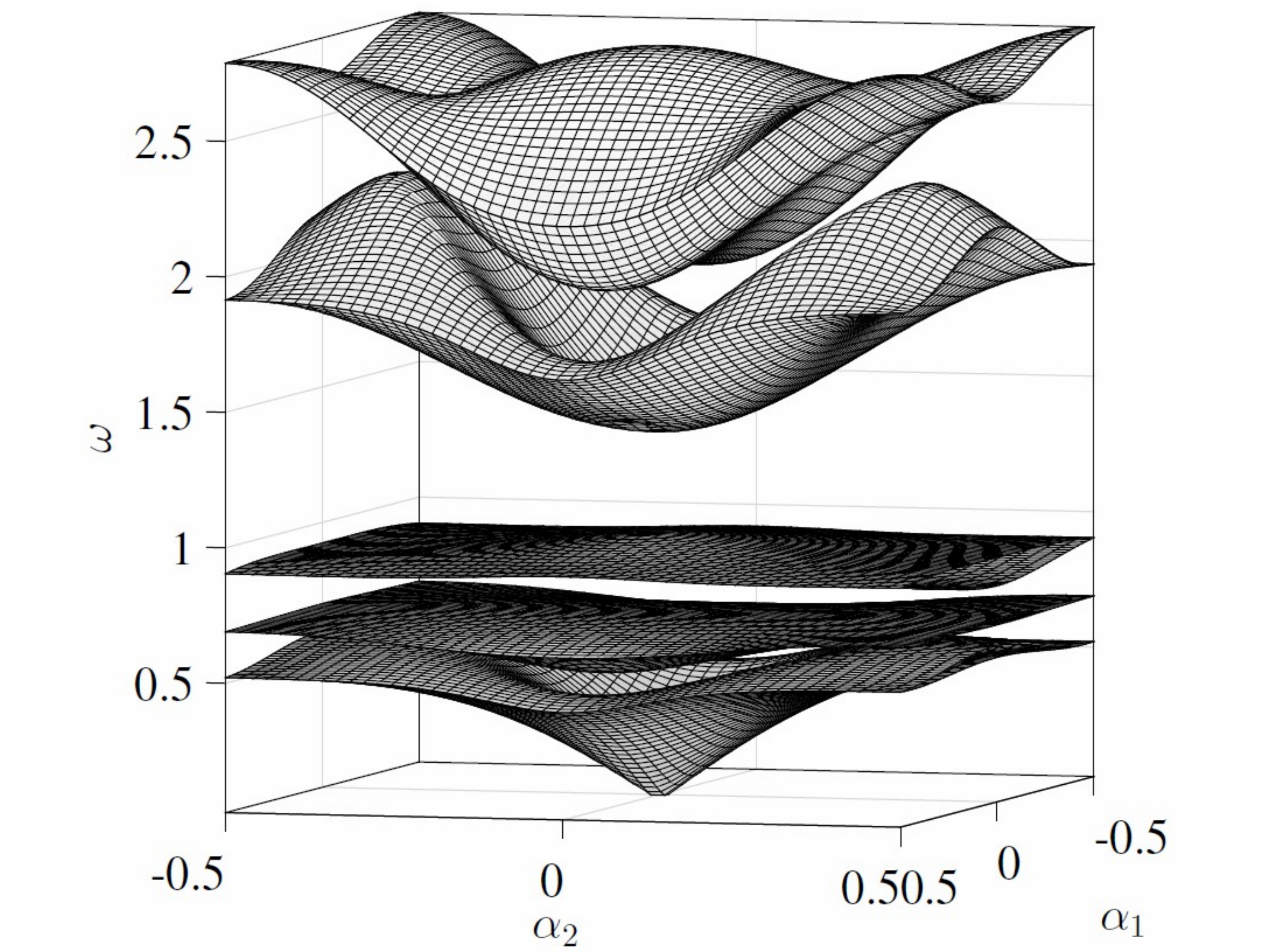}
		\caption{}
	\end{subfigure}
	\begin{subfigure}[t]{0.48\linewidth}
		\includegraphics[width=\linewidth]{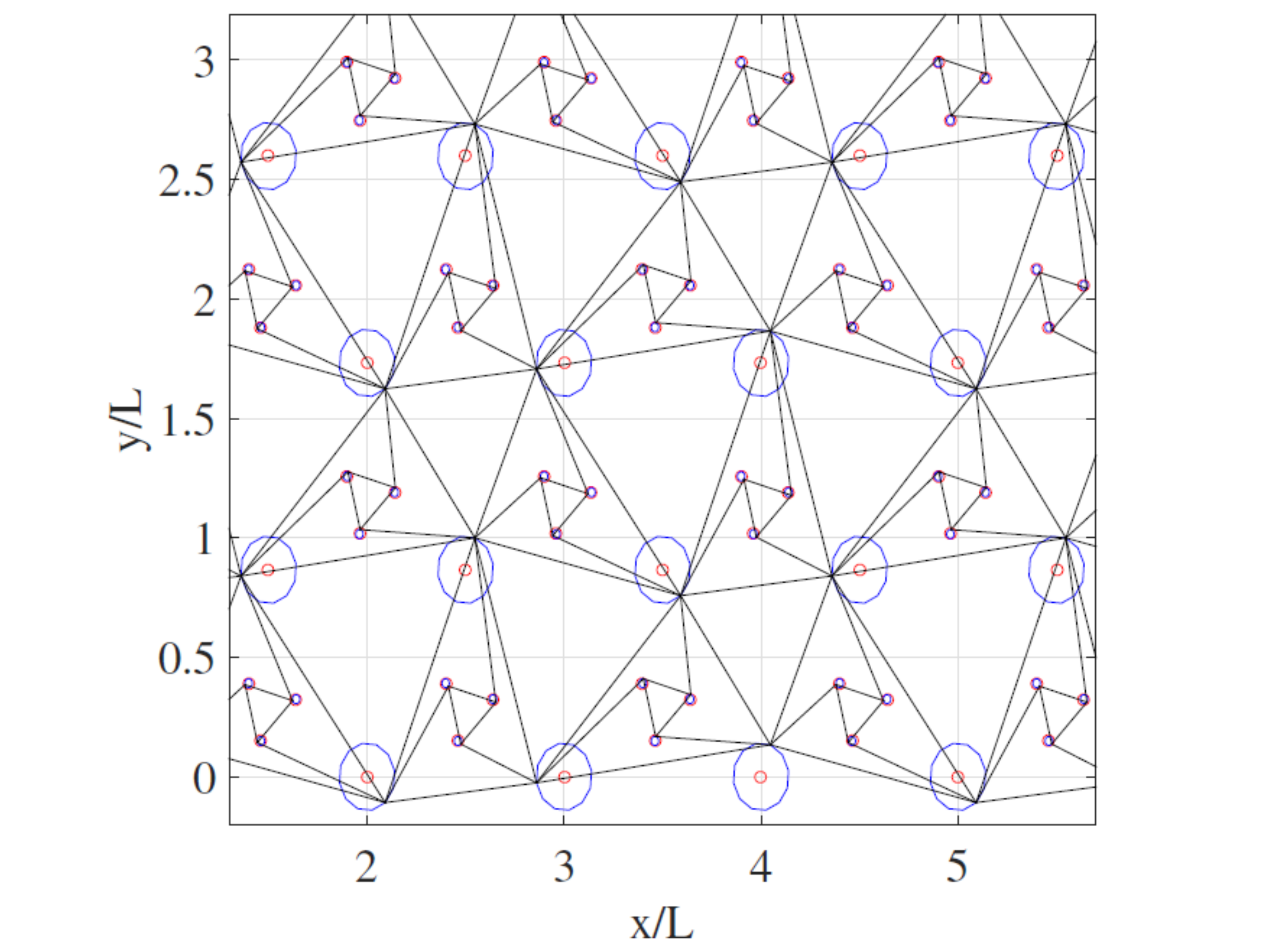}
		\caption{}
	\end{subfigure}
	\begin{subfigure}[t]{0.48\linewidth}
		\includegraphics[width=\linewidth]{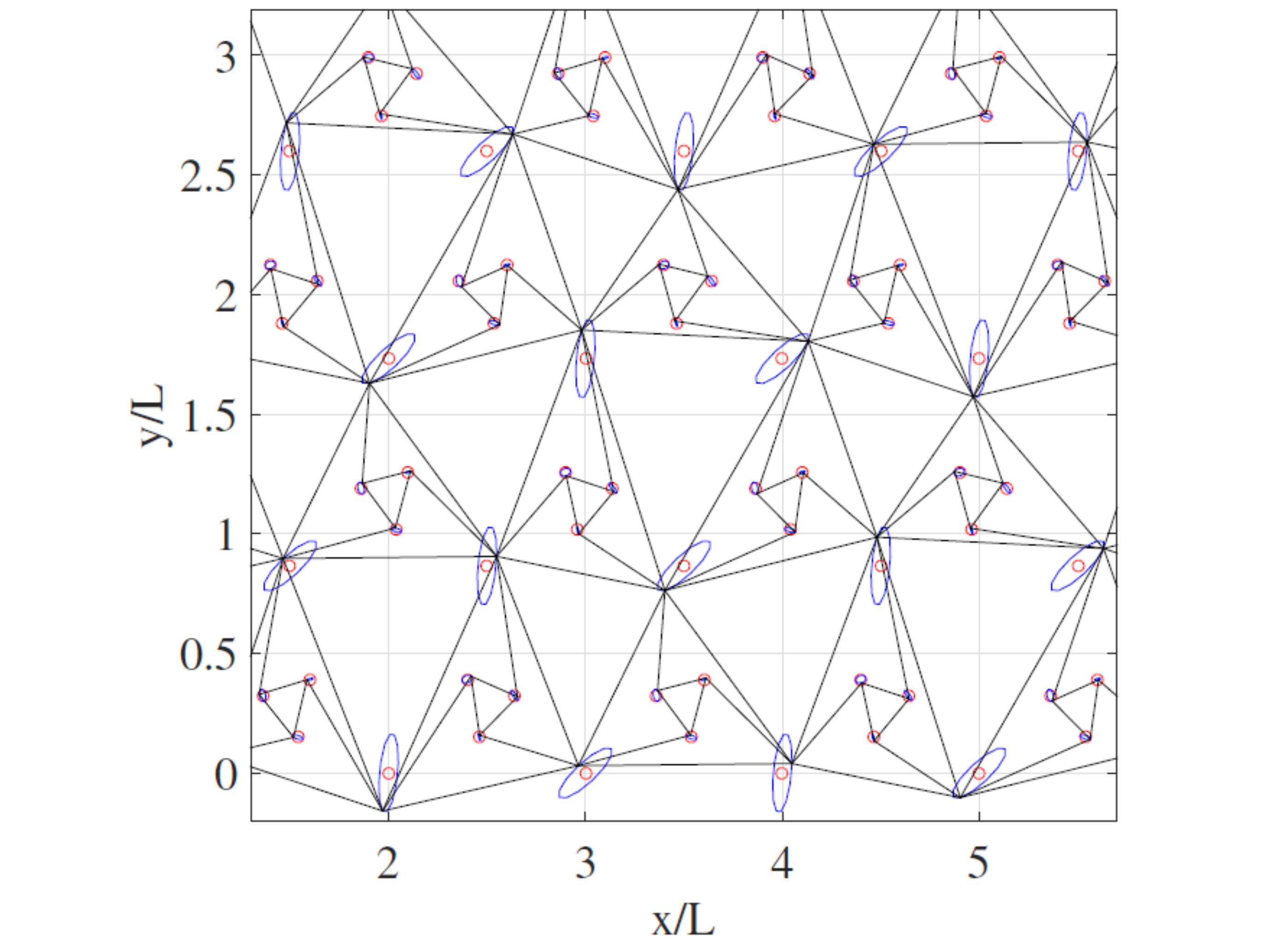}
		\caption{}
	\end{subfigure}
	\caption{\label{fig:surf_angles_macro-cell} Panels (a) and (b) are dispersion surfaces for two different macro-cells. Panel (a) corresponds to a  macro-cell whose resonators are rotated of the same angle $\vartheta_0=\vartheta_{\rm max}$. In Panel (b), a resonator is rotated of an angle $\vartheta_0=\vartheta_{\rm max}$ and the other one of an angle  $\vartheta_0=-\vartheta_{\rm max}$,  as shown in Fig. \ref{fig:system_MC}(a). The remaining parameters are fixed as listed in Table \ref{tab:parameters}. Panels (c) and (d) are illustrations of Bloch waves at ${\rm M}'$ which correspond to the macro-cells considered in panels (a) and (b), respectively. Symbols are the same  as in Fig. \ref{fig:mode_shapes}. In panel (c)  and (d) the Bloch frequencies at ${\rm M}'$ are $\omega\approx 2.75~{\rm rad/s}$  and   $\omega\approx 2.82~{\rm rad/s}$, respectively. }
\end{figure}
%\begin{figure}
%	\centering
%	\includegraphics[width=1\textwidth]{fig_eigs_mc_same_angle.pdf}\hfill  \\
%		$\rm \,\,\,\,\,\,\,\,\,\,\,\,\,\,\,\,\,\,\,\,\,\,\,\,\,\,\,\,\,\,\,\,\,\,\,\,\,\,\,\,\,\,\,\,\,\,\,\,\,\,\,\,\,\,\,\,\,\,\,\,\,\,\,\,\,\,\,\,\,\,\,\,\,(a)\,\,\,\,\,\,\,\,\,\,\,\,\,\,\,\,\,\,\,\,\,\,\,\,\,\,\,\,\,\,\,\,\,\,\,\,\,\,\,\,\,\,\,\,\,\,\,\,\,\,\,\,\,\,\,\,\,\,\,\,\,\,\,\,\,\,\,\,\,\,\,\,\,$ 
%	\includegraphics[width=1\textwidth]{fig_eigs_mc_opposite_angle.pdf}\hfill \\
%		$\rm \,\,\,\,\,\,\,\,\,\,\,\,\,\,\,\,\,\,\,\,\,\,\,\,\,\,\,\,\,\,\,\,\,\,\,\,\,\,\,\,\,\,\,\,\,\,\,\,\,\,\,\,\,\,\,\,\,\,\,\,\,\,\,\,\,\,\,\,\,\,\,\,\,(b)\,\,\,\,\,\,\,\,\,\,\,\,\,\,\,\,\,\,\,\,\,\,\,\,\,\,\,\,\,\,\,\,\,\,\,\,\,\,\,\,\,\,\,\,\,\,\,\,\,\,\,\,\,\,\,\,\,\,\,\,\,\,\,\,\,\,\,\,\,\,\,\,\,$ 
%	\caption{\label{fig:eigs_angles_macro-cell} }
%\end{figure}
%

Thus far we have only considered uniform lattices where the tilting angle of each resonator is identical.
In the present section, we examine the effects of allowing the titling angle to vary from one resonator to the next.
In particular, we consider a configuration where adjacent cells in the same row have TIRs rotated by the same angle but in the opposite direction, as shown in Fig.~\ref{fig:system_MC}.
Consequently, the vibrational modes may involve counter-current rotations of resonators in the neighbouring cells of the structure.

The variation in angle can be conveniently accommodated by introducing the macro-cell illustrated in Fig.~\ref{fig:system_MC}, where the primitive lattice vectors are marked (c.f. Eq.~\eqref{eq:t_unit_vects}).
In this case, the unit cell comprises two masses $m$ (black solid dots) belonging to a trapezoidal unit cell and two  triangular resonators of side $\ell$.
The same assumptions made for the resonator introduced in Sec. \ref{sec:bloch} are applied here, i.e. that $c_o/c_{\ell o} \gg 1$ and $c_o/c_\ell \gg1$.
The two masses are located at the basis vectors
\begin{equation}\label{eq:basis-vectors-mc}
\bm{d}_0=
\vec{0}\qquad{\rm and}\qquad
\bm{d}_1=L
\begin{pmatrix}
1\\0
\end{pmatrix}, 
\end{equation}
respectively. Similarly, the rest positions of the centres of mass of the resonators are 
\begin{equation}\label{eq:cms-mc}
\tilde{\bm{r}}_{{\rm cm},i}=\tilde{\bm{r}}_{\rm cm}+\bm{d}_i,
\end{equation} 
where the vector $\tilde{\bm{r}}_{\rm cm}$ was introduced in Eq. (\ref{eq:rcm_eq_t}), and the vectors ${\bm d}_i$ with $i=\{0,1\}$ are given in Eq. (\ref{eq:basis-vectors-mc}).
The left resonator in Fig. \ref{fig:system_MC} is tilted of an angle $\vartheta_{0}$ and the right one of an angle $-\vartheta_0$, as illustrated.
Since our aim is to compare the dispersion properties arising from the the unit cells in Figs \ref{fig:system_t}(a) and \ref{fig:system_MC}, it is necessary to choose a common region in which to perform the comparison. 

In this section, we present results for the frequency surfaces as a function of the Bloch vector  
\begin{equation}\label{eq:k_in_mc_BZ}
{\bm k} = {\alpha}_1 \bm{K}_1 + \alpha_2 \bm{K}_2,
\end{equation}
where $(\alpha_1,\alpha_2)\in[-1/2,1/2]^2$ and 
\begin{equation}
\bm{K}_1=\frac{1}{2}\bm{G}_1~~~~~{\rm   and   }~~~~~ \bm{K}_2=\frac{1}{2}\bm{G}_1+\bm{G}_2, 
\end{equation}
with ${\bm G}_1$ and ${\bm G}_2$ given in Eq. (\ref{eq:G_unit_vects}). In particular, Fig.~\ref{fig:surf_angles_macro-cell} shows the dispersion surfaces over the set of Bloch wave vectors \eqref{eq:k_in_mc_BZ} and eigenmodes for two cases.
Panel (a) shows the dispersion surfaces for the case when the two resonators in the macro-cell, shown in Fig.~\ref{fig:system_MC}, are tilted by the same angle $\vartheta_{0}=\vartheta_{\rm max}$.
In panel (b), the two resonators are rotated by opposite angles.
The physical parameters used are listed in Table \ref{tab:parameters}.
We note the partial band gap, between the two optical surfaces, which opens when the two resonators are tilted in opposite directions.

In the subsection \ref{subsec:M}, we highlighted some features of Bloch-Floquet waves at point $\rm M$.
The corresponding point for the macro-cell shown in Fig.~\ref{fig:system_MC} is 
\begin{equation}\label{eq:M_prime}
{\rm M}'= \frac{\pi}{3L} 
\begin{pmatrix}
1 \\
\sqrt{3}
\end{pmatrix}.
\end{equation}
When the resonators in the macro-cell are rotated in the same direction, it is natural to expect the same Bloch eigenmodes at $\rm M'$ that we would get at $\rm M$ for the single cell.
Fig. \ref{fig:surf_angles_macro-cell}(c) shows the displacement amplitude field at $\rm M'$ for the structure whose dispersion surface is given in Fig. \ref{fig:surf_angles_macro-cell}(a).
We observe that the eigenmode shown in Fig.~\ref{fig:surf_angles_macro-cell}(c) is equivalent, up to an arbitrary phase shift, to that shown in Fig.~\ref{fig:mode_shapes}(b) for the single resonator cell corresponding to the point $\rm M$.
Fig.~\ref{fig:surf_angles_macro-cell}(d) shows the Bloch eigenmode at $\rm M'$ corresponding to a macro-cell with resonators rotated in opposite directions; the corresponding dispersion surfaces are shown in Fig.~\ref{fig:surf_angles_macro-cell}(b).
In this case, we observe that the orbits described by the triangular lattice notes are elliptic.
As in the previous case of single resonator unit cells, 
the resonators do not contribute to the propagation of elastic Bloch waves.
\begin{table}[h!]
	\centering
	\begin{tabular}{@{}llllllr@{}} \toprule%\cmidrule(r){1-6}
		&$c_\ell$ 					   &						$c_{\ell o}$& $L$ 				   &     $\ell$ &                  $m$ &    $ m_o$                     \\ 
		\midrule
		TLR																 &9 ${\rm N/m}$            &                  1.35 $c_{\ell}$ & 1 ${\rm m}$     &$1/4~L$ &  1 ${\rm Kg}$      &         1 ${\rm Kg}$  \\ 
		TL                      									      &9 ${\rm N/m}$            &                               & 1 ${\rm m}$     &              &  4 ${\rm Kg}$    		&     \\ 
		\bottomrule
	\end{tabular}
	\caption{\label{tab:parameters_comsol}Physical parameters used in ${\rm COMSOL\, Multiphysics}^{\circledR}$.}
\end{table}
\section{Transmission of elastic waves through slabs of TIRs \label{sec:transmission}}
\begin{figure}
	\centering
	\begin{subfigure}[t]{0.48\linewidth}
		\includegraphics[width=\linewidth]{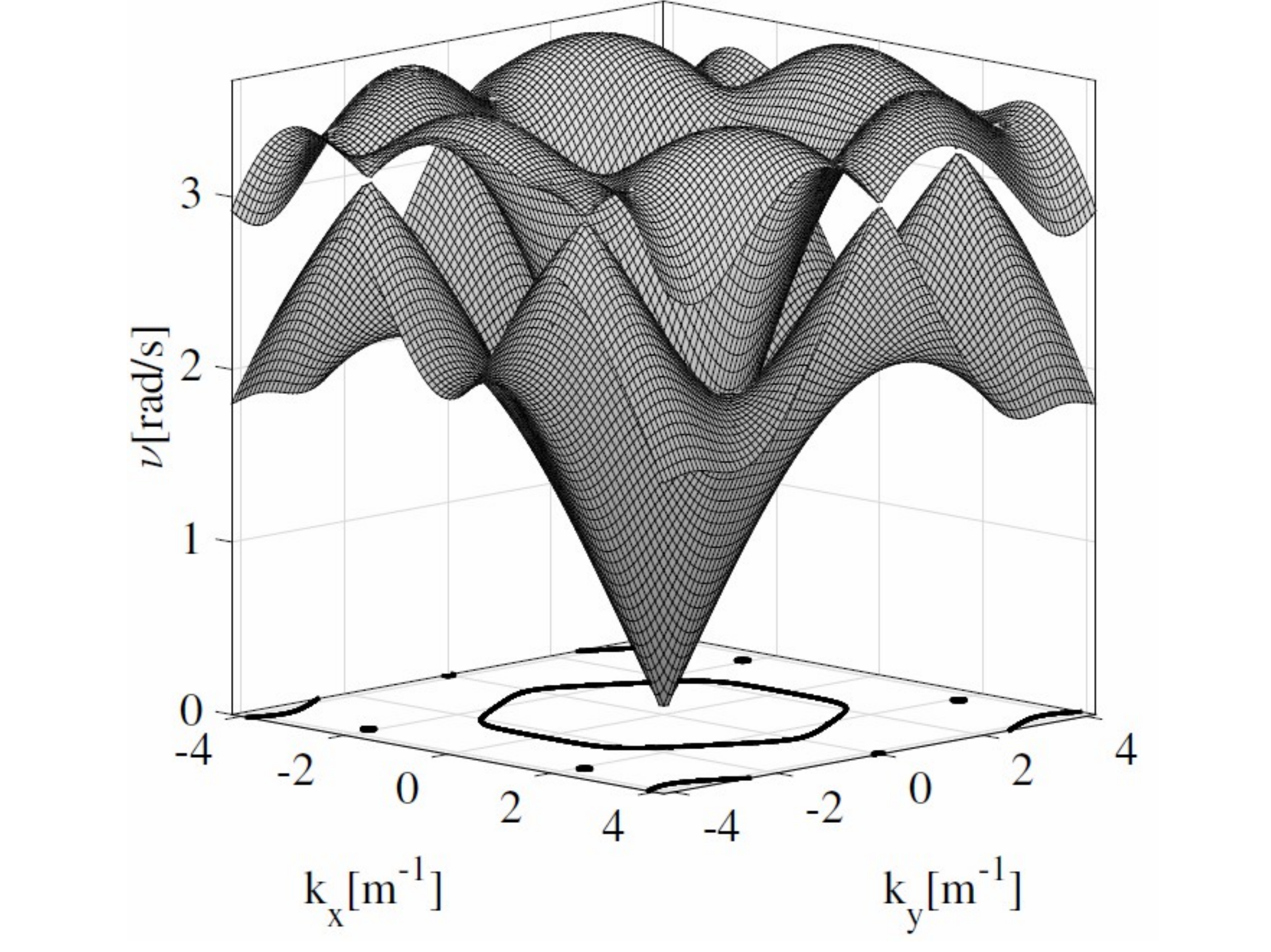}
		\caption{}
	\end{subfigure}
	\begin{subfigure}[t]{0.48\linewidth}
		\includegraphics[width=\linewidth]{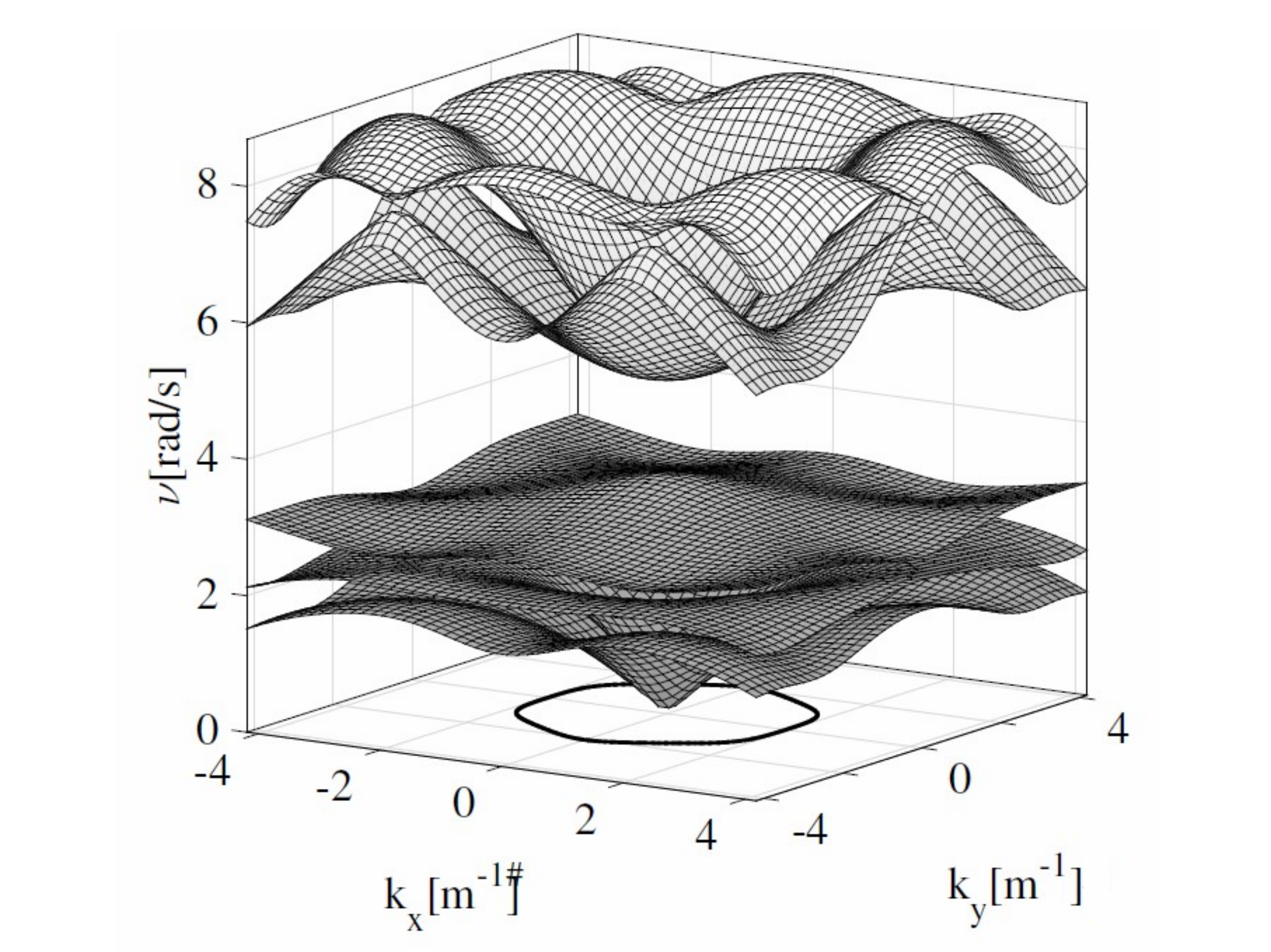}
		\caption{}
	\end{subfigure}
	\\
	\vspace{\baselineskip}
	\begin{subfigure}[t]{0.48\linewidth}
		\includegraphics[width=\linewidth]{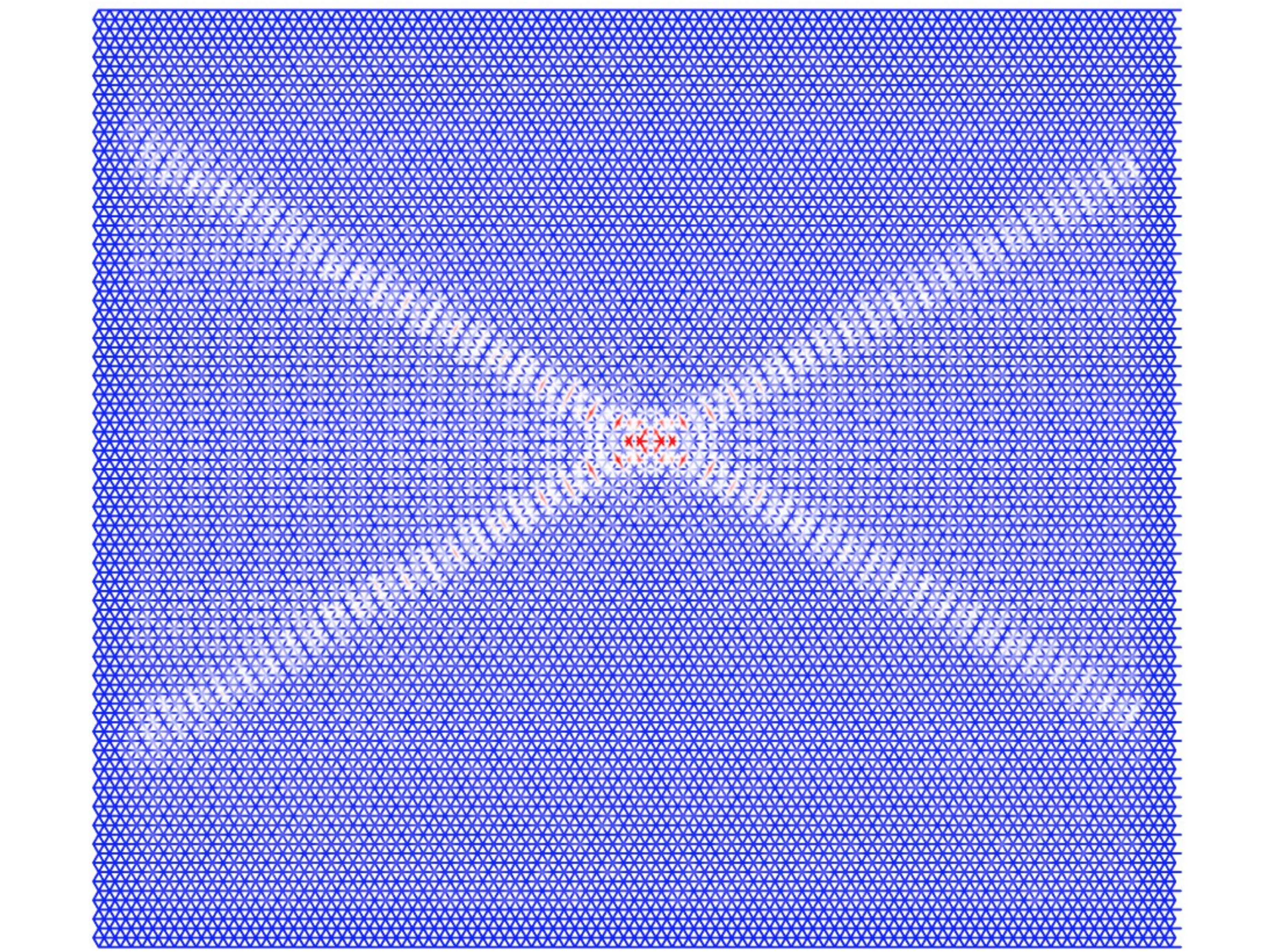}
		\caption{}
	\end{subfigure}
	\begin{subfigure}[t]{0.48\linewidth}
		\includegraphics[width=\linewidth]{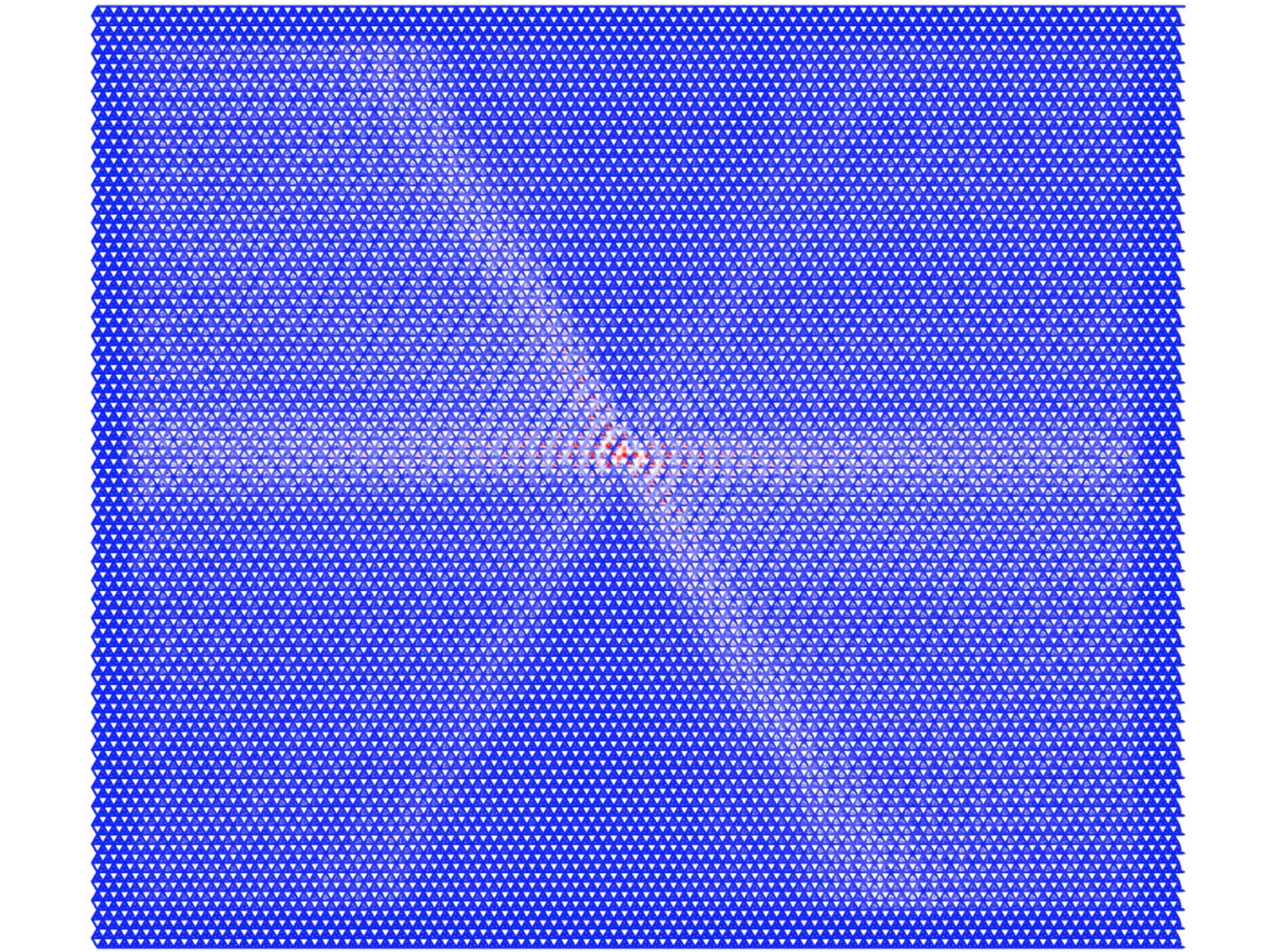}
		\caption{}
	\end{subfigure}
	\caption{\label{fig:harm_loading} Panel (a) and (b) are dispersion surfaces for a monatomic triangular lattice (TL) and a triangular lattice with TIRs (TLR), respectively. The slowness contours at $\nu=\pi~ {\rm rad/s}$ are also reported as projections onto the plane $\nu=0$. The parameters used are listed in Table \ref{tab:parameters_comsol}.  Panel (c) and (d)  are  elastic responses of the TL and TLR, respectively considered in panels (a) and (b).  The source of the excitation has amplitude as in Eq. (\ref{eq:force_amplitude}), with $F=0.1~{\rm N}$ and angular frequency $\nu=\pi~{\rm rad/s}$.}
\end{figure}
\begin{figure}
	\centering
	\begin{subfigure}[t]{0.48\linewidth}
		\includegraphics[width=\linewidth]{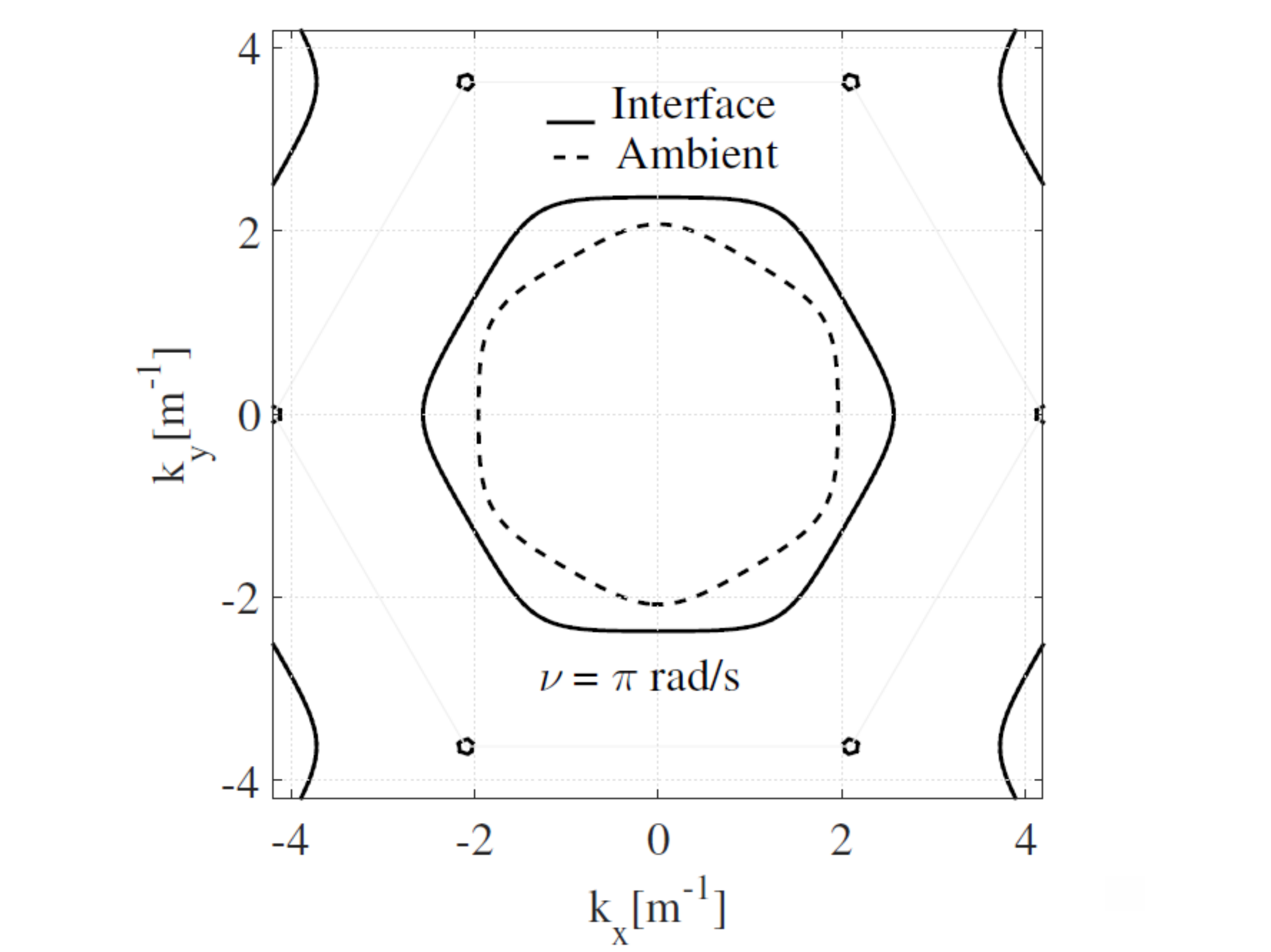}
		\caption{}
	\end{subfigure}
	\begin{subfigure}[t]{0.48\linewidth}
		\includegraphics[width=\linewidth]{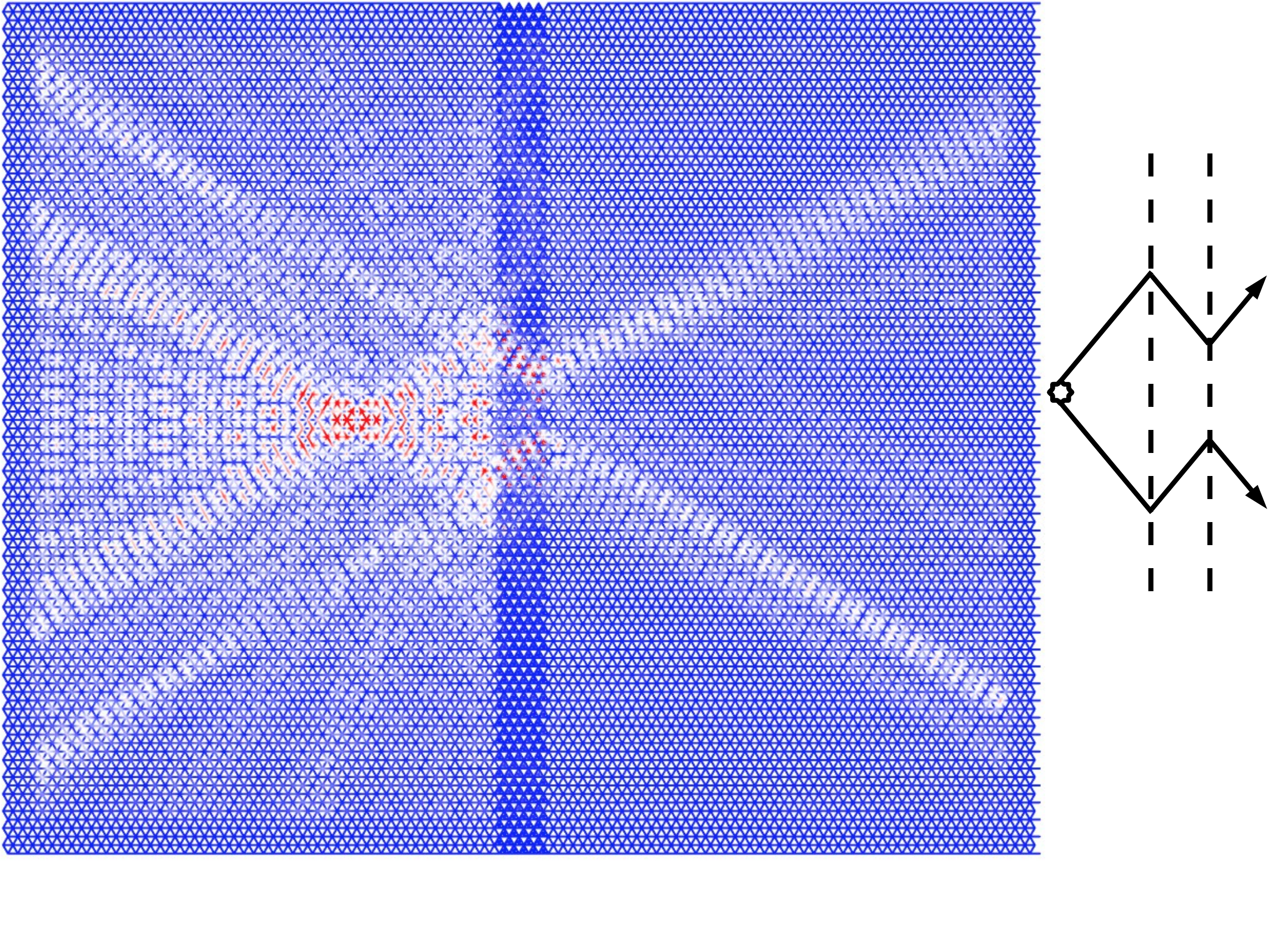}
		\caption{}
	\end{subfigure}
	\\
	\vspace{\baselineskip}
	\begin{subfigure}[t]{0.48\linewidth}
		\includegraphics[width=\linewidth]{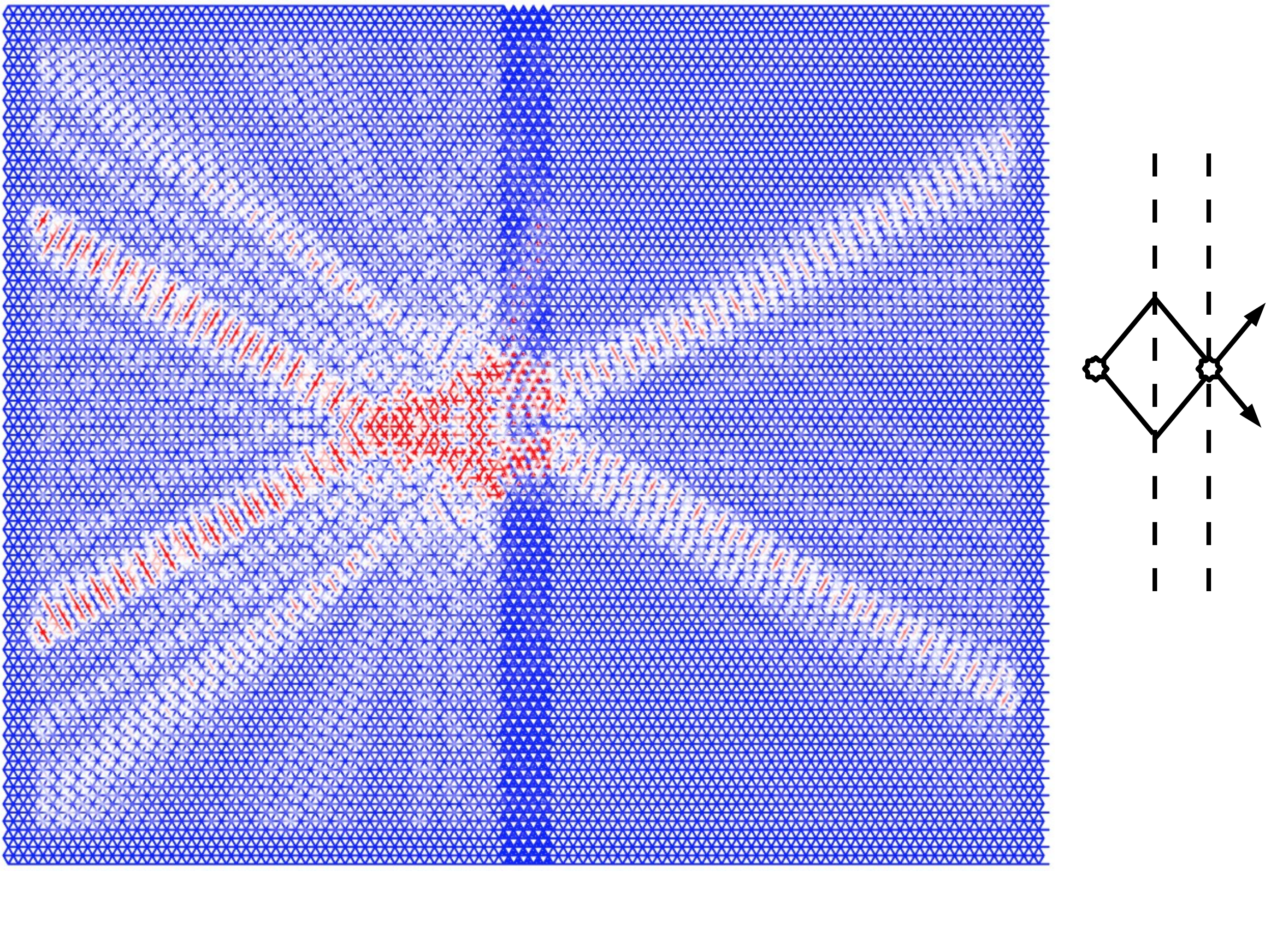}
		\caption{}
	\end{subfigure}
	\begin{subfigure}[t]{0.48\linewidth}
		\includegraphics[width=\linewidth]{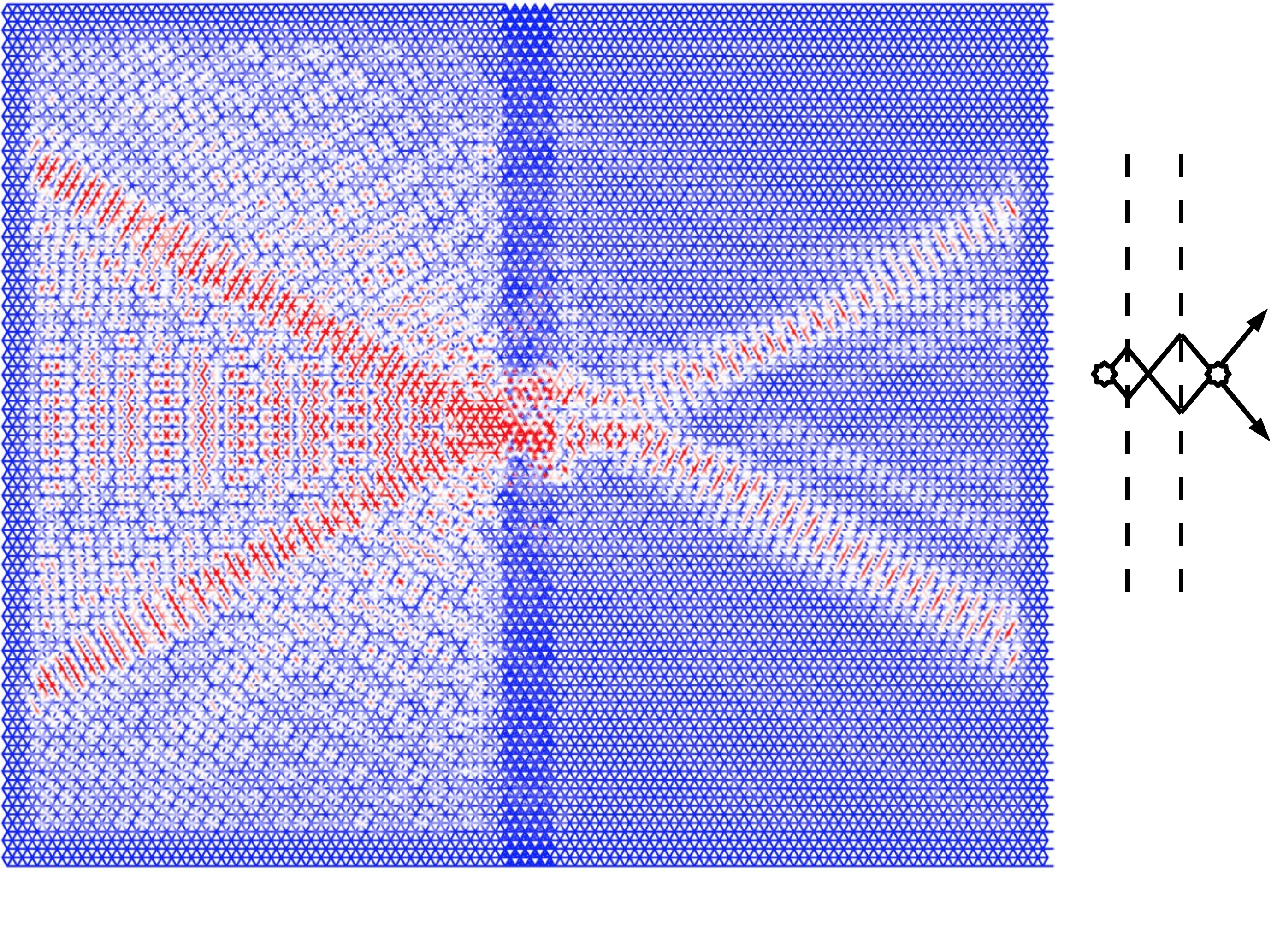}
		\caption{}
	\end{subfigure}
	\caption{\label{fig:negative_refraction}Panel (a): A comparison of slowness curves presented in Fig. \ref{fig:harm_loading}; the slowness curves are plotted at the same  frequency as the one of the harmonic excitation, that is $\nu=\pi~{\rm rad/s}$.  Panel (b), (c) and (d): Numerical simulations of a negative refractive flat lense containing TIRs surrounded by a TL; the physical parameters for the TL and TLR are fixed as in Table \ref{tab:parameters_comsol}. The time-harmonic force is introduced in  Eq. (\ref{eq:force_amplitude}) and has amplitude $F=0.1~{\rm N}$ and angular frequency $\nu=\pi~{\rm rad/s}$. In panels (b), (c) and (d), the distance of the point source away from from the left interface is 20, 10 and 1 lattice sites.  }
\end{figure}

In this section we focus on the scattering of in-plane elastic waves by a slab of TIRs, embedded in a triangular elastic lattice.
Although the scattering problem is formally different from the Bloch-Floquet problem, it has been shown~\cite{Colquitt2011,Colquitt_2012} that there exists a connection between the two classes of problems and that the dispersive properties of the infinite lattice can be used to make qualitative predictions regarding scattering problems.
In the present section, we are primarily concerned with the range of frequencies corresponding to the novel branch of the dispersion surfaces which is associated with the chiral effects of the resonator.

We start by examining the forced problem for the triangular lattice both with, and without, TIRs.
Similar problems have been extensively studied in the literature, with the classical reference text being the book by Maradudin \emph{et al.}~\cite{Maradudin1963}.
We also mention the papers by Martin~\cite{Martin_2006} and Movchan and Slepyan~\cite{Movchan_2009}, which analyse the properties of the dynamic Green's functions for a square lattice in the pass and stop band, respectively.
The dynamic Green's functions for square and triangular elastic lattices were examined by Colquitt  \emph{et al.} in~\cite{Colquitt_2012}, with a particular emphasis on the resonances associated with dynamic anisotropy and \emph{primitive waveforms}.
These primitive waveforms have also been examined in the papers by Langley~\cite{Langley_JSV_197_447_1996}, Ruzzene \emph{et al.}~\cite{Ruzzene2003}, Ayzenberg-Stepanenko and Slepyan~\cite{Ayzenberg-Stepanenko2008}, and Osharovich \emph{et al.}~\cite{Osharovich2010}, among others.

The Green's function for a lattice with triangular periodicity can be written in the form of a Fourier integral over the first Brillouin zone
\begin{equation}\label{eq:green_RS}
{\bm G}_{mn}(\omega) = \frac{\sqrt{3}L^2}{8\pi^2}\int_{{\bm k}\in {\rm BZ}} {\rm d}{\bm k}~{\bm g}_{mn}({\bm k},\omega)e^{-i{\bm k}\cdot(n{\bm t}_1+m{\bm t}_2)},
\end{equation} 
where ${\bm t}_1$ and ${\bm t}_2$ have been introduced in Eq. (\ref{eq:t_unit_vects}), and $m$ and $n$ are integers. The influence of the rigid TIRs is captured by the Fourier transformed Green's function 
\begin{equation}\label{eq:green_FS}
{\bm g}_{mn}({\bm k},\omega)=\left(\hat{\Sigma}_{\bm k}-\omega^2\hat{\cal M}\right)^{-1}{\bm F},
\end{equation}
where $\hat{\Sigma}_{\bm k}$ is given in Eq. \eqref{eq:Sigma_k} and $\hat{{\cal M}}$ in Eq. \eqref{eq:mass_BF_rigid}.
In Eq. (\ref{eq:green_FS}), ${\bm F}$ represents the amplitude of the time-harmonic forcing applied at the origin.
Solutions of Eq. (\ref{eq:green_RS}) are obtained using ${\rm COMSOL\, Multiphysics}^{\circledR}$.  
The numerical parameters used for the illustrative computations presented in this section, are listed in Table \ref{tab:parameters_comsol}.
Figs. \ref{fig:harm_loading} (a) and \ref{fig:harm_loading}(b) so the dispersion surfaces for a triangular lattice and a triangular lattice with TIRs, respectively.
The slowness contours at $\nu=\pi ~{\rm rad/s}$ are also shown as projections onto the $\nu=0$ plane.
Figs. \ref{fig:harm_loading}(c) and \ref{fig:harm_loading}(d) show the magnitude of the displacement field generated by an harmonic point load.
In particular, panels (c) and (d) show the fields for a triangular lattice and a triangular lattice with TIRs, respectively.
The source of the excitation is a point, time-harmonic loading exerted at the centre of the computational window.
The orientation of the applied load to chosen to be parallel to the horizontal $x$-axis and of unit amplitude, such that
\begin{equation}\label{eq:force_amplitude}
{\bm F} =F\left(1,0,0,0,0\right)^{\rm T}.
\end{equation}
The angular frequency of the excitation is $\nu=\pi~{\rm rad/s}$.
Figs \ref{fig:harm_loading}(c) and \ref{fig:harm_loading}(d) show the expected strong dynamic anisotropy.
In particular, panel (c) shows a cross-like propagation pattern with the arms of the cross at $\pi/6$ radians with respect to the horizontal direction.
This  is consistent with the slowness contour in panel (a) with the waves propagating in the directions normal to the  hexagon-like slowness contour.
No wave propagation is observed in the vertical direction because the source is polarised  in the ${x}$ direction.
Panel (d) shows a similar star-like shape for the displacement field.
The slowness contour shown in panel (b) arises from the mode associated with the TIR.
Again, we observe that the displacement field shown in panel (d) is consistent with the associated slowness contour and waves propagate in the directions parallel to normals of the slowness contours.

We now proceed to consider the scattering problem associated with a thin strip of triangular lattice with TIRs embedded within an ambient triangular lattice.
Fig. \ref{fig:negative_refraction}(a) shows the slowness curves which also appear in Figs \ref{fig:harm_loading}(a) and \ref{fig:harm_loading}(b), at $\nu=\pi$.
We observe that the two hexagonal slowness contours are rotated, by $\pi/6$, with respect to each other.
Additionally, the group velocity associated associated with these slowness contours is negative (positive) for the lattice with TIRs (ambient lattice), resulting in a negative (positive) effective mass.
These two properties can be used to design a flat lens capable of utilising negative refraction to focus elastic waves in a triangular lattice.
This effect is illustrated in Fig. \ref{fig:negative_refraction}(b).
The excitation source is the same as the one used to generate Fig. \ref{fig:harm_loading}(c) and the parameter values are described in Table \ref{tab:parameters_comsol}.

At the chosen frequency $\nu=\pi~{\rm rad/s}$, exciting the triangular lattice with a point source results in a cross-like displacement field, as already discussed for Fig. \ref{fig:harm_loading}(c);
the angle between the beams and the $x$-axis is $\pi/6$.
This is consistent with the slowness contours for the ambient lattice shown in Fig. \ref{fig:negative_refraction}(a).
The slowness contours for the interface are rotated by $\pi/6$ with respect to those for the ambient lattice, as shown in Fig. \ref{fig:negative_refraction}(a).
This explains the negative refraction evident at the interfaces between the triangular lattice and the lattice with TIRs, shown in panel (b).
A virtual image of the source can be observed on the right side of the ambient lattice.
The formation of the virtual image is illustrated in the inset diagram.
Panel (b) is obtained locating the point source twenty lattice spaces away from the left interface of the lens.
In panel (c), we move the source closer (ten lattice spaces).
We observe that the position of the virtual image moves closer to the right interface of the flat interface (see also inset diagram).
Finally, in panel (d), we locate the source one lattice site away from the flat lens. This results in the formation of a \emph{real} image of the source on the right side of the diagram, and demonstrates focussing of elastic waves by means of negative refraction.  
\section{Conclusions \label{sec:conclusions}}
In this paper we have introduced and studied a new class of chiral lattice systems which exhibit a range of interesting dynamic properties, including dynamic anisotropy, negative refraction, filtering and focusing of elastic waves.
The chiral lattice is created by the introduction of a periodic array of triangular resonators embedded within an infinite uniform triangular lattice; the chirality arises as a result of \emph{tilting} the resonator and thus breaking the mirror symmetry of the ambient lattice.
The introduction of the resonator results in the appearance of several additional modes that interact with the modes corresponding to the triangular lattice without resonators.
In particular, the resonators give rise to a novel rotationally dominant mode, which we refer to as the ``chiral branch''.
We use this chiral branch to design and implement a flat lens for mechanical waves in a triangular lattice.
Indeed, the chiral branch corresponds to the negative refraction and the flat lens illustrated in Fig. \ref{fig:negative_refraction}.

The effect of the resonators on the dispersive properties of the infinite chiral lattice system is examined in detail.
In particular, the tilting angle of the resonator is found to be a useful tuning parameter and can be used in the control and optimisation of pass and stop bands, as well as the creation of resonances in the band structure of the lattice.
The effects of the resonators are most significant in the dynamic regime and give rise to many interesting phenomena, such as dynamic anisotropy, spatial localisation and Dirac cones.
We show that the effective material properties of the lattice in the long-wave regime are unchanged by the presence of the resonator.
The effect of introducing a non-uniform tilting angle has also been examined and it has been shown that vortex-like modes can be obtained by alternating the tilting angle from one cell to the next.

%In section \ref{sec:model_resonator}, we have shown that a non-tilted resonator considered in the tension and compression scheme is statically undetermined. W have also demonstrated that a finite bending stiffness of the links cures the degeneracy. Therefore, a lattice whose links a structure whose links 
\section*{Acknowledgements}
D.T. gratefully acknowledges the People Programme (Marie Curie Actions) of the European Union's Seventh Framework Programme FP7/2007-2013/ under REA grant agreement number PITN-GA-2013- 606878.
A.B.M. and N.V.M. acknowledge the financial support of the EPSRC through programme grant EP/L024926/1.
\begin{appendices}
	\numberwithin{equation}{section}
	\section{Derivation of the stiffness matrices using soft resonators \label{sec:app_soft}}
	Here we derive for the Bloch-Floquet problem in Eq. (\ref{eq:Sigma_k_prime}) and obtain the stiffness matrix for a single hinged resonator in Eq. (\ref{eq:sigma_prime_hinged}) as a limiting case. 
	The displacements of the masses which compose the TIR (black solid dots in Fig. \ref{fig:system_t}(a)) obey the following Newton's equations 
	\begin{eqnarray}\label{eq:newton_u_i_t_soft}
	m_o\omega^2 {\bm u}^{({\bm n})}_{\rm 1}({\bm k}) &=&
	c_{\ell o}~\hat{\Pi}_1
	\left({{\bm u}}^{({\bm n})}_{1}({\bm k})- {{\bm u}}^{({\bm n}+\bf{e}_2)}_{0}({\bm k}) \right) +c_{o}\hat{\pi}_3\left({{\bm u}}^{({\bm n})}_{1}({\bm k})-{{\bm u}}^{({\bm n})}_{2}({\bm k})\right)+c_{o}\hat{\pi}_2\left({{\bm u}}^{({\bm n})}_{1}({\bm k})-{{\bm u}}^{({\bm n})}_{3}({\bm k})\right),\nonumber \\
	m_o\omega^2 {\bm u}^{({\bm n})}_{\rm 2}({\bm k}) &=&
	c_{\ell o}~\hat{\Pi}_2
	\left( {{\bm u}}^{({\bm n})}_{2}({\bm k})- {{\bm u}}^{({\bm n}+\bf{e}_1)}_{0}({\bm k}) \right) +c_{o}\hat{\pi}_3\left({{\bm u}}^{({\bm n})}_{2}({\bm k})-{{\bm u}}^{({\bm n})}_{1}({\bm k})\right)+c_{o}\hat{\pi}_1\left({{\bm u}}^{({\bm n})}_{2}({\bm k})-{{\bm u}}^{({\bm n})}_{3}({\bm k})\right),\nonumber \\
	m_o\omega^2 {\bm u}^{({\bm n})}_3({\bm k}) &=&
	c_{\ell o}~\hat{\Pi}_3
	\left( {{\bm u}}^{({\bm n})}_{3}({\bm k})- {{\bm u}}^{({\bm n})}_{0}({\bm k}) \right) +c_{o}\hat{\pi}_2\left({{\bm u}}^{({\bm n})}_3({\bm k})-{{\bm u}}^{({\bm n})}_{1}({\bm k})\right)+c_{o}\hat{\pi}_1\left({{\bm u}}^{({\bm n})}_3({\bm k})-{{\bm u}}^{({\bm n})}_2({\bm k})\right).\nonumber\\
	\end{eqnarray}
	The displacement for the nodal point  ${\bm n}$ of the triangular lattice (see empty circles in Fig. \ref{fig:system_t}(a)) is
	\begin{eqnarray}\label{eq:newton_u_0_t_soft}
	-m\omega^2{{\bm u}}^{({\bm n})}_{0}({\bm k}) &=&
	c_{ \ell}~ \hat{\tau}_1 
	\left( {{\bm u}}^{({\bm n}+\bf{e}_1)}_{0}({\bm k}) +{{\bm u}}^{({\bm n}-{\bf e}_{1})}_{0}({\bm k})-2{{\bm u}}^{({\bm n})}_{0}({\bm k})\right)\nonumber \\
	&+&c_{ \ell}~ \hat{\tau}_2 
	\left( {{\bm u}}^{({\bm n}+\bf{e}_2)}_{0}({\bm k}) +{{\bm u}}^{({\bm n}-{\bf e}_{2})}_{0}({\bm k})-2{{\bm u}}^{({\bm n})}_{0}({\bm k})\right)\nonumber \\
	&+&c_{ \ell}~ \hat{\tau}_3
	\left( {{\bm u}}^{({\bm n}+\bf{e}_1-\bf{e}_2)}_{0}({\bm k}) +{{\bm u}}^{({\bm n}+{\bf e}_{2}-{\bf e}_{1})}_{0}({\bm k})-2{{\bm u}}^{({\bm n})}_{0}({\bm k})\right)\nonumber \\
	&+&c_{\ell o}~ \hat{\Pi}_1
	\left( {{\bm u}}^{({\bm n}-\bf{e}_2)}_{1}({\bm k}) -{{\bm u}}^{({\bm n})}_{\rm 0}({\bm k})
	\right)\nonumber\\
	&+&c_{\ell o}~ \hat{\Pi}_2
	\left( {{\bm u}}^{({\bm n}-\bf{e}_1)}_{2}({\bm k}) -{{\bm u}}^{({\bm n})}_{\rm 0}({\bm k})
	\right)\nonumber\\
	&+&c_{\ell o}~ \hat{\Pi}_3
	\left( {{\bm u}}^{({\bm n})}_{3}({\bm k}) -{{\bm u}}^{({\bm n})}_{\rm 0}({\bm k})
	\right).
	\end{eqnarray}
	In Eqs. (\ref{eq:newton_u_i_t_soft}) and (\ref{eq:newton_u_0_t_soft}) we introduce the notation ${\bm e}_1=(1,0)^{\rm T}$ and ${\bm e}_2=(0,1)^{\rm T}$. Bloch-Floquet conditions for the displacements are 
	\begin{equation}\label{eq:BF_soft}
	{\bm u}_i^{({\bm n} +{\bm m})}(\bm k)=e^{i {\bm k}\cdot\hat{\cal T}{\bm m}}	{\bm u}_i^{({\bm n})}(\bm k),
	\end{equation}
	where $i=\{0,1,2,3\}$ and $\hat{\cal T }$ is the matrix whose columns coincide with the primitive vectors of the triangular lattice. After imposing the conditions (\ref{eq:BF_soft}), Eqs. (\ref{eq:newton_u_i_t_soft}) and (\ref{eq:newton_u_0_t_soft}) become
	\begin{eqnarray}\label{eq:newton_u_i_t__BF_soft}
	m_o\omega^2 {\bm u}_{\rm 1}({\bm k}) &=&
	c_{\ell o}~\hat{\Pi}_1
	\left({{\bm u}}_{1}({\bm k})- e^{i{\bm k}\cdot{\bm t}_2}{{\bm u}}_{0}({\bm k}) \right) +c_{o}\hat{\pi}_3\left({{\bm u}}_{1}({\bm k})-{{\bm u}}_{2}({\bm k})\right)+c_{o}\hat{\pi}_2\left({{\bm u}}_{1}({\bm k})-{{\bm u}}_{3}({\bm k})\right),\nonumber \\
	m_o\omega^2 {\bm u}_{\rm 2}({\bm k}) &=&
	c_{\ell o}~\hat{\Pi}_2
	\left( {{\bm u}}_{2}({\bm k})- e^{i{\bm k}\cdot{\bm t}_1}{{\bm u}}_{0}({\bm k}) \right) +c_{o}\hat{\pi}_3\left({{\bm u}}_{2}({\bm k})-{{\bm u}}^{({\bm n})}_{1}({\bm k})\right)+c_{o}\hat{\pi}_1\left({{\bm u}}_{2}({\bm k})-{{\bm u}}_{3}({\bm k})\right),\nonumber \\
	m_o\omega^2 {\bm u}_3({\bm k}) &=&
	c_{\ell o}~\hat{\Pi}_3
	\left( {{\bm u}}_{3}({\bm k})- {{\bm u}}_{0}({\bm k}) \right) +c_{o}\hat{\pi}_2\left({{\bm u}}_3({\bm k})-{{\bm u}}_{1}({\bm k})\right)+c_{o}\hat{\pi}_1\left({{\bm u}}_3({\bm k})-{{\bm u}}_2({\bm k})\right).\nonumber\\
	\end{eqnarray}
	and
	\begin{eqnarray}\label{eq:newton_u_0_t_soft_BF}
	-m\omega^2 {{\bm u}}_{0}({\bm k}) &=&
	2c_{ \ell}~ \hat{\tau}_1\left( \cos({\bm k}\cdot{\bm t}_1)-1\right){{\bm u}}_{0}({\bm k})\nonumber \\
	&+&2c_{ \ell}~ \hat{\tau}_2
	\left( \cos({\bm k}\cdot{\bm t}_2)-1\right){{\bm u}}_{0}({\bm k})\nonumber \\
	&+&2c_{ \ell}~ \hat{\tau}_3
	\left( \cos({\bm k}\cdot({\bm t}_1-{\bm t}_2))-1\right){{\bm u}}_{0}({\bm k})\nonumber \\
	&+&c_{\ell o}~ \hat{\Pi}_1
	\left(    e^{-i{\bm k}\cdot{\bm t}_2}   {{\bm u}}_{1}({\bm k})-{{\bm u}}_{\rm 0}({\bm k})
	\right)\nonumber \\
	&+&c_{\ell o}~ \hat{\Pi}_2
	\left(    e^{-i{\bm k}\cdot{\bm t}_1}   {{\bm u}}_{\rm 2}({\bm k})     -{{\bm u}}_{\rm 0}({\bm k})
	\right)\nonumber \\
	&+&c_{\ell o}~ \hat{\Pi}_3
	\left(    {{\bm u}}_{\rm 3}({\bm k})-{{\bm u}}_{\rm 0}({\bm k})
	\right).
	\end{eqnarray}
	From Eqs. (\ref{eq:newton_u_0_t_soft_BF}) and (\ref{eq:newton_u_0_t_soft_BF}) follows the secular equation (\ref{eq:secular_prime}) with the stiffness matrix given in Eq. (\ref{eq:Sigma_k_prime}). Moreover the limit $m\rightarrow+\infty$ in Eq. (\ref{eq:newton_u_0_t_soft_BF}) implies ${\bm u}_0({\bm k})=0$, that is hinged conditions, and the secular equation (\ref{eq:secular_hinged}) with stiffness matrix (\ref{eq:sigma_prime_hinged}) is obtained.
	\section{Calculation of the rotational frequency of a single rigid resonator with flexible links \label{sec:app_single_bending}}
	We  study a model for a single resonator of the type shown in Fig. \ref{fig:system_sr}(a). We allow the links between the hinges and the TIR to be flexible beams, with a given bending stiffness $B$.  Moreover, we assume that the longitudinal elongations of the links are negligible. The TIR is assumed to be a rigid-body. We fix the $y$-axis of a  Cartesian coordinate system to be parallel to the top link in Fig. \ref{fig:system_sr}(a), pointing downwards. Moreover, we fix the origin to coincide with the top hinge. By symmetry considerations, each link produces the same torque on the rigid resonator. Hence, without loss of generality, it is sufficient to study a single beam, \emph{e.g.} the one along the $y$-axis. The massless beam assumption implies that implies that the flexural displacement is
	\begin{equation}\label{eq:w}
	{\bm w}(y)=w(y)\hat{\bm x}= \left( {\cal A}_0+{\cal A}_1 y + {\cal A}_2 y^2 + {\cal A}_3 y^3\right) \hat{\bm x},
	\end{equation}
	where $\hat{\bm x}$ denote the unit vector along the $x$-axis and ${\cal A}_i$ are real constants to be determined from the conditions at the boundaries of the beam. Hinged junction conditions at $y=0$ read
	\begin{equation}
	w(0)=w''(0)=0\Longrightarrow{\cal A}_0={\cal A}_2=0.
	\end{equation}
	We denote by $\vartheta(t)$ the angular oscillation of the resonator and assume positive angles in the clockwise direction. At the junction between the resonator and the beam, clamped conditions apply. These conditions lead to 
	\begin{equation}\nonumber
	\left.w(y)\right|_{y=(L-\ell)/\sqrt{3}}\simeq \frac{\vartheta(t)\ell}{\sqrt{3}}\,\,\,\,\,\,{\rm and}\,\,\,\,\,\,\,\,\left.w'(y)\right|_{y=(L-\ell)/\sqrt{3}}=-\vartheta(t),
	\end{equation}
	which imply 
	\begin{equation}\label{eq:clamped}
	{\cal A}_3=-\frac{3}{2} \frac{L}{(L-\ell)^3}\vartheta(t)\,\,\,\,\,\,{\rm and}\,\,\,\,\,\,\,\,\,{\cal A }_1=\frac{L/2+\ell}{L-\ell} \vartheta(t).
	\end{equation}
	Finally, the Newton equation for the small angular oscillations of the resonator is 
	\begin{equation}
	I\frac{d^2\vartheta(t)}{dt^2}=-\frac{9\sqrt{3}LB}{(L-\ell)^2} \vartheta(t),
	\end{equation}
	which has harmonic solutions of angular frequency squared given in Eq. (\ref{eq:eigf_squared_bending}).
	\section{Derivation of the stiffness matrices using rigid resonators\label{sec:app_rigid}}
	Here we derive the stiffness matrices in Eq. (\ref{eq:Sigma_k}) and obtain the stiffness matrix for a single hinged resonators  whose eigenfrequencies are listed in Eq. (\ref{eq:eigs_rigid}). 
	%For convenience we rewrite here Eq. (\ref{eq:var_cm_t}) 
	%\begin{equation}\label{eq:app_var_cm_t}
	%{\bm u}_{i}^{({\bm n})}({\bm k})={\bm u}_{\rm cm}({\bm k})+ {\bm b}^{(1)}_i\vartheta^{({\bm n})}({\bm k})
	%\end{equation} 
	
	Time-harmonic propagation of in-plane elastic  waves around the equilibrium positions given in Eqs. (\ref{eq:u_eq_lat_t}) and (\ref{eq:u_eq_res_t}) is considered. The  displacements around the equilibrium positions   are identified by ${\bm u}^{({\bm n})}_{i}({\bm k})$, where the index $i=\{0,1,2,3\}$ runs over the masses of a given unit cell $({\bm n})$. The displacements of the masses which compose the TIR (black solid dots in Fig. \ref{fig:system_t}(a)) obey the following Newton's equations 
	\begin{eqnarray}\label{eq:newton_u_i_t}
	-m_o\omega^2 {\bm u}^{({\bm n})}_{\rm 1}({\bm k}) &=&
	c_{\ell o}~\hat{\Pi}_1
	\left( {{\bm u}}^{({\bm n}+\bf{e}_2)}_{0}({\bm k}) -{{\bm u}}^{({\bm n})}_{1}({\bm k})\right)-{\bf s}_{12}({\bm k}) +{\bf s}_{31}({\bm k}),\nonumber \\
	-m_o \omega^2{{\bm u}}^{({\bm n})}_{\rm 2}({\bm k})  &=&c_{\ell o}~\hat{\Pi}_2
	\left( {{\bm u}}^{({\bm n}+\bf{e}_1)}_{0}({\bm k}) -{{\bm u}}^{({\bm n})}_{2}({\bm k})\right)+{\bf s}_{12}({\bm k}) -{\bf s}_{23}({\bm k}),\nonumber \\
	-m_o \omega^2{{\bm u}}^{({\bm n})}_{\rm 3}({\bm k})  &=&c_{\ell o}~\hat{\Pi}_3
	\left( {{\bm u}}^{({\bm n})}_{0}({\bm k}) -{{\bm u}}^{({\bm n})}_{3}({\bm k})\right)+{\bf s}_{23}({\bm k}) -{\bf s}_{31}({\bm k}).
	\end{eqnarray}
	In Eq. (\ref{eq:newton_u_i_t}) we introduce the force ${\bf s}_{i j}({\bm k})$ on mass  $i$ due to mass $j\neq i$ with $i,j=\{1,2,3\}$. Since ${\bf s}_{ij}({\bm k})=-{\bf s}_{ji}({\bm k})$, it follows that the tension forces can be rewritten as 
	\begin{equation}\label{eq:tensions}
	{\bf s}_{12}({\bm k})=\frac{s_1({\bm k})}{\ell}(\tilde{\bf b}_2-\tilde{\bf b}_1), \,\,\,\,
	{\bf s}_{23}({\bm k})=\frac{s_2({\bm k})}{\ell}(\tilde{\bf b}_3-\tilde{\bf b}_2),\,\,\,\,
	{\bf s}_{31}({\bm k})=\frac{s_3({\bm k})}{\ell}(\tilde{\bf b}_1-\tilde{\bf b}_3),
	\end{equation} 
	where $s_i({\bm k})=|{\bf s}_{ij}({\bm k})|$ are the moduli of the tension forces along the equilibrium position of the trusses. Moreover, in Eq. (\ref{eq:newton_u_i_t}) we introduce the notation ${\bm e}_1=(1,0)^{\rm T}$ and ${\bm e}_2=(0,1)^{\rm T}$.
	
	The displacement for the nodal point  ${\bm n}$ of the triangular lattice (see empty circles in Fig. \ref{fig:system_t}(a)) is
	\begin{eqnarray}\label{eq:newton_u_0_t}
	-m\omega^2{{\bm u}}^{({\bm n})}_{0}({\bm k}) &=&
	c_{ \ell}~ \hat{\tau}_1 
	\left( {{\bm u}}^{({\bm n}+\bf{e}_1)}_{0}({\bm k}) +{{\bm u}}^{({\bm n}-{\bf e}_{1})}_{0}({\bm k})-2{{\bm u}}^{({\bm n})}_{0}({\bm k})\right)\nonumber \\
	&+&c_{ \ell}~ \hat{\tau}_2 
	\left( {{\bm u}}^{({\bm n}+\bf{e}_2)}_{0}({\bm k}) +{{\bm u}}^{({\bm n}-{\bf e}_{2})}_{0}({\bm k})-2{{\bm u}}^{({\bm n})}_{0}({\bm k})\right)\nonumber \\
	&+&c_{ \ell}~ \hat{\tau}_3
	\left( {{\bm u}}^{({\bm n}+\bf{e}_1-\bf{e}_2)}_{0}({\bm k}) +{{\bm u}}^{({\bm n}+{\bf e}_{2}-{\bf e}_{1})}_{0}({\bm k})-2{{\bm u}}^{({\bm n})}_{0}({\bm k})\right)\nonumber \\
	&+&c_{\ell o}~ \hat{\Pi}_1
	\left( {{\bm u}}^{({\bm n}-\bf{e}_2)}_{1}({\bm k}) -{{\bm u}}^{({\bm n})}_{\rm 0}({\bm k})
	\right)\nonumber\\
	&+&c_{\ell o}~ \hat{\Pi}_2
	\left( {{\bm u}}^{({\bm n}-\bf{e}_1)}_{2}({\bm k}) -{{\bm u}}^{({\bm n})}_{\rm 0}({\bm k})
	\right)\nonumber\\
	&+&c_{\ell o}~ \hat{\Pi}_3
	\left( {{\bm u}}^{({\bm n})}_{3}({\bm k}) -{{\bm u}}^{({\bm n})}_{\rm 0}({\bm k})
	\right).
	\end{eqnarray}
	We now eliminate the tension forces  from Eqs. (\ref{eq:newton_u_i_t}), by  
	summing term-by-term Eqs. (\ref{eq:newton_u_i_t}). Moreover, we use Eq. (\ref{eq:var_cm_t}). The result is
	\begin{eqnarray}\label{eq:newton_u_cm_t}
	-\omega^2M {{\bm u}}^{({\bm n})}_{\rm cm}({\bm k}) &=&
	c_{\ell o}~\hat{\Pi}_1
	\left( {{\bm u}}^{({\bm n}+\bf{e}_2)}_{0}({\bm k}) -{{\bm u}}^{({\bm n})}_{\rm cm}({\bm k})-\hat{\cal R}'_1\tilde{\bm b}_1\vartheta_{{\bm n}}({\bm k})
	\right)\nonumber \\
	&+&c_{\ell o}~ \hat{\Pi}_2
	\left( {{\bm u}}^{({\bm n}+\bf{e}_1)}_{0}({\bm k}) -{{\bm u}}^{({\bm n})}_{\rm cm}({\bm k})-\hat{\cal R}'_2\tilde{\bm b}_1\vartheta_{{\bm n}}({\bm k})
	\right)\nonumber\\
	&+&c_{\ell o}~\hat{\Pi}_3
	\left( {{\bm u}}^{({\bm n})}_{0}({\bm k}) -{{\bm u}}^{({\bm n})}_{\rm cm}({\bm k})-\hat{\cal R}'_3\tilde{\bm b}_1\vartheta_{{\bm n}}({\bm k})
	\right).
	\end{eqnarray}
	where $M=3m_o$ is the total mass of the TIR.  
	
	We now project each of the equations in (\ref{eq:newton_u_i_t}) on the vectors $\tilde{{\bf b}}_3-\tilde{{\bf b}}_2$,$\tilde{{\bf b}}_1-\tilde{{\bf b}}_3$ and $\tilde{{\bf b}}_2-\tilde{{\bf b}}_1$, respectively. By summing the resulting equations and performing algebraic simplifications, we  obtain 
	\begin{eqnarray}\label{eq:newton_theta_t}
	-\omega^2 I \vartheta_{{\bm n}}({\bm k})&=&- \frac{1}{\sqrt{3}}c_{\ell o}~ ({\tilde{ {\bf b} }}_3-{\tilde{ {\bf b} }}_2)^{\rm T}  \hat{\Pi}_1
	\left( {{\bm u}}^{({\bm n}+\bf{e}_2)}_{0}({\bm k}) -{{\bm u}}^{({\bm n})}_{\rm cm}({\bm k})-\hat{\cal R}'_1\tilde{\bm b}_1\vartheta_{{\bm n}}({\bm k})
	\right)\nonumber \\
	&-& \frac{1}{\sqrt{3}}c_{\ell o}~ ({\tilde{ {\bf b} }}_1-{\tilde{ {\bf b} }}_3)^{\rm T} \hat{\Pi}_2
	\left( {{\bm u}}^{({\bm n}+\bf{e}_1)}_{0}({\bm k}) -{{\bm u}}^{({\bm n})}_{\rm cm}({\bm k})-\hat{\cal R}'_2\tilde{\bm b}_1\vartheta_{{\bm n}}({\bm k})
	\right)\nonumber\\
	&-& \frac{1}{\sqrt{3}}c_{\ell o}~({\tilde{ {\bf b} }}_2-{\tilde{ {\bf b} }}_1)^{\rm T}\hat{\Pi}_3
	\left( {{\bm u}}^{({\bm n})}_{0}({\bm k}) -\hat{\cal R}'_3\tilde{\bm b}_1\vartheta_{{\bm n}}({\bm k}),
	\right)
	\end{eqnarray}
	where $I=3m_ob^2=m_{o}\ell^2$ is the moment of inertia with respect to the centre of mass reference frame.

	Bloch-Floquet conditions on the displacement fields are
	\begin{equation}\label{eq:BF}
	{{\bm u}}^{({\bm n}+{\bm m})}_{\rm cm}({\bm k}) = e^{i{\bm k}\cdot\hat{\cal T}\bm{m}}{{\bm u}}^{({\bm n})}_{\rm cm}({\bm k}) ,\,\,\,\,{{\bm u}}^{({\bm n}+{\bm m})}_{\rm 0}({\bm k}) = e^{i{\bm k}\cdot\hat{\cal T}{\bm m}}{{\bm u}}^{({\bm n})}_{\rm 0}({\bm k}),\,\,\,\,{\vartheta}_
	{{\bm n}+{\bm m}}({\bm k}) = e^{i{\bm k}\cdot\hat{\cal T}{\bm m}}{\vartheta}_
	{{\bm n}}({\bm k}),
	\end{equation}
	where the matrix $\hat{\cal T}$ has as column vectors the unit vectors of the lattice and ${\bm m}$ and ${\bm n}$ are integer vectors. 
	
	By using the Bloch-Floquet conditions in Eq. (\ref{eq:BF}), Eqs. (\ref{eq:newton_u_0_t}), (\ref{eq:newton_u_cm_t}),  and (\ref{eq:newton_theta_t}) become respectively
	\begin{eqnarray}\label{eq:newton_u_0_t_BF}
	-m\omega^2 {{\bm u}}_{0}({\bm k}) &=&
	2c_{ \ell}~ \hat{\tau}_1\left( \cos({\bm k}{\bm t}_1)-1\right){{\bm u}}_{0}({\bm k})\nonumber \\
	&+&2c_{ \ell}~ \hat{\tau}_2
	\left( \cos({\bm k}\cdot{\bm t}_2)-1\right){{\bm u}}_{0}({\bm k})\nonumber \\
	&+&2c_{ \ell}~ \hat{\tau}_3
	\left( \cos({\bm k}\cdot({\bm t}_1-{\bm t}_2))-1\right){{\bm u}}_{0}({\bm k})\nonumber \\
	&+&c_{\ell o}~ \hat{\Pi}_1
	\left(    e^{-i{\bm k}\cdot{\bm t}_2}   {{\bm u}}_{\rm cm}({\bm k})     +    e^{-i{\bm k}\cdot{\bm t}_2}\hat{\cal R}'_1\tilde{\bm b}_1\vartheta({\bm k})-{{\bm u}}_{\rm 0}({\bm k})
	\right)\nonumber \\
	&+&c_{\ell o}~ \hat{\Pi}_2
	\left(    e^{-i{\bm k}\cdot{\bm t}_1}   {{\bm u}}_{\rm cm}({\bm k})     +    e^{-i{\bm k}\cdot{\bm t}_1}\hat{\cal R}'_2\tilde{\bm b}_1 \vartheta({\bm k})-{{\bm u}}_{\rm 0}({\bm k})
	\right)\nonumber \\
	&+&c_{\ell o}~ \hat{\Pi}_3
	\left(    {{\bm u}}_{\rm cm}({\bm k})     +   \hat{\cal R}'_3\tilde{\bm b}_1 \vartheta({\bm k})-{{\bm u}}_{\rm 0}({\bm k})
	\right),
	\end{eqnarray}
	\begin{eqnarray}\label{eq:newton_u_cm_t_BF}
	-M \omega^2{\bm u}_{\rm cm}({\bm k}) &=&
	c_{\ell o}~\hat{\Pi}_1
	\left( {e^{i{\bm k}{\bm t}_2}{\bm u}}_{0}({\bm k}) -{{\bm u}}_{\rm cm}({\bm k})-\hat{\cal R}'_1\tilde{\bm b}_1\vartheta({\bm k})
	\right)\nonumber \\&+&
	c_{\ell o}~\hat{\Pi}_2 
	\left( {e^{i{\bm k}{\bm t}_1}{\bm u}}_{0}({\bm k}) -{{\bm u}}_{\rm cm}({\bm k})-\hat{\cal R}'_2\tilde{\bm b}_1\vartheta({\bm k})
	\right)\nonumber\\&+&
	c_{\ell o}~ \hat{\Pi}_3\left( {{\bm u}}_{0}({\bm k}) -{{\bm u}}_{\rm cm}({\bm k})-\hat{\cal R}'_3\tilde{\bm b}_1\vartheta({\bm k})
	\right)\,\,\,\,\,,
	\end{eqnarray}
	and 
	\begin{eqnarray}\label{eq:newton_theta_t_BF}
	\omega^2I\vartheta({\bm k})&=& \frac{1}{\sqrt{3}}c_{\ell o}~ ({\tilde{ {\bf b} }}_3-{\tilde{ {\bf b} }}_2)^{\rm T} ~\hat{\Pi}_1
	\left( {e^{i{\bm k}{\bm t}_2}{\bm u}}_{0}({\bm k}) -{{\bm u}}_{\rm cm}({\bm k})-\hat{\cal R}'_1\tilde{\bm b}_1\vartheta({\bm k})
	\right)\nonumber \\
	&+& \frac{1}{\sqrt{3}}c_{\ell o}~ ({\tilde{ {\bf b} }}_1-{\tilde{ {\bf b} }}_3)^{\rm T} ~\hat{\Pi}_2
	\left( {e^{i{\bm k}{\bm t}_1}{\bm u}}_{0}({\bm k}) -{{\bm u}}_{\rm cm}({\bm k})-\hat{\cal R}'_2\tilde{\bm b}_1\vartheta({\bm k})
	\right)\nonumber\\
	&+& \frac{1}{\sqrt{3}}c_{\ell o}~ ({\tilde{ {\bf b} }}_2-{\tilde{ {\bf b} }}_1)^{\rm T}  ~\hat{\Pi}_3
	\left( {{\bm u}}_{0}({\bm k}) -{{\bm u}}_{\rm cm}({\bm k})-\hat{\cal R}'_3\tilde{\bm b}_1\vartheta({\bm k})
	\right),
	\end{eqnarray}
	where we suppress the cell index $({\bm n})$ after using the Bloch-Floquet conditions (\ref{eq:BF}). Eq. (\ref{eq:newton_u_0_t_BF}), (\ref{eq:newton_u_cm_t_BF}) and (\ref{eq:newton_theta_t_BF}) imply the matrix equation  (\ref{eq:secular_BF}). In particular, by using 
	\begin{equation}
	\hat{\cal R}_3-\hat{\cal R}_2=-\sqrt{3}\hat{\cal R}_1',\,\,\,\,
	\hat{\cal R}_2-\hat{\cal R}_1=-\sqrt{3}\hat{\cal R}_3'\,\,\,{\rm and}\,\,\,\hat{\cal R}_1-\hat{\cal R}_3=-\sqrt{3}\hat{\cal R}_2',
	\end{equation}
	in Eq. \eqref{eq:newton_theta_t_BF}, it follows the expression for  the stiffness matrix (\ref{eq:Sigma_k}). In addition, the limit $m\rightarrow+\infty$ of Eq. ({\ref{eq:secular_BF}}Eqs. (\ref{eq:newton_u_0_t_BF}) leads the eigenvalue problem for a single hinged resonator (see Fig.  \ref{fig:system_sr}(a)), whose natural frequencies are given in Eq. (\ref{eq:eigs_rigid}). 
	\section{Derivation of the effective group velocities \label{app:taylor}}
	The determinant of the matrix equation (\ref{eq:secular_BF}) is a polynomial function of fifth degree in the variable $\omega^2$, \emph{i.e.}
	\begin{equation}\label{eq:polynom_dispersion}
	{\mathcal D}(\bm{k},\omega)=\sum_{j=0}^5 {\mathcal D}^{(2j)}({\bm k})\cdot(\omega)^{2j},
	\end{equation}  
	where the coefficients ${\mathcal D}^{(2j)}({\bm k})$ are analytical functions of $\bm k$. From  analyticity it follows that the order in which the ${\bm k}\rightarrow0$ and $\omega\rightarrow0$ is performed do not affect the result.
	We deliberately assume $\vartheta_0\neq0$, \emph{i.e.} a statically determined lattice. In Eq. (\ref{eq:polynom_dispersion}) we retain only the terms up to fourth order in $\omega$ and  expand  ${\mathcal D}^{(2j)}({\bm k})$  in Maclaurin series for small ${\bm k}$ around $\Gamma$. To leading order, the coefficients of Eq.  (\ref{eq:disp_eq_taylor}) are 
	\begin{align}\label{eq:coeff_Gamma}
	{\cal D}_{\Gamma}^{(0)}({\bm k}) & =-\frac{3^5}{2^8}\frac{L^6 \ell^2}{\ell_r^2} c_\ell^2c_{\ell o}^3\sin^2(\vartheta_0)\left| {\bm k}\right|^4, \nonumber \\
	{\cal D}_{\Gamma}^{(2)}({\bm k})&=\frac{3^3}{2^3}\frac{L^4\ell^2}{\ell_r^2}c_{\ell}c_{\ell o}^3\sin^2(\vartheta_0)(m + 3m_o)\left|{\bm k}\right|^2,\,\,\,\,{\rm and}\nonumber\\
	{\cal D}_{\Gamma}^{(4)}({\bm k}) & = %-\frac{3^3}{2^4}\frac{\ell^2L^2}{\ell_r^2}c_{\ell o}^2m_o[( 2\ell_r^2+8 L^2 \sin^2\vartheta_0) c_\ell m
	%+ 6\ell_r^2 c_l m_o +(  \ell_r^2 - L^2 +10L^2 \sin^2\vartheta_0)c_{\ell o} m_o ](k_x^2+k_y^2)\nonumber\\
	-\frac{3^2}{2^2}\frac{L^2\ell^2}{\ell_r^2}  c_{\ell o}^3 \sin^2\vartheta_0(m+3m_o)^2.
	\end{align}
	In Eqs. (\ref{eq:coeff_Gamma}), the index ``$\Gamma$" indicates  Taylor expansion in ${\bm k}$ around $\Gamma$.The effective long-waves group velocities in Eq. \eqref{eq:eff_v} follow from
	\begin{equation}v^2=\frac{1}{2|{\bm k}|^2{\cal D}_{\Gamma}^{(4)}({\bm k})}\left(-{\cal D}_{\Gamma}^{(2)}({\bm k})\pm\sqrt{({\cal D}_{\Gamma}^{(2)}({\bm k}))^2-4{\cal D}_{\Gamma}^{(4)}({\bm k}){\cal D}_{\Gamma}^{(0)}({\bm k})}\right).
	\end{equation}
\end{appendices}

\end{document}